\documentclass[11pt,a4paper]{article}
\usepackage{jheppub,epstopdf,bbm} 

\usepackage{amsmath,amssymb,epsfig,graphicx}

\makeatletter
\newcommand{\fmslash}[2][0mu]{
\mathchoice
    {\fmsl@sh\displaystyle{#1}{#2}}%
        {\fmsl@sh\textstyle{#1}{#2}}%
    {\fmsl@sh\scriptstyle{#1}{#2}}%

    {\fmsl@sh\scriptscriptstyle{#1}{#2}}}
\newcommand{\fmsl@sh}[3]{%
\m@th\ooalign{$\hfil#1\mkern#2/\hfil$\crcr$#1#3$}}

\newcommand{\tr}{\hbox{tr}}

\title{\center{Quantum duality under the $\theta$-exact\\ Seiberg-Witten map}}

\author[a,1]{Carmelo P. Mart\'{\i}n
\note{carmelop@fis.ucm.es}}
\affiliation[a]{Departamento de F\'{\i}sica Te\'{o}rica I, Facultad de Ciencias F\'{\i}sicas,
Universidad Complutense de Madrid, 28040-Madrid, Spain}
\author[b,c,2]{Josip Trampetic
\note{josip@irb.hr}}
\affiliation[b]{Rudjer Bo\v skovi\' c Institute, Division of Experimental Physics, P.O.Box 180, HR-10002 Zagreb, Croatia}
\affiliation[c]{Max-Planck-Institut f\"ur Physik, (Werner-Heisenberg-Institut), F\"ohringer Ring 6, D-80805 M\"unchen, Germany}
\author[d,3]{ and  Jiangyang You
\note{youjiangyang@gmail.com}}
\affiliation[c]{Rudjer Bo\v skovi\' c Institute, Divisions of Theoretical Physics, P.O.Box 180, HR-10002 Zagreb, Croatia}

\abstract{We show that in the perturbative regime defined by the coupling constant,  the $\theta$-exact Seiberg-Witten (SW) map applied to the noncommutative U(N) Yang-Mills--with or without Supersymmetry--gives an ordinary gauge theory which is, at the quantum   level, dual to the former. We do so by using the on-shell DeWitt effective action and dimensional regularization. We explicitly compute the one-loop two-point function  contribution to the on-shell DeWitt effective action of the  ordinary U(1) theory furnished by the $\theta$-exact Seiberg-Witten map. We find that the non-local UV divergences
found in the propagator in the Feynman gauge all but disappear, so that they are not physically relevant. We also show that the quadratic noncommutative IR divergences are gauge-fixing independent and go away in the Supersymmetric version of the U(1) theory.}

\keywords{Non-Commutative Geometry, Supersymmetry, Photon Physics}

\begin{document}

\maketitle

\section{Introduction}

It was found in \cite{Seiberg:1999vs} that noncommutative U(N) Yang-Mills  theories admit  two dual formulations at the classical level: one, in terms of noncommutative gauge fields, the other by using ordinary gauge fields. One moves between these two dual descriptions of the same theory by employing the so-called Seiberg-Witten (SW) map. This map maps ordinary U(N) gauge fields into noncommutative U(N) gauge fields and viceversa \cite{Mehen:2000vs,Jurco:2000fb}. Whether this duality persists at the quantum level is still an open issue, even at the one-loop level. Indeed, a key feature of noncommutative field theories defined by using noncommutative fields is the UV/IR mixing phenomenon unveiled in \cite{Martin:1999aq,Minwalla:1999px,Hayakawa:1999yt,Hayakawa:1999zf}; a phenomenon which cannot be seen by defining the action of the would-be dual ordinary theory as the expansion in the noncommutativity matrix, $\theta^{\mu\nu}$,  furnished by the Seiberg-Witten map, with the latter constructed as a formal power series in $\theta^{\mu\nu}$. It was not known how to reproduce the UV/IR mixing effect  in the formulation of noncommutative gauge theory  in terms of ordinary fields until the paper in \cite{Schupp:2008fs} was issued. In accord with the very essence of the perturbative coupling constant description of the quantum field theory our approach to the Seiberg-Witten map issue, is to built the SW map by using the expansion in terms of the coupling constant \cite{Schupp:2008fs,Trampetic:2015zma} (and no expansion in $\theta^{\mu\nu}$ is carried out). Thus, the $\theta^{\mu\nu}$ dependence of the perturbative, in the coupling constant, definition of the theory is treated in an exact way and, then, the UV/IR mixing effect pops up. The occurrence of the UV/IR mixing phenomenon in both these quantum field theories gives strong support to the idea that they are dual descriptions of the same underlying quantum field theory, at least in the perturbative  regime defined by the coupling constant. And yet, in U(1) Yang-Mills theory, the UV divergent part of the two-point function of the noncommutative gauge field is local, whereas the UV divergent bit of the two-point function of the ordinary theory obtained by using the $\theta$-exact Seiberg-Witten map contains unusual $\theta$-dependent nonlocal contributions, at least in the Feynman gauge. These nonlocal contributions where unearthed in \cite{Horvat:2011bs, Horvat:2013rga, Horvat:2015aca}, and their existence casts doubts on the truth of the quantum duality conjecture at hand. Of course, UV divergent contributions to the two-point function are, in general,  gauge dependent; so to decide whether the duality conjecture is right or wrong, there remains to be seen whether or not those nonlocal terms are really gauge dependent; since the gauge dependent contributions are not physically relevant.

It is known that in the $\theta$-unexpanded noncommutative nonsupersymmetric gauge theory defined in terms of the noncommutative fields, the noncommutative quadratic IR divergence induced by UV/IR mixing signals an IR instability \cite{Armoni:2001uw};  and that this IR instability can be cured by making the theory supersymmetric, since supersymmetry removes the corresponding quadratic noncommutative IR divergences \cite{Zanon:2000nq, Ruiz:2000hu}. Furthermore, it was shown in \cite{Martin:2008xa} that if the noncommutative fields carry a linear realization of Supersymmetry their ordinary duals under the Seiberg-Witten map carry a nonlinear realization of Supersymmetry. Hence, it is far from trivial that the Supersymmetry cancelation mechanism between the one-loop noncommutative quadratic IR divergences coming from bosonic and fermionic degrees of freedom works when the classical noncommutative theory is formulated, first, in terms of the ordinary fields and then quantized. And yet, it has been shown in \cite{Martin:2016zon} that the Supersymetry cancelation mechanism just mentioned works for all the two-point functions when we have ${\cal N}=1, 2$ and $4$ Supersymmetry. This result gives further robustness to the quantum duality conjecture between the formulation in terms of ordinary fields and the description in terms of noncommutative fields. However, the nonlocal UV divergent structure still persists after introducing Supersymmetry into the game. But, by using two different gauge-fixing terms, it was shown in  \cite{Martin:2016zon} that the nonlocal  UV divergent contributions are gauge dependent and, therefore, it could be possible to remove them. This is unlike the noncommutative quadratic IR divergences which do not change with the gauge-fixing term as proved in \cite{Ruiz:2000hu}, in the noncommutative field description, and in \cite{Martin:2016zon}, in the ordinary field formulation, respectively.

Now, since it is known that the on-shell DeWitt effective action is independent of the gauge-fixing term used to define the path integral, thus this action can be used to compute the S matrix elements. Hence, by using on-shell DeWitt effective action \cite{DeWitt,Kallosh:1974yh,DeWitt:1980jv,DeWitt:1988fm}, one could settle the question of the physical relevance of the nonlocal UV divergent terms found in the two-point functions of the noncommutative theory formulated in terms of the ordinary fields, and as a bonus obtain a complete proof of the gauge-fixing independence of the UV/IR mixing phenomenon and also the cancelation of the noncommutative quadratic IR divergences achieved by introducing Supersymmetry.

Let $S_{\rm NCYM}\big[\hat{A}_{\mu}\big]$ denote the classical action of noncommutative U(N) Yang-Mills theory, where $\hat{A}_{\mu}$ is a noncommutative gauge field configuration. Let  $\hat\Gamma_{\rm DeW}\big[\hat{B}_\mu\big]$ stand for the on-shell DeWitt effective action of noncommutative field theory whose action is $S_{\rm NCYM}\big[\hat{A}_{\mu}\big]$. Let $\Gamma_{\rm DeW}\big[B_\mu\big]$ be the symbol for the on-shell DeWitt effective action of  the ordinary --i.e., defined in terms of ordinary fields-- U(N) gauge theory whose classical action is  $S_{\rm NCYM}\Big[\hat{A}_{\mu}\big[A_\mu\big]\Big]$, $\hat{A}_{\mu}\big[A_\mu\big]$ being the Seiberg-Witten map that relates the ordinary U(N) gauge field $A_{\mu}$ with the noncommutative U(N) gauge field $\hat{A}_{\mu}$. Then, the purpose of this paper is to show that, at any loop order and in dimensional regularization,
\begin{equation}
\hat\Gamma_{\rm DeW}\Big[\hat{B}_\mu\big[B_\mu\big]\Big]\,=\,\Gamma_{\rm DeW}\big[B_\mu\big],
\label{generalequiv}
\end{equation}
where $\hat{B}_{\mu}[B_\mu]$ stands for the Seiberg-Witten map between the ordinary U(N) gauge field $B_{\mu}$ with the noncommutative U(N) gauge field
$\hat{B}_{\mu}$. Thus proving that the conjecture that Seiberg-Witten map provides two dual formulations of the same underlying quantum theory is true for the noncommutative U(N) Yang-Mills theories, with or without Supersymmetry.

Let us warn the reader that to avoid any clashes with unitarity \cite{Gomis:2000zz}, we shall always consider $\theta^{\mu\nu}$ --the noncommutativity matrix-- such that $\theta^{0i}=0$, $i=1,2,3$.

The layout of this paper is as follows. In section 2, we introduce the on-shell DeWitt action for noncommutative U(N) Yang-Mills theory and the corresponding ordinary U(N) gauge theory defined by means of the $\theta$-exact Seiberg-Witten map. In the same  section, we also discuss the effect and some properties of the Seiberg-Witten map applied to the ordinary background-field splitting. We establish the quantum equivalence of the field theories defined in the previous section by performing the appropriate changes of variables in the path integral in section 3, while in section 4, we check the conclusion reached in section 3 by direct computation --i.e., by using the Feynman rules (FR) derived from classical action-- of the one-loop two-point function of the on-shell DeWitt action of dual ordinary U(1) gauge theory: we show by explicit computation that, in particular, all the ugly non-local UV divergences --see \cite{Martin:2016zon}-- that occur in the propagator in the Feynman gauge all but go away. The noncommutative quadratic IR divergences remain unless the noncommutative theory is supersymmetric with ${\cal N}=1, 2$ and $4$ Supersymmetry. Of course, there is no UV divergence nor any noncommutative IR divergence (quadratic or logarithmic) for ${\cal N} = 4$ U(1) super Yang-Mills. We have included appendices to help understanding the central body of the paper.

\section{Seiberg-Witten map and DeWitt effective action}

In this and next section we provide an expanded proof/review of the equivalence between DeWitt effective actions in terms of noncommutative and ordinary fields via $\theta$-exact Seiberg-Witten map indicated in~\cite{Martin:2016hji}.

\subsection{DeWitt action and the path integral in terms of noncommutative fields}

Let $\hat A_\mu=\hat B_\mu+\hbar^{\frac{1}{2}}\hat Q_\mu$ be the standard splitting  of the noncommutative  U(N) gauge field $\hat A_\mu$ in a noncommutative background $\hat B_\mu$ and a noncommutative quantum field $\hat Q_\mu$. We shall assume that $\hat B_\mu$ satisfies the classical noncommutative equations of motion (EOM), which read $\hat D_\mu \hat F^{\mu\nu}=0$. Then the on-shell DeWitt effective action \cite{DeWitt:1988fm}, $\hat\Gamma_{\rm DeW}\big[\hat B_\mu\big]$, is given by the following path integral
\begin{equation}
e^{\frac{i}{\hbar}\hat\Gamma_{\rm DeW}\big[\hat B_\mu\big]}=\int d\hat Q_\mu^a d\hat C^a d\hat{\bar C}^a d\hat F^a\; e^{\frac{i}{\hbar}S_{\rm NCYM}\big[\hat B_\mu+\hbar^{\frac{1}{2}}\hat Q_\mu\big]+i S_{\rm BFG}\big[\hat B_\mu,\hat Q_\mu,\hat F,\hat{\bar C},\hat C\big]}.
\label{effectivenoncommutative}
\end{equation}
The gauge-fixing term $S_{\rm BFG}\big[\hat B_\mu,\hat Q_\mu,\hat F,\hat{\bar C},\hat C\big]$ is\footnote{As usual, we use $\rm tr$ to denote trace over the Lie algebra generators, while $\rm Tr$ for functional trace.}
\begin{equation}
\begin{array}{l}
{S_{\rm BFG}\big[\hat B_\mu,\hat Q_\mu,\hat F,\hat{\bar C},\hat C\big]=\frac{\hbar^{\frac{1}{2}}}{g^2}\int \tr\, \hat\delta_{\rm BRS}\, \hat{\bar C}\left(\alpha\hat F+ \hat D_\mu\big[\hat B_\mu\big]\hat Q^\mu\right)}\\[4pt]
{\phantom{S_{\rm BFG}\left[\hat B_\mu,\hat Q_\mu,\hat F,\hat{\bar C},\hat C\right]}
=\frac{1}{g^2}\int \tr\, \left(\alpha\hat F\star\hat F- \hat{\bar C}\hat D_\mu\big[\hat B_\mu\big]\hat D^\mu\big[\hat B_\mu+\hbar^{\frac{1}{2}}\hat Q_\mu\big]\hat C
\right)}
\end{array}
\label{2.2}
\end{equation}
where $\hat\delta_{\rm BRS}$ stands for the noncommutative BRS operator, which acts on the noncommutative fields as follows:
\begin{gather}
\hat\delta_{\rm BRS}\hat B_\mu=0,\;\;\hat\delta_{\rm BRS}\hat Q_\mu=\hbar^{-\frac{1}{2}} \hat D_\mu\big[\hat B_\mu+\hbar^{\frac{1}{2}}\hat Q_\mu\big]\hat C,\:
\nonumber\\
\hat\delta_{\rm BRS}\hat C=-i\hat C\star\hat C,\:
\hat\delta_{\rm BRS}\hat{\bar C}=\hbar^{-\frac{1}{2}}\hat F,\:
\hat\delta_{\rm BRS}\hat F=0.
\label{2.3}
\end{gather}
Although --as shown in \cite{Ichinose:1992np} by using BRS techniques for ordinary theories, a proof which remains valid in the case at hand-- $\hat\Gamma_{\rm DeW}\big[\hat B_\mu\big]$ does not depend on the choice of gauge-fixing term, we have chosen the background field gauge (BFG) for convenience.

\subsection{DeWitt effective action of the dual classical ordinary theory}

The next task will be the background field quantization of the ordinary theory with action
\begin{equation*}
S_{\rm NCYM}[A_\mu]=-\frac{1}{4g^2}\int \tr\left(\hat F_{\mu\nu}\Big[\hat A_\mu\big[A_\mu\big]\Big]\hat F^{\mu\nu}\Big[\hat A_\mu\big[A_\mu\big]\Big]\right),
\end{equation*}
where $\hat A_\mu[A_\mu]$ is the $\theta$-exact Seiberg-Witten map which expressed the noncommutative field $\hat A_\mu$ in terms of its ordinary counterpart $A_\mu$. But before carrying out  the background field quantization, we need to discuss --since we are dealing with a nonlinear map-- how the Seiberg-Witten map acts on the background-field-quantization splitting, $A_\mu=B_\mu+\hbar^{\frac{1}{2}}Q_\mu$, of the ordinary gauge field; this we do next.

\subsubsection{Seiberg-Witten map and the background-field splitting}

Let $T^a$ denote the generators of U(N) in the fundamental representation
\begin{equation}
\tr\, T^a T^b=\delta^{ab},\:\:\big[T^a,T^b\big]=if^{abc}T^c.
\end{equation}

Respectively, here $\hat A_\mu=\hat A_\mu^a T^a$ is the noncommutative gauge field, the $A_\mu=A_\mu^a T^a$ is the ordinary gauge field, and the $\hat C=\hat C^a T^a$ is the noncommutative ghost field, while the $C=C^a T^a$ is the ordinary ghost field, all in terms of components fields. The BRS transformations of $\hat A_\mu$, $\hat C$, $A_\mu$ and $C$ read
\begin{gather}
\hat\delta_{\rm BRS}\hat A_\mu=\hat D_\mu\big[\hat A_\mu\big]\hat C
=\partial_\mu\hat C+i\big[\hat A_\mu\stackrel{\star}{,}\hat C\big],\:\:
\hat\delta_{\rm BRS}\hat C=-i\hat C\star\hat C,\:\:
\\
\delta_{\rm BRS} A_\mu=D_\mu\big[A_\mu\big] C=\partial_\mu C+i\big[A_\mu, C\big],\:\:
\delta_{\rm BRS} C=- iC\cdot C.
\end{gather}
The Seiberg-Witten (SW) map
\begin{gather}
\hat A_\mu=\hat A_\mu\big[A_\mu,\theta\big],\:\:
\hat C=\hat C\big[A_\mu,C,\theta\big],
\end{gather}
is a solution to the following equations
\begin{gather}
\hat\delta_{\rm BRS}\hat A_\mu=\delta_{\rm BRS}\hat A_\mu\big[A_\mu,\theta\big],\:\:
\hat\delta_{\rm BRS}\hat C=\delta_{\rm BRS}\hat C\big[A_\mu,C,\theta\big].
\label{SWeqs}
\end{gather}

One can expand the Seiberg-Witten map $\theta$-exactly -see \cite{Martin:2012aw, Martin:2015nna}:
\begin{gather}
\hat A_\mu\big[A_\mu,\theta\big](x)=A_\mu(x)+\sum\limits_{n=2}^\infty \mathcal{A}_\mu^{(n)}(x),
\label{SW1}
\\
\hat C\big[A_\mu,C,\theta\big](x)=C(x)+\sum\limits_{n=1}^\infty \mathcal{C}^{(n)}(x),
\label{SW2}
\end{gather}
where
\begin{gather}
\begin{split}
\mathcal{A}_\mu^{(n)}(x)=\int\prod\limits_{i=1}^n\frac{d^4 p_i}{(2\pi)^4} e^{i\left(\sum\limits_{i=1}^n p_i\right)x}
&\mathfrak{A}^{(n)}_\mu\big[(a_1,\mu_1,p_1),......,(a_n,\mu_n,p_n);\theta\big]
\\&\cdot\tilde A_{\mu_1}^{a_1}(p_1)......\tilde A_{\mu_n}^{a_n}(p_n),
\end{split}
\label{SW3}
\\
\begin{split}
\mathcal{C}^{(n)}(x)=\int\prod\limits_{i=1}^n\frac{d^4 p_i}{(2\pi)^4} e^{i\left(p+\sum\limits_{i=1}^n p_i\right)x}
&\mathfrak{C}^{(n)}\big[(a_1,\mu_1,p_1),......,(a_n,\mu_n,p_n);(a,p);\theta\big]
\\&\cdot \tilde A_{\mu_1}^{a_1}(p_1)......\tilde A_{\mu_n}^{a_n}(p_n)C^a(p),
\end{split}
\label{SW4}
\end{gather}
$\mathfrak{A}_\mu^{(n)}$ and $\mathfrak{C}^{(n)}$ are totally symmetric under the permutations with respect to the set of the parameter-triples $\left\{(a_i,\mu_i,p_i)|i=1,...,n\right\}$, which have the property --of key importance in our later discussion-- that only the momenta which are not contracted with
$\theta^{\mu\nu}$ build up polynomials which  never occur in the denominator \cite{Martin:2012aw, Martin:2015nna}.

Now, let us introduce the ordinary background-field splitting
\begin{equation}
A_\mu=B_\mu+\hbar^{\frac{1}{2}}Q_\mu,
\label{ordinarysplit}
\end{equation}
where $B_\mu$ is the background field and $Q_\mu$ the quantum fluctuation. Substituting (\ref{ordinarysplit}) into (\ref{SW1}-\ref{SW4}), one gets
\begin{gather}
\hat A_\mu\big[B_\mu+\hbar^{\frac{1}{2}}Q_\mu,\theta\big]=\hat B_\mu\big[B_\mu,\theta\big]+\hbar^{\frac{1}{2}}\hat Q_\mu\big[B_\mu,Q_\mu,\hbar,\theta\big],
\label{Qdefinition}
\\
\hat C\big[B_\mu+\hbar^{\frac{1}{2}}Q_\mu,C,\theta\big]=\hat C\big[B_\mu,C,\theta\big]+\hbar^{\frac{1}{2}}\hat C^{(1)}\big[B_\mu,Q_\mu,C,\hbar,\theta\big],
\label{Cdefinition}
\end{gather}
where
\begin{equation}
\begin{split}
\hat B_\mu\big[B_\mu,\theta\big]=B_\mu+\sum\limits_{n=2}^\infty\int\prod\limits_{i=1}^n\frac{d^4 p_i}{(2\pi)^4} e^{i\left(\sum\limits_{i=1}^n p_i\right)x}
&\mathfrak{A}^{(n)}_\mu\big[(a_1,\mu_1,p_1),......,(a_n,\mu_n,p_n);\theta\big]
\\&\cdot\tilde B_{\mu_1}^{a_1}(p_1)......\tilde B_{\mu_n}^{a_n}(p_n),
\end{split}
\label{Bhat}
\end{equation}
\begin{equation}
\begin{split}
\hat Q_\mu\big[B_\mu,Q_\mu,\hbar,\theta\big]=Q_\mu+&\sum\limits_{n=2}^\infty\int\prod\limits_{i=1}^n\frac{d^4 p_i}{(2\pi)^4} e^{i\left(\sum\limits_{i=1}^n p_i\right)x}
\mathfrak{A}^{(n)}_\mu\big[(a_1,\mu_1,p_1),......,(a_n,\mu_n,p_n);\theta\big]
\\\cdot \Big(\sum\limits_{m=1}^n\,\hbar^{\frac{m-1}{2}}\,&\frac{n!}{m!(n-m)!}\,\tilde Q_{\mu_1}^{a_1}(p_1)....\tilde Q_{\mu_m}^{a_m}(p_m)\tilde B_{\mu_{m+1}}^{a_m+1}(p_{m+1})....\tilde B_{\mu_{n}}^{a_{n}}(p_{n})\Big),
\end{split}
\label{Qhat}
\end{equation}

\begin{equation}
\begin{split}
\hat C\big[B_\mu,C,\theta\big]=&C(x)+\sum\limits_{n=1}^\infty \int\prod\limits_{i=1}^n\frac{d^4 p_i}{(2\pi)^4} e^{i\left(p+\sum\limits_{i=1}^n p_i\right)x}
\\&\cdot
\mathfrak{C}^{(n)}\big[(a_1,\mu_1,p_1),......,(a_n,\mu_n,p_n);(a,p);\theta\big]
\tilde B_{\mu_1}^{a_1}(p_1)......\tilde B_{\mu_n}^{a_n}(p_n)C^a(p),
\end{split}
\label{chat}
\end{equation}
\begin{equation}
\begin{split}
\hat C^{(1)}\big[B_\mu,Q_\mu,C,\hbar,\theta\big]&=\sum\limits_{n=1}^\infty\int\prod\limits_{i=1}^n\frac{d^4 p_i}{(2\pi)^4} e^{i\left(p+\sum\limits_{i=1}^n p_i\right)x}
\\&\cdot
\mathfrak{C}^{(n)}\big[(a_1,\mu_1,p_1),......,(a_n,\mu_n,p_n);(a,p);\theta\big]
\\
\cdot\Big(\sum\limits_{m=1}^n\,\hbar^{\frac{m-1}{2}}\,&\frac{n!}{m!(n-m)!}\,\tilde Q_{\mu_1}^{a_1}(p_1)....\tilde Q_{\mu_m}^{a_m}(p_m)\tilde B_{\mu_{m+1}}^{a_{m+1}}(p_{m+1})....\tilde B_{\mu_{n}}^{a_{n}}(p_{n})C^a(p)\Big).
\end{split}
\end{equation}
In the previous equations the convention $\tilde B_{\mu_{n+1}}^{a_{n+1}}(p_{n+1})=1$ is assumed.

Let us stress that $\hat B_\mu\big[B_\mu,\theta\big]$ and $\hat C\big[B_\mu,C,\theta\big]$ are standard Seiberg-Witten maps, i.e., are solutions to the equations in (\ref{SWeqs}), when in the latter $A_\mu$ has been replaced with $B_\mu$. However, $Q_\mu\big[B_\mu,Q_\mu,\hbar,\theta\big]$ and $C^{(1)}\big[B_\mu,Q_\mu,C,\hbar,\theta\big]$ are not standard Seiberg-Witten maps, for they are solutions to
\begin{equation}
\begin{array}{l}
{\hat\delta_{\rm BRS}\hat Q_\mu=\delta_{\rm BRS}\hat Q_\mu\big[B_\mu,Q_\mu,\hbar,\theta\big],}\\[8pt]
{\hat\delta_{\rm BRS}\hat C}={\delta_{\rm BRS}\hat C\big[B_\mu+\hbar^{\frac{1}{2}}Q_\mu,C,\theta\big]},
\end{array}
\label{SWqbg}
\end{equation}
where the splitting of $\hat C\big[B_\mu+\hbar^{\frac{1}{2}}Q_\mu,C,\theta\big]$ is given/defined in (\ref{Cdefinition}), 
and
\begin{equation}
\begin{array}{l}
{\delta_{\rm BRS} B_\mu=0},\\[8pt]
{\delta_{\rm BRS} Q_\mu=\hbar^{-\frac{1}{2}} D_\mu\big[B_\mu+\hbar^\frac{1}{2} Q_\mu\big]C,}\\[8pt]
{\delta_{\rm BRS} C=-i C\cdot C,}\\[8pt]
{\hat\delta_{\rm BRS} \hat Q_\mu=\hbar^{-\frac{1}{2}} \hat D_\mu\big[\hat B_\mu+\hbar^\frac{1}{2}\hat Q_\mu\big]\hat C,}\\[8pt]
{\hat\delta_{\rm BRS}\hat C\big[B_\mu+\hbar^{\frac{1}{2}}Q_\mu,C,\theta\big]=-i\hat C\star\hat C}.
\end{array}
\label{BRStrans}
\end{equation}
In view of equation (\ref{SWqbg}), $\hat Q_\mu\left[B_\mu,Q_\mu,\hbar,\theta\right]$ can be called the Seiberg-Witten map of the quantum field $\hat Q_\mu$ in the presence of the background field $\hat B_\mu$.

That $\hat Q_\mu\big[B_\mu,Q_\mu,\hbar,\theta\big]$ satisfies  (\ref{SWqbg}) is a consequence of the fact that it is defined in terms of $\hat A_\mu\big[A_\mu,\theta\big]$ as done in  (\ref{Qdefinition}) and that $\hat A_\mu\big[A_\mu,\theta\big]$, along with $\hat C\big[A_\mu,C,\theta\big]$, solves the Seiberg-Witten equations in (\ref{SWeqs}). Indeed,
\begin{equation*}
\begin{split}
\hbar^\frac{1}{2} \hat Q_\mu\big[B_\mu,Q_\mu,\hbar,\theta\big]=&\hat A_\mu\big[ B_\mu+\hbar^\frac{1}{2} Q_\mu,\theta\big]-\hat B_\mu\big[B_\mu,\theta\big]\:\:\Longrightarrow
\\
\hbar^\frac{1}{2}\delta_{\rm BRS}\hat Q_\mu\big[B_\mu,Q_\mu,\hbar,\theta\big]=&\delta_{\rm BRS}\hat A_\mu\big[ B_\mu+\hbar^\frac{1}{2} Q_\mu,\theta\big]-\delta_{\rm BRS}\hat B_\mu\big[B_\mu,\theta\big]
\\=&\delta_{\rm BRS}\hat A_\mu\big[ B_\mu+\hbar^\frac{1}{2} Q_\mu,\theta\big]
\\=&\hat D_\mu\Big[\hat A_\mu\big[ B_\mu+\hbar^\frac{1}{2} Q_\mu,\theta\big]\Big]\hat C\big[B_\mu+\hbar^\frac{1}{2} Q_\mu,C,\theta\big].
\end{split}
\end{equation*}

Using the results displayed above, one can show that
\begin{gather}
\delta^2_{\rm BRS}\hat Q_\mu\big[B_\mu,Q_\mu,\hbar,\theta\big]=\hat\delta^2_{\rm BRS}\hat Q_\mu=0,
\label{nil1}
\\
\delta^2_{\rm BRS}\hat C\big[B_\mu+\hbar^{\frac{1}{2}}Q_\mu,C,\theta\big]=\hat\delta^2_{\rm BRS}\hat C=0.
\label{nil2}
\end{gather}

Before closing this subsection, for later use we shall show that both above terms,
$\mathfrak{A}^{(n)}_\mu\big[(a_1,\mu_1,p_1),......,(a_n,\mu_n,p_n);\theta\big]$ and $\mathfrak{C}^{(n)}\big[(a_1,\mu_1,p_1),......,(a_n,\mu_n,p_n);(a,p);\theta\big]$ defining all Seiberg-Witten maps introduced in this section, are linear combinations of functions of the type
\begin{equation}
\mathbb{Q}(p_1,....,p_n)\cdot \mathbb{K}(p_i\theta p_j),
\label{Pstructure}
\end{equation}
where $\mathbb{Q}(p_1,....,p_n)$ is a monomial of the momenta $p_i$ and $\mathbb{K}(p_i\theta p_j)$ is a function of the variables $p_i\theta p_j$, $i,j=1...n$, only. We use well known notation $q\theta k=q_\mu\theta^{\mu\nu}k_{\nu}$.

Let us begin with  $\mathfrak{A}^{(n)}_\mu\big[(a_1,\mu_1,p_1),......,(a_n,\mu_n,p_n);\theta\big]$. In \cite{Martin:2012aw}, it was shown that the $\theta$-exact Seiberg-Witten map can be constructed by setting $h=1$ in the following formal series
\begin{equation}
A_{\mu}[a_\rho;h\theta]=\sum_{n=1}^{\infty}\,{\cal A}^{(n)}_\mu[a_\rho;h\theta],
\end{equation}
where ${\cal A}^{(n)}_\mu[a_\rho;h\theta]$ is of order $n$ in the number of classical fields and it is given by the recursive solution to  the following set of equations
\begin{equation}
\begin{array}{l}
{{\cal A}^{(1)}_{\mu}[a_{\rho};h\theta]=a_\mu, \forall h,}\\[12pt]
{{\cal A}^{(2)}_{\mu}[a_{\rho};h\theta]=\int_{0}^{h}\,dt\,\Big(\frac{1}{2}\,\theta^{ij}\{{\cal A}^{(1)}_{i},\partial_{j}{\cal A}^{(1)}_{\mu}\}_{\star_{t}}
-\frac{1}{4}\,\theta^{ij}\{{\cal A}^{(1)}_{i},\partial_{\mu}{\cal A}^{(1)}_{j}\}_{\star_{t}}\Big),}\\[12pt]
{{\cal A}^{(3)}_{\mu}[a_{\rho};h\theta]=\int_{0}^{h}\,dt\,\Big(
\frac{1}{2}\,\theta^{ij}\{{\cal A}^{(1)}_{i},\partial_{j}{\cal A}^{(2)}_{\mu}[a_{\rho};t\theta]\}_{\star_{t}}
+\frac{1}{2}\,\theta^{ij}\{{\cal A}^{(2)}_{i}[a_{\rho};t\theta],\partial_{j}{\cal A}^{(1)}_{\mu}\}_{\star_{t}} }\\[8pt]
{\phantom{{\cal A}^{(3)}_{\mu}[a_{\rho};h\theta]=\int_{0}^{h}\,dt\,\Big(}-\frac{1}{4}\,\theta^{ij}\{{\cal A}^{(2)}_{i}[a_{\rho};t\theta],\partial_{\mu}{\cal A}^{(1)}_{j}\}_{\star_{t}}
-\frac{1}{4}\,\theta^{ij}\{{\cal A}^{(1)}_{i},\partial_{\mu}{\cal A}^{(2)}_{j}[a_{\rho};t\theta]\}_{\star_{t}} }\\[8pt]
{\phantom{{\cal A}^{(3)}_{\mu}[a_{\rho};h\theta]=\int_{0}^{h}\,dt\,\Big(}+\frac{i}{4}\,\theta^{ij}\{{\cal A}^{(1)}_i,[{\cal A}^{(1)}_j,{\cal A}^{(1)}_\mu]_{\star_{t}}\}_{\star_{t}}, \Big),}\\[12pt]
{...........}\\[12pt]
{{\cal A}^{(n)}_{\mu}[a_{\rho};h\theta]=\int_{0}^{h}\,dt\,\Big(
\frac{1}{2}\,\theta^{ij}\sum_{m_1+m_2=n}\{{\cal A}^{(m_1)}_{i},\partial_{j}{\cal A}^{(m_2)}_{\mu}\}_{\star_{t}}}
\\[12pt]
{\phantom{{\cal A}^{(n)}_{\mu}[a_{\rho};h\theta]=\int_{0}^{h}\,dt\,\Big(
\frac{1}{2}\,\theta^{ij}}
-\frac{1}{4}\,\theta^{ij}\sum_{m_1+m_2=n}\{{\cal A}^{(m_1)}_{i},\partial_{\mu}{\cal A}^{(m_2)}_{j}\}_{\star_{t}}}
\\[12pt]
{\phantom{{\cal A}^{(n)}_{\mu}[a_{\rho};h\theta]=\int_{0}^{h}\,dt\,\Big(
\frac{1}{2}\,\theta^{ij}}
+\frac{i}{4}\,\theta^{ij}\sum_{m_1+m_2+m_3=n}\{{\cal A}^{(m_1)}_i,[{\cal A}^{(m_2)}_j,{\cal A}^{(m_3)}_\mu]_{\star_{t}}\}_{\star_{t}}\Big),\;\forall n>3.}
\end{array}
\label{recursive}
\end{equation}
Above ${\cal A}^{(m_i)}_\mu$ is a shorthand for $ {\cal A}^{(m_i)}_\mu[a_\rho;h\theta]$.

Now, it is easily seen by inspection of the formulae given in \cite{Martin:2012aw} that, indeed, ${\cal A}^{(1)}_{\mu}[a_{\rho};h\theta]$ and ${\cal A}^{(2)}_{\mu}[a_{\rho};h\theta]$ are, after setting h=1, the Fourier transforms of a linear combination of functions of the type displayed in (\ref{Pstructure}) multiplied by one or two ordinary gauge fields, respectively. Further, in ${\cal A}^{(1)}_{\mu}[a_{\rho};h\theta]$ and ${\cal A}^{(2)}_{\mu}[a_{\rho};h\theta]$,  $h$ only occurs in exponentials of the type
\begin{equation}
e^{\pm i\frac{h}{2} \sum\limits_{(i_1,i_2)} p_{i_1}{\theta}\,p_{i_2}}.
\label{exponentialfactor}
\end{equation}
Notice that  in ${\cal A}^{(1)}_{\mu}[a_{\rho};h\theta]$ and ${\cal A}^{(2)}_{\mu}[a_{\rho};h\theta]$ there is no polynomial dependence in $h$, but, we shall allow for the possibility that for higher $n$ there is a cancelation among phase factors that gives rise upon integration over $t$ to positive powers of $h$.  Before we go on, let us recall that, for all integers $s\geq 0$, we have
\begin{equation}
\int_{0}^h\,dt\;t^s\,e^{At}=\frac{e^{Ah}}{A}\sum\limits_{k=0}^{s}\,(-1)^{2s-k}\frac{s!}{(s-k)!A^k}h^{s-k}-(-1)^s\frac{s!}{A^{s+1}}.
\end{equation}
Next, let us assume that, for all $m<n$ we have that, ${\it a})$   ${\cal A}^{(m)}_{\mu}[a_{\rho};h\theta]$ is, for $h=1$, the Fourier transform of a linear combination of functions of the type displayed in (\ref{Pstructure}) multiplied by $m$ ordinary gauge fields and that, ${\it b})$ the $h$-dependence in
${\cal A}^{(m)}_{\mu}[a_{\rho};h\theta]$ only occurs through functions of the form
\begin{equation}
h^\alpha\,e^{\pm i\frac{h}{2} \sum\limits_{(i_1,i_2)
} p_{i_1}{\theta}\,p_{i_2}}\quad{\rm or}\quad h^\beta,
\end{equation}
with $\alpha\geq 0$ and $\beta \geq 0$ being integers. Then, last equation in (\ref{recursive}) tell us that ${\it a})$ and ${\it b})$  hold for ${\cal A}^{(n)}_{\mu}[a_{\rho};h\theta]$, so that mathematical induction  leads to the conclusion that ${\it a})$ and ${\it b})$ also hold for any $n$; which in turn implies that  $\mathfrak{A}^{(n)}_\mu\big[(a_1,\mu_1,p_1),......,(a_n,\mu_n,p_n);\theta\big]$ is a linear combination of functions of the type (\ref{Pstructure}), for whatever value of $n$.

It is plain that the same kind of reasoning can be carried out to show that
\begin{equation}
\mathfrak{C}^{(n)}\big[(a_1,\mu_1,p_1),......,(a_n,\mu_n,p_n);(a,p);\theta\big],
\end{equation}
is also a linear combination of functions of the type (\ref{Pstructure}).

\subsubsection{DeWitt effective action for the ordinary fields}

We are now ready to quantize the classical ordinary U(N) gauge theory which is dual, under the $\theta$-exact Seiberg-Witten map, to the noncommutative U(N) Yang-Mills theory. To quantize the ordinary theory in question, we shall use the background-field splitting; so the classical action that defines the ordinary theory reads
\begin{equation}
S_{\rm NCYM}\big[B_\mu+\hbar^\frac{1}{2} Q_\mu\big]=-\frac{1}{4g^2}\int \tr\left(\hat F_{\mu\nu}\big[\hat B_\mu+\hbar^\frac{1}{2}\hat Q_\mu\big]\hat F^{\mu\nu}\big[\hat B_\mu+\hbar^\frac{1}{2}\hat Q_\mu\big]\right),
\label{2.33}
\end{equation}
with $\hat B_\mu=\hat B_\mu\big[B_\mu\big]$ and $\hat Q_\mu=\hat Q_\mu\big[B_\mu,Q_\mu,\hbar,\theta\big]$ are the Seiberg-Witten map --standard and in the presence of a background-- introduced in section 2.2.1.

Let us first introduce two extra fields,  $\hat F= \hat F^a T^a$, $\hat{\bar C}=\hat{\bar C}^a T^a$, on which the ordinary $\delta_{\rm BRS}$, and noncommutative $\hat \delta_{\rm BRS}$, operators act by the following definition:
\begin{equation}
\delta_{\rm BRS}\hat{\bar C}=\hat\delta_{\rm BRS}\hat{\bar C}=\hbar^{-\frac{1}{2}} \hat F,\delta_{\rm BRS} \hat F=\hat\delta_{\rm BRS} \hat F=0,\delta_{\rm BRS}^2\hat{\bar C}=\hat\delta_{\rm BRS}^2\bar C=\delta_{\rm BRS}^2 \hat F=\hat\delta_{\rm BRS}^2 \hat F=0.
\label{BRS3}
\end{equation}
Here $\hat{\bar C}$ is a Grassmann field and $\hat F$ is a boson field. Recall that $T^a$ is U(N) generator in the fundamental representation.

We shall assume from now on that $B_\mu$ is a solution to the classical equation of motion of the theory with action $S_{\rm NCYM}\left[B_\mu\right]$, as defined previously. Then, as shown in  the appendix A, $B_\mu$ satisfies:
\begin{equation}
\hat D_{\mu}\big[\hat B_\mu[B_\mu]\big]\hat F^{\mu\nu}\big[\hat B_\mu[B_\mu]\big]=0.
\label{NCYMEOM}
\end{equation}
The on-shell DeWitt action, $\Gamma_{\rm DeW}\left[B_\mu\right]$, of ordinary theory now reads
\begin{equation}
e^{\frac{i}{\hbar}\Gamma_{\rm DeW}\big[B_\mu\big]}=\int d Q_\mu^a d C^a d{\hat{\bar C}}^a d \hat F^a\; e^{\frac{i}{\hbar}S_{\rm NCYM}\left[B_\mu+\hbar^{\frac{1}{2}}Q_\mu\right]+i  S_{\rm gf}\left[ B_\mu, Q_\mu,\hat F,\hat{\bar C},C\right]},
\label{effectiveaction}
\end{equation}
where $S_{\rm gf}\big[ B_\mu, Q_\mu,\hat F,\hat{\bar C},C\big]$ is the gauge-fixing term, which is BRS-exact --thus benefits the BRS quantization method:
\begin{equation*}
S_{\rm gf}\big[ B_\mu, Q_\mu,\hat F,\Bar C,C\big]= \delta_{\rm BRS}\,X_{\rm gf}\big[ B_\mu, Q_\mu,\hat F,\hat{\bar C},C\big].
\end{equation*}
Here $\delta_{\rm BRS}$ is the ordinary BRS operator which acts on the fields $B_\mu$, $Q_\mu$ as defined in (\ref{BRStrans}) and on $\hat{\bar C}$ and $\hat F$ as defined in (\ref{BRS3}).
The $X_{\rm gf}\left[ B_\mu, Q_\mu,\hat F,\hat{\bar C},C\right]$ is an arbitrary functional --with ghost number -1-- of the fields, which can be expressed as formal series of the fields.

Taking into account that $\delta_{\rm BRS} ^2=0$, when acting on $B_\mu$, $Q_\mu$, $C$, $\hat{\bar C}$ and $\hat F$, respectively, one concludes that $\delta_{\rm BRS} S_{\rm gf}=0$. Hence, the results presented in \cite{Ichinose:1992np} also apply here, so that $\Gamma_{\rm DeW}\big[B_\mu\big]$ does not depend on the
$X_{\rm gf}\big[ B_\mu, Q_\mu,\hat F,\Bar C,C\big]$ that one chooses. For instance, one may choose the standard background field gauge of the ordinary fields, i.e.,
\begin{equation*}
X_{gf}=\frac{\hbar^\frac{1}{2}}{g^2}\int\,\tr\,\bar C\left(\alpha \hat F+ D_\mu\Big[ A_\mu\big[B_\mu+\hbar^{\frac{1}{2}}Q_\mu\big]\Big] C \right),
\end{equation*}
but this gauge-fixing term will not suit our purpose. We shall choose the following term
\begin{equation}
S_{\rm gf}=\delta_{\rm BRS}\;\frac{\hbar^\frac{1}{2}}{g^2}\int\,\hat{\bar C}\left(\alpha \hat F+\hat D_\mu\Big[\hat A_\mu\big[B_\mu+\hbar^{\frac{1}{2}}Q_\mu\big]\Big]\hat C\big[B_\mu,Q_\mu,C\big]\right),
\label{2.34}
\end{equation}
instead. Note that $\hat D_\mu\left[\hat A_\mu\left[B_\mu+\hbar^{\frac{1}{2}}Q_\mu\right]\right]$ is the noncommutative covariant derivative. Now, taking into account   (\ref{BRStrans}) and (\ref{BRS3}), one concludes that our choice of gauge-fixing term reads
\begin{equation}
S_{\rm gf}=\frac{1}{g^2}\int\tr\left(\alpha \hat F^2+\hat F\hat D_\mu\big[\hat B_\mu\big]\hat Q^{\mu}-\hat{\bar C}\hat D_\mu\big[\hat B_\mu\big]\hat D^\mu\big[\hat B_\mu+\hbar^{\frac{1}{2}}\hat Q_\mu\big]\hat C\right),
\label{gaugefix}
\end{equation}
which is the gauge-fixing term corresponding to the NC BFG.

\section{Establishing equivalence by changing variables in the path integral}

Let $\hat Q_\mu^a=tr(\hat Q_\mu\,T^a)$ and $\hat C^a=tr(\hat C\,C^a)$,  where $\hat Q_\mu$ and $\hat C$ are given by the Seiberg-Witten map in (\ref{Qhat}) and (\ref{chat}), respectively. Let $J_1[B^a,Q^a]$ and $J_2[B^a,Q^a]$ be the following Jacobian determinants
\begin{equation}
\begin{array}{l}
{J_1[B^a, Q^a]\,=\,\det\frac{\delta\hat Q^a_\mu(x)}{\delta Q^b_\nu(y)}\,=\,\exp\,{\rm Tr}\,\ln\Big(\frac{\delta\hat Q^a_\mu(x)}{\delta Q^b_\nu(y)}\Big)},\\[8pt]
{ J_2[B^a,Q^a]\,=\,\det\frac{\delta\hat C^a(x)}{\delta C^b(y)}\,=\,\exp\,{\rm Tr}\,\ln\Big(\frac{\delta\hat C^a(x)}{\delta C^b(y)}\Big).}
 \end{array}
 \label{jacobdet}
 \end{equation}
By changing variables in the path integral in (\ref{effectiveaction}): $C^a \to \hat C^a$ and $Q_\mu^a\to\hat Q_\mu^a$, we obtain
\begin{equation}
\begin{array}{l}
{e^{\frac{i}{\hbar}\Gamma_{\rm DeW}\left[B_\mu\right]}=\int d\hat Q_\mu^a d \hat C^a d\hat{\bar C}^a d \hat F^a\;\Big\{J^{-1}_1[B,Q]\,J_2[B,Q]\,}\\[8pt]
{\phantom{e^{\frac{i}{\hbar}\Gamma_{\rm DeW}\left[B_\mu\right]}=\int d\hat Q_\mu^a d \hat C^a d\hat\bar C^a d \hat F^a\;}
\cdot e^{\frac{i}{\hbar}S_{\rm NCYM}\left[\hat B_\mu+\hbar^{\frac{1}{2}}\hat Q_\mu\right]+i  S_{\rm gf}\left[\hat B_\mu, \hat Q_\mu,\hat F,\hat{\bar C},\hat C\right]}\Big\},}
\end{array}
\label{effectiveactionchanged}
\end{equation}
where $\hat B_\mu$ and $\hat Q_\mu$ are expressed in terms of $B_\mu$ and $Q_\mu$ and
\begin{equation*}
S_{\rm gf}\big[\hat B_\mu, \hat Q_\mu,\hat F,\hat{\bar C},\hat C\big]= S_{\rm BFG}\big[\hat B_\mu,\hat Q_\mu,\hat F,\hat{\bar C},\hat C\big].
\end{equation*}

To continue we start with the following {\it proposition}: \\
{\it If $J_1[B, Q]= 1$ and $J_2[B,Q]=1$, then the right hand side of (\ref{effectiveactionchanged}) equals the right hand side of (\ref{effectivenoncommutative}), leading to (\ref{generalequiv}), that is:
$
\Gamma_{\rm DeW}\big[B_\mu\big]=\hat\Gamma_{\rm DeW}\big[\hat B_\mu[B_\mu]\big].
$}

This result is valid on-shell since $\hat B_\mu[B_\mu]$ satisfies the noncommutative Yang-Mills equations (\ref{NCYMEOM}), and the reason is the on-shell uniqueness of DeWitt effective action ~\cite{Kallosh:1974yh,Ichinose:1992np}.

In summary, if we are able to show that {\it proposition} holds, that would prove that both theories defined in terms of noncommutative fields and in terms of ordinary fields, through the Seiberg-Witten map, have the same on-shell DeWitt effective action and, therefore, they are dual --i.e., they are different descriptions of the same underlying theory-- to each other at the quantum level. We shall show below that, indeed, in dimensional regularization, and in the perturbative regime defined by the coupling constant, our {\it proposition} holds.

\subsection{No one-loop two-point contribution coming from $J_1\left[B,Q\right]$ nor $J_2\left[B,Q\right]$.}

Before we plunge into the general proof that {\it proposition} holds, we shall show that by employing dimensional regularization the one-loop two-point contribution to
\begin{equation}
\ln\, J_1[B,Q]\,=\,{\rm Tr}\,\ln\Big(\frac{\delta\hat Q^a_\mu(x)}{\delta Q^b_\nu(y)}\Big),
\label{thelog}
\end{equation}
vanishes. By working out this simple instance, we shall acquaint ourselves with the techniques that  we shall employ in the general case, as well as the type of dimensionally regularized integrals one has to face.

From \eqref{Qhat} and
\begin{equation}
\frac{\delta Q^a_\mu(p)}{\delta Q^b_\nu(y)}=e^{-ipy}\delta^a_b\delta^\nu_\mu,
\label{functionaldelta}
\end{equation}
one obtains
\begin{equation}
\begin{array}{l}
{\frac{\delta\hat Q^a_\mu(x)}{\delta Q^b_\nu(y)}=\delta^a_b\delta^\nu_\mu\delta(x-y)+
\frac{\delta}{{\delta Q^b_\nu(y)}}\Big\{\sum\limits_{n=2}^\infty\int\prod\limits_{i=1}^n\frac{d^4 p_i}{(2\pi)^4} e^{i\left(\sum\limits_{i=1}^n p_i\right)x}}\\[8pt]
{\cdot n\,\tr\left(T^a\mathfrak{A}^{(n)}_\mu\left[(a_1,\mu_1,p_1),......,(a_n,\mu_n,p_n);\theta\right]\right)
\tilde B_{\mu_1}^{a_1}(p_1)......\tilde B_{\mu_{n-1}}^{a_{n-1}}(p_{n-1})\tilde Q_{\mu_n}^{a_n}(p_n)\Big\}+O(\hbar^{\frac{1}{2}})}\\[8pt]
{=\delta^a_b\delta^\mu_\nu\delta(x-y)+{\cal M}_1\left[B\right]^{a\nu}_{b\mu}(x,y)+{\cal M}_2\left[B\right]^{a\nu}_{b\mu}(x,y),+O(B^3)+O(\hbar^{\frac{1}{2}})},
\end{array}
\end{equation}
with
\begin{equation}
\begin{array}{l}
{{\cal M}_1\left[B\right]^{a\nu}_{b\mu}(x,y)=
2\int\frac{d^4 p_1}{(2\pi)^4} e^{i p_1 x}e^{-ip_2(x-y)}\,\tr\left(T^a\mathfrak{A}^{(2)}_\mu\left[(a_1,\mu_1,p_1);(b,\nu,p_2);\theta\right]\right)\tilde B_{\mu_1}^{a_1}(p_1)},\\[8pt]
{{\cal M}_2\left[B\right]^{a\nu}_{b\mu}(x,y)=3\int\frac{d^4 p_1}{(2\pi)^4}\frac{d^4 p_2}{(2\pi)^4} e^{i (p_1+p_2) x}e^{ip_3(x-y)}}
\\[8pt]
{\cdot\tr\left(T^a\mathfrak{A}^{(3)}_\mu\left[(a_1,\mu_1,p_1),(a_2,\mu_2,p_2);
(b,\nu,p_3);\theta\right]\right)\tilde B_{\mu_1}^{a_1}(p_1)\tilde
B_{\mu_2}^{a_2}(p_2)}.
\end{array}
\end{equation}
Now, substituting the previous results in (\ref{thelog}), one gets
\begin{equation}
\begin{array}{l}
{{\rm Tr}\,\ln\Big(\frac{\delta\hat Q^a_\mu(x)}{\delta Q^b_\nu(y)}\Big)={\rm Tr}\,\ln\Big(\delta^a_b\delta^\mu_\nu\delta(x-y)+{\cal M}_1\left[B\right]^{a\nu}_{b\mu}(x,y)+{\cal M}_2\left[B\right]^{a\nu}_{b\mu}(x,y)\Big)+
O(B^3)+O(\hbar^{\frac{1}{2}})}\\[8pt]
{={\rm Tr}\,\ln\left(\mathbbm{1}+{\cal M}_1+{\cal M}_2\right)+O(B^3)+O(\hbar^{\frac{1}{2}})={\rm Tr}\sum\limits_{k=1}^\infty\frac{(-)^{k+1}}{k}({\cal M}_1+{\cal M}_2)^k+O(B^3)+O(\hbar^{\frac{1}{2}})}\\[8pt]
{={\rm Tr}\,{\cal M}_1+   {\rm Tr}\,{\cal M}_2+{\rm Tr}\,{\cal M}_1\,{\cal M}_1+O(B^3)+O(\hbar^{\frac{1}{2}})},
\end{array}
\label{logexpan}
\end{equation}
where
\begin{equation}
\begin{array}{l}
{{\rm Tr}\,{\cal M}_1=\int d^4x\,{\cal M}_1\left[B\right]^{a\mu}_{a\mu}(x,x)}\\[8pt]
{\phantom{ {\rm Tr}\,{\cal M}_1}
=\int\frac{d^{2\omega }p_1}{(2\pi)^{2\omega}}(2\pi)^{2\omega}\delta(p_1)B_{\mu_1}^{a_1}(p_1)\int\frac{d^{2\omega}q}{(2\pi)^{2\omega}}\tr\left(T^a\mathfrak{A}^{(2)}_\mu\left[(a_1,\mu_1,p_1),(a,\mu,q);\theta\right]\right)},
\end{array}
\label{trM1}
\end{equation}
and
\begin{equation}
\begin{array}{l}
{{\rm Tr}\,{\cal M}_2=\int d^4x\,{\cal M}_2\left[B\right]^{a\mu}_{a\mu}(x,x)=\int\frac{d^{2\omega} p_1}{(2\pi)^{2\omega}}\int\frac{d^{2\omega }p_2}{(2\pi)^{2\omega}}(2\pi)^{2\omega}\Big\{\delta(p_1+p_2)B_{\mu_1}^{a_1}(p_1)B_{\mu_2}^{a_2}(p_2)}\\[8pt]
{\phantom{ {\rm Tr}\,{\cal M}_2=\int\frac{d^{2\omega} p_1}{(2\pi)^{2\omega}}\int\frac{d^{2\omega }p_2}{(2\pi)^{2\omega}}(2\pi)^{2\omega}}
\cdot3\int\frac{d^{2\omega}q}{(2\pi)^{2\omega}}\tr\left(T^a\mathfrak{A}^{(3)}_\mu\left[(a_1,\mu_1,p_1),(a_2,\mu_2,p_2),(a,\mu,q);\theta\right]\right)\Big\}},
\end{array}
\label{trM2}
\end{equation}
and
\begin{equation}
\begin{array}{l}
{{\rm Tr}\,{\cal  M}_1{\cal M}_1=\int d^{2\omega}x\int d^4y\,{\cal M}_1\left(B\right)^{a\mu}_{a'\mu'}(x,y){\cal M}_1\left(B\right)^{a'\mu'}_{a\mu}(y,x)}\\[8pt]
{=\int\frac{d^{2\omega }p_1}{(2\pi)^{2\omega}}\int\frac{d^{2\omega }q_1}{(2\pi)^{2\omega}}(2\pi)^{2\omega}\Big\{\delta(p_1+q_1)\tilde B_{\mu_1}^{a_1}(p_1)\tilde B_{\nu_1}^{b_1}(q_1)}\\[8pt]
{ \cdot4\int\frac{d^{2\omega }q}{(2\pi)^{2\omega}}\tr\left(T^a\mathfrak{A}^{(2)}_{\mu_2}\left[(a_1,\mu_1,p_1),(a_2,\mu,q);\theta\right]\right)
 \cdot\tr\left(T^{a_2}\mathfrak{A}^{(2)}_\mu\left[(b_1,\nu_1,p_1),(a,\mu_2,p_1+q);\theta\right]\right)\Big\}}.
\end{array}
\label{trM1M1}
\end{equation}

Let us show now that in dimensional regularization ${\rm Tr}\,{\cal M}_1=0$. 
The term 
$\mathfrak{A}^{(2)}_\mu\left[(a_1,\mu_1,p_1),(a,\mu,q);\theta\right]$ can be obtained from $\mathbb{A}^{(2)}_\mu$ in section III of ref. \cite{Martin:2012aw}:
\begin{equation}
\begin{split}
\mathfrak{A}^{(2)}_\mu\left[(a_1,\mu_1,p_1),(a_2,\mu_2,p_2);\theta\right]=&\frac{1}{2}\Big(\mathbb{A}^{(2)}_\mu\left[(a_1,\mu_1,-p_1),(a_2,\mu_2,-p_2);\theta\right]
\\&+\mathbb{A}^{(2)}_\mu\left[(a_2,\mu_2,-p_2),(a_1,\mu_1,-p_1);\theta\right]\Big).
\end{split}
\end{equation}
Hence, in dimensional degularization the loop integral --the integral over q-- (in ${\rm Tr}\,{\cal M}_1$ --see (\ref{trM1})--) reads
\begin{equation}
\begin{split}
&\int\frac{d^{2\omega}q}{(2\pi)^{2\omega}}\tr\left(T^a\mathfrak{A}^{(2)}_\mu\left[(a_1,\mu_1,p_1),(a,\mu,q));\theta\right]\right)
\\
&=\frac{1}{2}\frac{d^{2\omega}q}{(2\pi)^{2\omega}}\tr\left(\mathbb{A}^{(2)}_\mu\left[(a_1,\mu_1,-p_1),(a,\mu,q);\theta\right]
+\mathbb{A}^{(2)}_\mu\left[(a,\mu,q),(a_1,\mu_1,-p_1);\theta\right]\right)
\\
&=-\frac{1}{4}\tr\int\frac{d^{2\omega}q}{(2\pi)^{2\omega}}\theta^{ij}(2q_j\delta_i^{\mu_1}\delta^\mu_\mu-q_\mu\delta_i^{\mu_1}\delta_j^\mu)
\left(T^aT^{a_1}T^a\frac{e^{-\frac{i}{2}q\theta p}-1}{q\theta p}-T^aT^aT^{a_1}\frac{e^{\frac{i}{2}q\theta p}-1}{q\theta p}\right)
\\
&=0,
\end{split}
\end{equation}
since
\begin{gather}
\int\frac{d^{2\omega}q}{(2\pi)^{2\omega}}\frac{q_{\mu_1}... q_{\mu_r}}{q\theta p}=0,
\:\:
\int\frac{d^{2\omega}q}{(2\pi)^{2\omega}}q_{\mu_1}... q_{\mu_r}\frac{e^{i\xi q\theta p}}{q\theta p}=0,\forall \xi.
\end{gather}
One may actually use $\delta(p_1)$ to further simplify the argument, as only the second vanishing identity above would be needed. We have included the discussion of the vanishing of the previous type of integrals due to the employment of  dimensional regularization  in the appendix B.

Next, by integrating out the Dirac delta function, $\delta(p_1+p_2)$ in (\ref{trM2}), one comes to the conclusion that to work out ${\rm Tr}{\cal M}_2$, one has to compute the following dimensionally regularized integral
\begin{equation}
\int\frac{d^{2\omega}q}{(2\pi)^{2\omega}}\tr\left(T^a\mathfrak{A}^{(3)}_\mu\big[(a_1,\mu_1,p_1),(a_2,\mu_2,-p_1),(a,\mu,q);\theta\big]\right),
\label{M2int}
\end{equation}
where $\mathfrak{A}^{(3)}_\mu\left[(a_1,\mu_1,p_1),(a_2,\mu_2,-p_1),(a,\mu,q);\theta\right]$ is obtained from $\mathbb{A}^{(3)}_\mu$ in equation (3.1) of ref.\cite{Martin:2012aw} by appropriate symmetrization. By expressing $\mathfrak{A}^{(3)}_\mu\left[(a_1,\mu_1,p_1),(a_2,\mu_2,-p_1),(a,\mu,q);\theta\right]$ in terms
of $\mathbb{A}^{(3)}_\mu$, one concludes that the integral in (\ref{M2int}) is a linear combination of the following types of dimensionally regularized integrals:
\begin{gather*}
\int\frac{d^{2\omega}q}{(2\pi)^{2\omega}}\mathbb{Q}(q)\,\mathbb{I}(-p_1,p_1,q,\theta),\;
\int\frac{d^{2\omega}q}{(2\pi)^{2\omega}}\mathbb{Q}(q)\,\mathbb{I}(-p_1,-q,p_1,\theta),\;
\\
\int\frac{d^{2\omega}q}{(2\pi)^{2\omega}}\mathbb{Q}(q)\,\mathbb{I}(-q,-p_1,p_1,\theta),\;
\int\frac{d^{2\omega}q}{(2\pi)^{2\omega}}\mathbb{Q}(q)\,\mathbb{I}(p_1,p_1,-q,\theta),\;
\\
\int\frac{d^{2\omega}q}{(2\pi)^{2\omega}}\mathbb{Q}(q)\,\mathbb{I}(p_1,-q,p_1,\theta),\;
\int\frac{d^{2\omega}q}{(2\pi)^{2\omega}}\mathbb{Q}(q)\,\mathbb{I}(-q,p_1,p_1,\theta);
\end{gather*}
\begin{gather*}
\int\frac{d^{2\omega}q}{(2\pi)^{2\omega}}\mathbb{Q}(q)\,\mathbb{F}(-p_1,p_1,q,\theta),\;
\int\frac{d^{2\omega}q}{(2\pi)^{2\omega}}\mathbb{Q}(q)\,\mathbb{F}(-p_1,-q,p_1,\theta),\;
\\
\int\frac{d^{2\omega}q}{(2\pi)^{2\omega}}\mathbb{Q}(q)\,\mathbb{F}(-q,-p_1,p_1,\theta),\:
\int\frac{d^{2\omega}q}{(2\pi)^{2\omega}}\mathbb{Q}(q)\,\mathbb{F}(p_1,p_1,-q,\theta),\;
\\
\int\frac{d^{2\omega}q}{(2\pi)^{2\omega}}\mathbb{Q}(q)\,\mathbb{F}(p_1,-q,p_1,\theta),\;
\int\frac{d^{2\omega}q}{(2\pi)^{2\omega}}\mathbb{Q}(q)\,\mathbb{F}(-q,p_1,p_1,\theta);
\end{gather*}
where $\mathbb{Q}(q)$ denotes symbolically a monomial in q (i.e. $\mathbb{Q} \equiv  q_{\mu_1}... q_{\mu_r}$), and
\begin{gather}
\mathbb{I}(p_1,p_2,p_3,\theta)=(p_2\theta p_3)^{-1}\bigg[\frac{e^{-\frac{i}{2}(p_1\theta p_2+p_1\theta p_3+p_2\theta p_3)}-1}{p_1\theta p_2+p_1\theta p_3+p_2\theta p_3}-\frac{e^{-\frac{i}{2}p_1\theta(p_2+p_3)}-1}{p_1\theta(p_2+p_3)}\bigg],
\\
\mathbb{F}(p_1,p_2,p_3,\theta)=\frac{e^{-\frac{i}{2}(p_1\theta p_2+p_1\theta p_3+p_2\theta p_3)}-1}{p_1\theta p_2+p_1\theta p_3+p_2\theta p_3}.
\end{gather}
Hence
\begin{gather*}
\mathbb{I}(-p_1,p_1,q,\theta)=\frac{e^{\frac{i}{2}q\theta p_1}-1}{(q\theta p_1)^2}-\frac{i}{2q\theta p_1},\:\mathbb{I}(-p_1,-q,p_1,\theta)=\frac{(e^{\frac{i}{2}q\theta p_1}-1)^2}{2(q\theta p_1)^2},\:
\\
\mathbb{I}(-q,-p_1,p_1,\theta)=-\frac{1}{8},\;\mathbb{I}(p_1,p_1,-q,\theta)=\frac{(e^{-\frac{i}{2}q\theta p_1}-1)^2}{2(q\theta p_1)^2},\:
\\
\mathbb{I}(p_1,-q,p_1,\theta)=\frac{e^{-\frac{i}{2}q\theta p_1}-1}{(q\theta p_1)^2}+\frac{i}{2q\theta p_1},\:\mathbb{I}(q,-p_1,p_1,\theta)=-\frac{1}{8},
\\
\mathbb{F}(-p_1,p_1,q,\theta)=-\frac{i}{2},\:\mathbb{F}(-p_1,-q,p_1,\theta)=\frac{1-e^{iq\theta p_1}}{2 q\theta p_1},\:
\\
\mathbb{F}(-q,-p_1,p_1,\theta)=-\frac{i}{2},\;\mathbb{F}(p_1,p_1,-q,\theta)=\frac{e^{-iq\theta p_1}-1}{2 q\theta p_1},\:
\\
\mathbb{F}(p_1,-q,p_1,\theta)=-\frac{i}{2},\:\mathbb{F}(q,-p_1,p_1,\theta)=-\frac{i}{2}.
\end{gather*}
Putting it all together one reaches the conclusion that all the integrals listed above are of the type
\begin{equation}
\int\frac{d^{2\omega}q}{(2\pi)^{2\omega}}\frac{e^{iq\theta p}}{(q\theta p_1)^{n_1}......(q\theta p_r)^{n_r}},
\end{equation}
which vanish -see appendix B for details-- in dimensional regularization. We have thus shown that ${\rm Tr}\,{\cal M}_2=0$, in dimensional regularization.

Let us finally show that ${\rm Tr}\,{\cal M}_1 {\cal M}_1=0$ in dimensional regularization. The loop integral over q, contributing to ${\rm Tr}\,{\cal M}_1{\cal M}_1$, runs -as seen from (\ref{trM1M1})- is:
\begin{equation}
\begin{split}
&\int\frac{d^{2\omega }q}{(2\pi)^{2\omega}}\tr\left(T^a\mathfrak{A}^{(2)}_{\mu_2}\big[(a_1,\mu_1,p_1),(a_2,\mu,q);\theta\big]\right)
\tr\left(T^{a_2}\mathfrak{A}^{(2)}_\mu\big[(b_1,\nu_1,p_1),(a,\mu_2,p_1+q);\theta\big]\right);
\end{split}
\end{equation}
but this integral vanishes since it is, again, a linear combination of integrals of the type
\begin{equation}
\int\frac{d^{2\omega }q}{(2\pi)^{2\omega}}\mathbb{Q}\left(\frac{e^{\frac{i}{2}q\theta p_1}-1}{q\theta p_1}\right)\left(\frac{e^{\pm\frac{i}{2}q\theta p_1}-1}{q\theta p_1}\right).
\end{equation}
However, these integrals --appendix B-- are equal to zero in dimensional regularization.

In summary, we have just shown that in dimensional regularization ${\rm Tr}\,{\cal M}_1=0$, ${\rm Tr}\,{\cal M}_2=0$ and ${\rm Tr}\,{\cal M}_1 {\cal M}_1=0$, and, hence --see (\ref{logexpan}) and (\ref{thelog})-- one obtains
\begin{equation*}
\ln\, J_1[B,Q]= 0\,+\,O(B^3)\,+\,O(\hbar^{\frac{1}{2}}).
\end{equation*}

Finally, since the same types of integral contribute to $J_2[B,Q]$ it is plain that
\begin{equation*}
\ln\, J_2[B,Q]= 0\,+\,O(B^3)\,+\,O(\hbar^{\frac{1}{2}}),
\end{equation*}
also holds, and therefore, the one-loop two-point contribution to $\Gamma_{\rm DeW}[B_\mu]$ does not receive contributions  neither from $J_1[B,Q]$ nor from $J_2[B,Q]$.

Later in this paper a head-on --i.e., by using the Feynman rules for the ordinary fields and not changing variables in the path integral--  computation of the same two-point function will be performed. We are now ready to show that there are no nontrivial contribution either to  $J_1[B,Q]$ or to $J_2[B,Q]$.

\subsection{Triviality of the full Jacobian determinants}

It is shown in the appendix B that
\begin{equation}
\begin{array}{l}
{\frac{\delta\hat Q^a_\mu(x)}{\delta Q^b_\nu(y)}\,=\,\frac{1}{\hbar^{\frac{1}{2}}}\,\frac{\delta\hat A^a_\mu(x)}{\delta Q^b_\nu(y)}}\\[8pt]
{=\delta^a_b\delta^\nu_\mu\,\delta(x-y)+\sum\limits_{n=2}^{\infty}\,
\int\prod\limits_{i=1}^n\frac{d^4 p_i}{(2\pi)^4} e^{i\left(\sum\limits_{i=1}^{n-1} p_i\right)x}\,e^{ip_n (x-y)}
{\cal M}^{(n)\,a\,\nu}_{\phantom{(n)\,}b\,\mu}(p_1,p_2,....p_{n-1};p_n;\theta)},
\end{array}
\label{dethQQ}
\end{equation}
where
\begin{equation}
\begin{array}{l}
{{\cal M}^{(n)\,a\,\nu}_{\phantom{(n)\,}b\,\mu}(p_1,p_2,....p_{n-1};p_n;\theta)}\\[8pt]
=n\,\tr\Big(T^a{\mathfrak{A}^{(n)}_\mu\left[(a_1,\mu_1,p_1),...,(a_{n-1},\mu_{n-1},p_{n-1}),(b,\nu,p_n);\theta\right]\Big)
\tilde A_{\mu_1}^{a_1}(p_1)......\tilde A_{\mu_{n-1}}^{a_{n-1}}(p_{n-1})}.
\end{array}
\label{Mcaldef}
\end{equation}
Note  that the definition of splitting in the momentum space reads
$\tilde A_{\mu_i}^{a_i}(p_i)=\tilde B_{\mu_i}^{a_i}(p_i)+\hbar^{\frac{1}{2}}\tilde Q_{\mu_i}^{a_i}(p_i)$ for all $i$.
Now let's first define total momenta $l_i$, $i=1,..,m+1$, as the following sums 
\begin{equation*}
l_1=\sum\limits_{i_1=1}^{n_1-1}\, p_{1,i_1},\:\: l_2=\sum\limits_{i_2=1}^{n_2-1}\,p_{2,i_2},\:.....\:, \:\: l_{m}=\sum\limits_{i_{m}=1}^{n_{m}}\,p_{m,i_{m}}, \:\: l_{m+1}=\sum\limits_{i_{m+1}=1}^{n_{m+1}}\,p_{m+1,i_{m+1}},
\end{equation*}
then, by taking into account (\ref{dethQQ}) and carrying out a lengthy straightforward computation --see appendix B for details-- one gets
\begin{equation}
\begin{array}{l}
{\ln\,J_1[B,Q]={\rm Tr}\ln\,\left(\frac{\delta\hat Q^a_\mu(x)}{\delta Q^b_\nu(y)}\right)}
\\[8pt]
={\sum\limits_{n=2}^{\infty}\int\prod\limits_{i=1}^{n-1}\frac{d^4 p_i}{(2\pi)^4}\,\delta(\sum\limits_{i=1}^{n-1}p_i)\,\int\frac{d^4 q}{(2\pi)^4}\,
{\cal M}^{(n)\,a\,\mu}_{\phantom{(n)\,}a\,\mu}\left(p_1,p_2,....,p_{n-1};q;\theta\right)}
\\[8pt]
{+\sum\limits_{m=1}^{\infty}\frac{(-1)^{m}}{m+1}\sum\limits_{n_1=2}^{\infty}\cdots\cdots\sum\limits_{n_{m+1}
=2}^{\infty}
\int\prod\limits_{i_1=1}^{n_1-1}\frac{d^4 p_{1,i_1}}{(2\pi)^4}\cdots\cdots\int\prod\limits_{i_{m+1}=1}^{n_{m+1}-1}\frac{d^4 p_{m+1,i_{m+1}}}{(2\pi)^4}\;\delta\left(\sum\limits_{i=1}^{m+1}l_i\right)}
\\[8pt]
{\int\frac{d^4 q}{(2\pi)^4}\Big[
{\cal M}^{(n_1)\,a\,\mu_1}_{\phantom{(n_1)\,}a_1\,\mu}\left(p_{1,1},p_{1,2},....,p_{1,n_1-1};q;\theta\right)
\cdot{\cal M}^{(n_2)\,a_1\,\mu_2}_{\phantom{(n_2)\,}a_2\,\mu_1}\left(p_{2,1},p_{2,2},....,p_{2,n_2-1};q-l_2;\theta\right)}
\\[8pt]
{\phantom{\int\frac{d^4 q}{(2}\Big[}
\cdot{\cal M}^{(n_3)\,a_2\,\mu_3}_{\phantom{(n_3)\,}a_3\,\mu_2}\left(p_{3,1},p_{3,2},....,p_{3,n_3-1};q-l_2-l_3;\theta\right)}
\\[8pt]
{\phantom{\int\frac{d^4 q}{(2}\Big[}\cdots\cdots\cdots}
\\[8pt]
{\phantom{\int\frac{d^4 q}{(2}\Big[}
\cdot{\cal M}^{(n_{m})\,a_{m-1}\,\mu_m}_{\phantom{(n_{m})\,}a_m\,\,\,\,\,\,\,\mu_{m-1}}\left(p_{m,1},p_{m,2},....,p_{m,n_{m}-1};q-\sum\limits_{i=2}^{m}l_i;\theta\right)}\\[8pt]
{\phantom{\int\frac{d^4 q}{(2}\Big[}
\cdot{\cal M}^{(n_{m+1})\,a_m\,\mu}_{\phantom{(n_{m+1}\,}a\,\,\,\mu_m}\left(p_{m+1,1},p_{m+1,2},....,p_{m+1, n_{m+1}-1};q-\sum\limits_{i=2}^{m+1}l_i;\theta\right)\Big]}.
\end{array}
\label{genexpln}
\end{equation}
\begin{figure}
\begin{center}
\includegraphics[width=14cm]{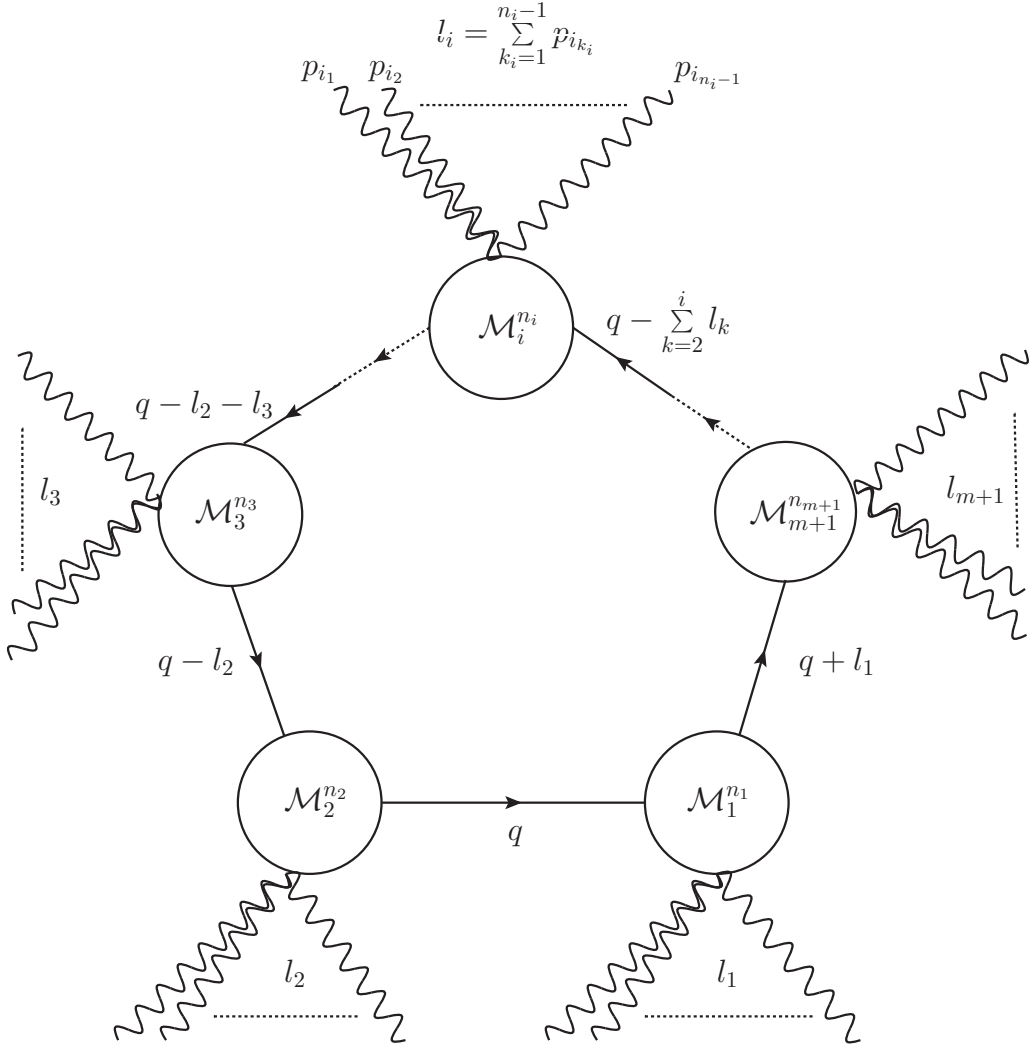}
\end{center}
\caption{The one-loop diagram interpretation/ilustration of (\ref{genexpln}): Each circle corresponds one $\mathcal M^{(n_i)}$, wavy lines denote the gauge field operators, either background or quantum, within the $\mathcal M^{(n_i)}$. The $l_i$'s are then just the total momentum brought in by these field operators. The solid line flows in each circle gives the assignment of $q-\sum\limits_k l_k$ into the corresponding $\mathcal M^{(n_i)}$ in (\ref{genexpln}).}
\label{fig:InterpretTrlnJ1}
\end{figure}
The general structure of the master integral (\ref{genexpln}) above can be visualized by a 1-loop diagram, as given in Fig.~\ref{fig:InterpretTrlnJ1}.

Now, in a view of previous equations (\ref{Mcaldef}) and (\ref{genexpln}), to complete computation of $\ln\,J_1[B,Q]$, one has to work out the following dimensionally regularized type of integrals over the internal momenta $q^\mu$:
\begin{equation}
\begin{array}{l}
{\mathfrak{V}=\int\frac{d^D q}{(2\pi)^D}\Big\{\tr\Big(T^a{\mathfrak{A}_\mu^{(n_1)}}\left[(b_{1,1},\nu_{1,1},p_{1,1}),.....,(b_{1,n_1-1},\nu_{1,n_1-1},p_{1,n_1-1}),(a_1,\mu_1,q);\theta\right]\Big)}\\[8pt]
{\cdot\tr\Big(T^{a_1}{\mathfrak{A}_{\mu_1}^{(n_2)}}\left[(b_{2,1},\nu_{2,1},p_{2,1}),.....,(b_{2,n_2-1},\nu_{2,n_2-1},p_{2,n_2-1}),(a_2,\mu_2,q-l_2);\theta\right]\Big)}\\[8pt]
{\cdot\tr\Big(T^{a_2}{\mathfrak{A}^{(n_3)}}_{\mu_2}\left[(b_{3,1},\nu_{3,1},p_{3,1}),.....,(b_{3,n_3-1},\nu_{3,n_3-1},p_{3,n_3-1}),(a_3,\mu_3,q-l_2-l_3);\theta\right]\Big)}\\[8pt]
{\cdots\cdots\cdots}\\[8pt]
{\cdot\tr\Big(T^{a_{m-1}}{\mathfrak{A}^{(n_{m})}}_{\mu_{m-1}}\big[(b_{m,1},\nu_{m,1},p_{m,1}),....,}\\[8pt]
{\quad\quad\quad\quad\quad\quad\quad\quad\quad
b_{m,n_{m}-1},\nu_{m,n_{m}-1},p_{m,n_{m}-1}),(a_m,\mu_m,q-\sum\limits_{i=2}^{m} l_i);\theta\big]\Big)}
\\[8pt]
{\cdot\tr\Big(T^{a_m}{\mathfrak{A}^{(n_{m+1})}}_{\mu_m}\big[(b_{m+1,1},\nu_{m+1,1},p_{m+1,1}),....,}\\[8pt]
{\quad\quad\quad\quad\quad\quad\quad\quad\quad
(b_{m+1,n_{m+1}-1},\nu_{m+1,n_{m+1}-1},p_{m+1,n_{m+1}-1}),(a,\mu,q-\sum\limits_{i=2}^{m+1} l_i);\theta\big]\Big)\Big\}.}
\end{array}
\end{equation}
However, according to the discussion at the end of subsection 2.2.1, the previous integral is a linear combination of integrals of the type
\begin{equation}
\mathfrak{I}\,=\,\int\frac{d^D q}{(2\pi)^D}\,\mathbb{Q}(q)\,\mathbb{I}(q\theta k_i,k_i\theta k_j),
\end{equation}
where $\mathbb{Q}(q)=q^{\rho_1}q^{\rho_2}\cdots q^{\rho_n}$, $q\theta k_i=q_\mu\theta^{\mu\nu}k_{i\nu}$, $i=1,.....,s$ 
and $k_i\theta k_j=k_{i\mu}\theta^{\mu\nu}k_{j\nu}$, $i,j=1,.....,s$. Here $n$ and $s$ run over all relevant momenta other than $q$, in general. It is important to stress that $\mathbb{Q}(q)$ is a monomial on $q^{\rho}$ and that the function $\mathbb{I}$ in the integrand of the previous integral is a function of the variables $q\theta k_i$ and $k_i\theta k_j$ only, and, hence, as shown in the appendix C, one concludes that
\begin{equation}
\mathfrak{I}=0\quad{\rm and}\quad \mathfrak{V}=0.
\end{equation}
By substituting $\mathfrak{V}=0$ in (\ref{genexpln}) one obtains that in dimensional regularization the following logarithm vanishes: $\ln\,J_1[B,Q]=0$, and
\begin{equation}
J_1[B,Q]=1.
\label{J_1=1}
\end{equation}
It is plain that identical lines of arguments apply to $J_2[B,Q]$ as well. Thus as expected, our {\it proposition} has been proven.

\subsection{Incorporating adjoint matter}

The $\theta$-exact Seiberg-Witten map for  scalar or fermion fields in the adjoint reads --see \cite{Martin:2015nna}:
\begin{equation}
\begin{array}{l}
{\hat \Phi\left[A_\mu,\Phi,\theta\right](x)=\Phi(x)+\sum\limits_{n=1}^\infty \mathcal{F}^{(n)}(x),}\\[8pt]
{\mathcal{F}^{(n)}(x)=\int\prod\limits_{i=1}^n\frac{d^4 p_i}{(2\pi)^4} e^{i\left(p+\sum\limits_{i=1}^n p_i\right)x}
\mathfrak{F}^{(n)}\left[(a_1,\mu_1,p_1),......,(a_n,\mu_n,p_n);(a,p);\theta\right]}\\[8pt]
{\phantom{\mathfrak{F}^{(n)} =\int\prod\limits_{i=1}^n\frac{d^4 p_i}{(2\pi)^4} e^{i\left(\sum\limits_{i=1}^n p_i\right)x}}
\cdot \tilde A_{\mu_1}^{a_1}(p_1)......\tilde A_{\mu_n}^{a_n}(p_n){\Phi}^a(p)},
\end{array}
\end{equation}
where $\Phi=\Phi^a T^a$ denotes an ordinary scalar or fermion field transforming under the adjoint of U(N). Now, taking into account the recursive equations -see section III of ref. \cite{Martin:2015nna}-- which yield $\hat \Phi\left[A_\mu,\Phi,\theta\right](x)$ have similar $\theta$ and momentum structure to the one for $\hat A_\mu\left[A_\mu,\theta\right]$, it is easy to see that $\mathfrak{F}^{(n)}\left[(a_1,\mu_1,p_1),......,(a_n,\mu_n,p_n);(a,p);\theta\right]$  is also a linear combination of functions of the type in (\ref{Pstructure}). Hence, one also concludes that the Jacobian determinant for the
change of variables $\Phi^a \to \hat \Phi^a(x)=\tr (\hat \Phi(x) T^a)=\tr(T^a\hat \Phi\left[A_\mu,\Phi,\theta\right](x))$ in the path integral over the fields $\Phi^a$ is one, i.e.:
\begin{equation}
\det \frac{\delta \hat\Phi^a(x)}{\delta \Phi^a(y)}=1.
\end{equation}
Hence the inclusion of matter fields in the adjoint --and for that matter any type of matter fields-- does not change the conclusion that we have reached above for gauge fields, i.e., that the $\theta$-exact Seiberg-Witten map associates every quantum field theory, with  gauge group U(N) and formulated in terms of noncommutative fields, to an ordinary gauge theory, with gauge group U(N), which is dual to the former at the quantum level, because, indeed, they do have the same on-shell DeWitt effective action.

Note that the massless tadpole integrals also vanish in the dimensional-reduction scheme, which preserves supersymmetry manifestly in the one-loop. Therefore our conclusion here should be also valid for the noncommutative supersymmetric Yang-Mills theories (NCSYMs).

\section{Testing the quantum equivalence by direct computation: the one-loop two-point function}

We choose to test the formal equivalence established in the last section by computing explicitly one-loop quantum correction to the quadratic part of the effective action in the noncommutative U(1) gauge theory prior to and after the Seiberg-Witten map. At this specific order, the general equivalence relation \eqref{generalequiv} reduces to a much simpler relation
\begin{equation}
\begin{split}
\int\frac{d^D p}{(2\pi)^D}\tilde B_\mu(-p)\Gamma^{\mu\nu}(p)\tilde B_\nu(p)=&\int\frac{d^D p}{(2\pi)^D}\tilde{\hat B}_\mu[\tilde B_\mu(-p)]\hat\Gamma^{\mu\nu}(p)\tilde{\hat B}_\nu[\tilde B_\mu(p)]
\\=&\int\frac{d^D p}{(2\pi)^D}\tilde B_\mu(-p)\hat\Gamma^{\mu\nu}(p)\tilde B_\nu(p),
\end{split}
\label{quadraticequiv}
\end{equation}
when $\tilde B_\nu(p)$ is placed on-shell, because only the zeroth order of the SW map counts here.

We start by reviewing the standard procedure for computing the DeWitt effective action of $\rm U(1)$ gauge theory perturbatively  in the background field formalism/method (BFM)~\cite{Kallosh:1974yh,DeWitt:1980jv}, which evaluates all 1-PI diagrams with all background field external legs and all integrand fields ($\hat Q_\mu,\hat{\bar C},\hat C,\hat F$) internal line using the following action $\hat S_{\rm loop}$ 
\begin{equation}
\hat S_{\rm loop}=S_{\rm gf}+S_{\rm NCYM}\big[\hat B_\mu+\hat Q_\mu\big]-S_{\rm NCYM}\big[\hat B_\mu\big]-\int \frac{\delta S_{\rm NCYM}\big[\hat B_\mu\big]}{\delta \hat B_\mu}\hat Q_\mu.
\end{equation}
Once the SW map is employed, one may choose to map the action above, making it
\begin{equation}
\begin{split}
S_{\rm loop}=&S_{\rm gf}\big[B_\mu,Q_\mu,\hat{\bar C}, C,\hat F\big]+S_{\rm NCYM}\big[\hat B_\mu[B_\mu]+\hat Q_\mu[Q_\mu,B_\mu]\big]
\\&-S_{\rm NCYM}\big[\hat B_\mu[B_\mu]\big]-\int \frac{\delta S_{\rm NCYM}\big[\hat B_\mu[B_\mu]\big]}{\delta \hat B_\mu}[B_\mu,Q_\mu]\hat Q_\mu[B_\mu,Q_\mu];
\end{split}
\label{sloop}
\end{equation}
or to map the classical gauge-fixed action then subtract the equations of motion with respect to the commutative/ordinary fields, i.e.
\begin{equation}
\begin{split}
S'_{\rm loop}=&S_{\rm gf}[B_\mu,Q_\mu,\hat{\bar C}, C,\hat F]
\\&+S_{\rm NCYM}\big[\hat B_\mu[B_\mu]+\hat Q_\mu[Q_\mu,B_\mu]\big]-S_{\rm NCYM}\big[\hat B_\mu[B_\mu]\big]-\int \frac{\delta S_{\rm NCYM}\big[\hat B_\mu[B_\mu]\big]}{\delta B_\mu} Q_\mu.
\end{split}
\end{equation}
These two actions are equivalent on-shell as long as the Seiberg-Witten map is invertible, as proven in the appendix A, yet they are but not identical to each other because of the additional field redefinition factor. We choose to proceed with $S_{\rm loop}$ in the computations presented below. As we will see soon, this choice leads to result directly identical to the computation using noncommutative fields, i.e.\footnote{Our prior computation in~\cite{Martin:2016zon} would actually correspond to the same evaluation but with $S'_{\rm loop}$, which is, because of the proof in the appendix A, equivalent to the results here on-shell.}
\begin{equation}
\hat\Gamma^{\mu\nu}(p)=\Gamma^{\mu\nu}(p).
\end{equation}


We are going to use the extended version of dimensional regularization scheme as in~\cite{Martin:2016zon}, which we know to be compatible with the prescriptions used in subsection 3.1. To simplify the computation we choose $\alpha=1$ and have the auxiliary field $F$ integrated out.

\subsection{Model definition}

As our first test we choose $S_{\rm gf}$ to be the background field gauge with respect to the noncommutative fields, i.e.
\begin{equation}
S_{\rm gf}=S_{\rm BFG}=\frac{1}{g^2}\int \tr\, \hat\delta_{\rm BRS}\, \hat{\bar C}\left(\alpha\hat F+ \hat D_\mu\big[\hat B_\mu\big]\hat Q^\mu\right).
\end{equation}
The U(1) theory version of \eqref{sloop} then reads
\begin{equation}
\begin{split}
S_{\rm U(1)_{\rm loop}}=&-\frac{1}{4g^2}\int \left(\hat D_\mu\big[\hat B_\mu\big] {\hat Q}_\nu-\hat D_\nu\big[\hat B_\mu\big] {\hat Q}_\mu\right)^2
-\frac{i}{2g^2}\int \hat F^{\mu\nu}\big[\hat B\big]\left[{\hat Q}_\mu\stackrel{\star}{,}{\hat Q}_\nu\right]
\\&-\frac{i}{2g^2}\int\left(\hat D_\mu\big[\hat B_\mu\big] {\hat Q}_\nu-\hat D_\nu\big[\hat B_\mu\big] {\hat Q}_\mu\right) \left[{\hat Q}_\mu\stackrel{\star}{,}{\hat Q}_\nu\right]+\frac{1}{4g^2}\int\left(\left[{\hat Q}_\mu\stackrel{\star}{,}{\hat Q}_\nu\right]\right)^2
\\&-\frac{1}{g^2}\int\left(\frac{1}{2}\left(\hat D_\mu\big[\hat B_\mu\big]{\hat Q}^{\mu}\right)^2+\bar C\hat D_\mu\big[\hat B_\mu\big]\hat D^\mu\big[\hat B\big]{\hat C}\right).
\end{split}
\label{S1U1}
\end{equation}
To perform the one-loop computation we must expand this action up to the $BBQQ$ order, which is worked out in details in the appendix D. in the end we get\footnote{We assume $g=\hbar=1$ from now on for simplicity, actually coupling is $g$ for $BQQ$ and $g^2$ for $BBQQ$. As a convention interactions with subindex $2$ are derived from $S_{\rm gf}$, while those with subindex $1$ are from the rest of $S_{\rm loop}$. We assume $\hat{\bar C}\equiv \bar C$ from now on, too.}
\begin{equation}
\begin{split}
S^{(1)}_{\rm U(1)}=&-\frac{1}{4}\int\big(\partial_\mu Q_\nu-\partial_\nu Q_\mu\big)^2-\frac{1}{2}\big(\partial_\mu Q^\mu\big)^2-{\bar C}\Box C
\\&+S_{BQQ}+S_{BBQQ}+S_{Bc\bar c}+S_{BBc\bar c}+\mathcal{O}(BBB),
\end{split}
\end{equation}
where
\begin{equation}
\begin{split}
S_{BQQ}=&S_{BQQ_1}+S_{BQQ_2},
\\
S_{BQQ_1}=&-\frac{1}{2}\int i B_{\mu\nu}\left[Q^{\mu}\stackrel{\star}{,}Q^{\nu}\right]
\\&
+Q_{\mu\nu}\theta^{ij}\left(B_{i\mu}\star_2 Q_{j\nu}+Q_{i\mu}\star_2 B_{j\nu}-B_i\star_2\partial_j Q_{\mu\nu}-Q_i\star_2\partial_j B_{\mu\nu}\right),
\end{split}
\label{S1U1}
\end{equation}
\begin{equation}
S_{BQQ_2}=-\int (\partial_\mu Q^\mu)(\partial^\nu\hat{\hat Q}^{(1)}_\nu)+i(\partial_\mu Q^\mu)\left[B_\mu\stackrel{\star}{,}Q^{\mu}\right],
\end{equation}
\begin{equation}
S_{BBQQ}=S_{BBQQ_1}+S_{BBQQ_2},
\end{equation}
\begin{equation}
\begin{split}
S_{BBQQ_1}=-\frac{1}{4}\int & \Big(\theta^{ij}\big(B_{i\mu}\star_2 Q_{j\nu}+Q_{i\mu}\star_2 B_{j\nu}-B_i\star_2\partial_j Q_{\mu\nu}-Q_i\star_2\partial_j B_{\mu\nu}\big)\Big)^2
\\&+4 Q^{\mu\nu} \partial_\mu \hat{\hat Q}^{(2)}_\nu+4i B^{\mu\nu}\left[ \hat{\hat Q}^{(1)}_\mu\stackrel{\star}{,}Q_\nu\right]+4iQ^{\mu\nu}\left[B_\mu \stackrel{\star}{,}\hat{\hat Q}^{(1)}_\nu\right]+{\rm irrelevant},
\end{split}
\end{equation}
\begin{equation}
\begin{split}
S_{BBQQ_2}=&\int -\frac{1}{2} \left(\partial^\mu \hat{\hat Q}^{(1)}_\mu\right)^2-(\partial_\nu Q^\nu)\left(\partial^{\mu}\hat{\hat Q}^{(2)}_\mu\right)
-i(\partial_\nu Q^\nu)\left[B_\mu \stackrel{\star}{,}\hat{\hat Q}^{(1)}_\mu\right]
\\&
-i\left[B_\mu\stackrel{\star}{,}Q^{\mu}\right]\left(\partial^\mu \hat{\hat Q}^{(1)}_\mu\right)+\frac{1}{2}\left(\left[B_\mu\stackrel{\star}{,}Q^{\mu}\right]\right)^2+{\rm irrelevant},
\end{split}
\end{equation}
and
\begin{equation}
S_{BC\bar C}=-\int \bar C\Box \hat{\hat C}^{(1)}+i\bar C\partial_\mu\left[ B^\mu\stackrel{\star}{,}C\right]+\bar C\left[ B^\mu\stackrel{\star}{,}\partial_\mu C\right],
\end{equation}
\begin{equation}
S_{BBC\bar C}=\int-i\bar C \partial_\mu\left[ B^\mu\stackrel{\star}{,}\hat{\hat C}^{(1)}\right] +i\left[ B^\mu\stackrel{\star}{,}\bar C\right]\partial^\mu \hat{\hat C}^{(1)}-\left[ B_\mu\stackrel{\star}{,}\bar C\right]\left[ B^\mu\stackrel{\star}{,} C\right]+{\rm irrelevant}.
\end{equation}
Note that we use $B_{\mu\nu}\equiv \partial_\mu B_\nu-\partial_\nu B_\mu$ and $Q_{\mu\nu}\equiv \partial_\mu Q_\nu-\partial_\nu Q_\mu$ in the equations above. Operators $\hat{\hat Q}_\mu$ and $\hat{\hat C}$ are defined in \eqref{Qhathatdefinition} and \eqref{Chathatdefinition}, respectively. Here and later ``irrelevant'' denotes those four-field-interaction terms which do not generate nontrivial nonlocal factor and/or denominator in the tadpole diagrams. Their contributions to tadpole is then zero under dimensional regularization because of the reasons given in the section 3 and the appendix C. They would still be needed for loop corrections to the three and higher point functions. The interaction Feynman rules are read out from the interactions listed above, and given in the appendix E.1.

\subsection{One-loop quantum corrections in the background field gauge}

\begin{figure}
\begin{center}
\includegraphics[width=6cm]{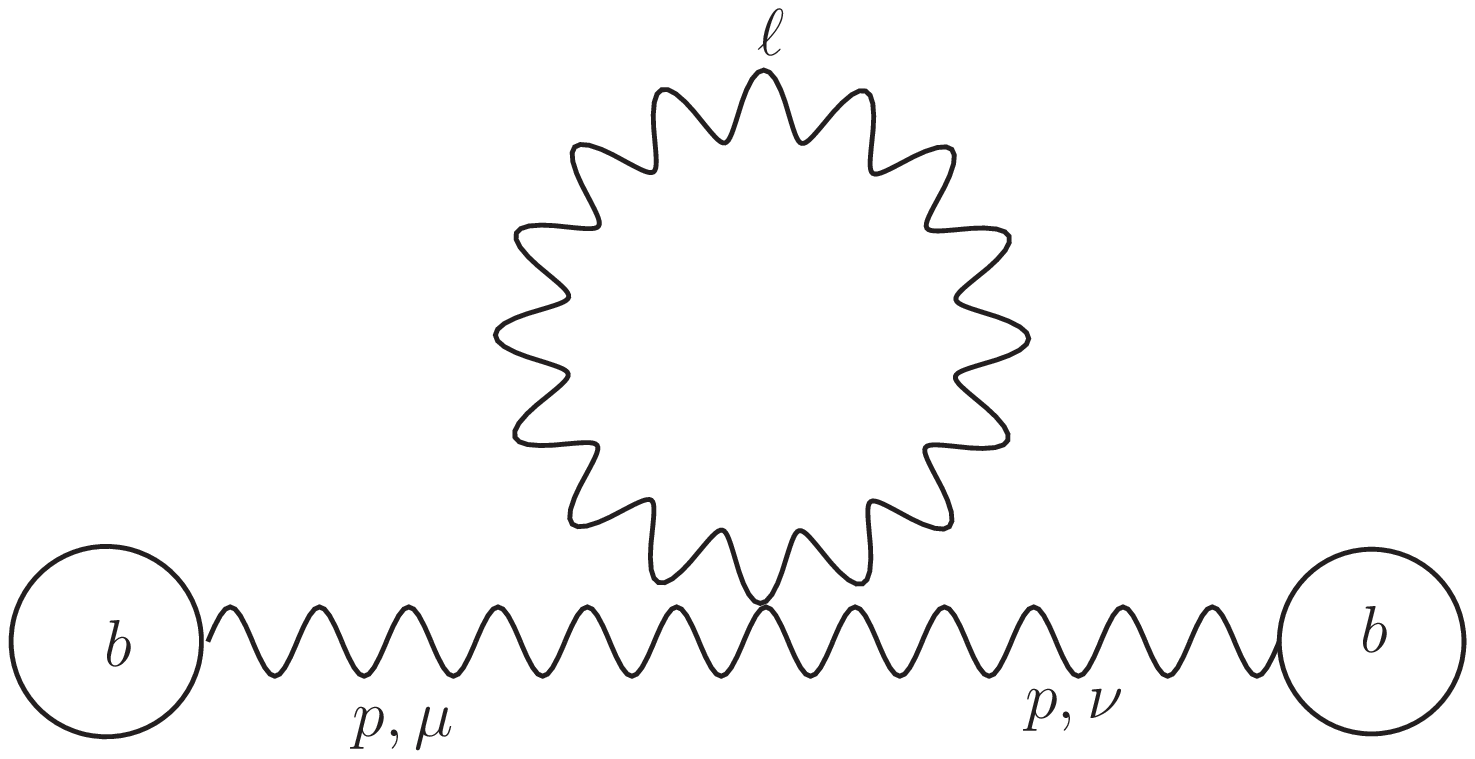}
\end{center}
\caption{Four-photon tadpole contribution to the photon two-point function $T_{\rm BFG_{\rm photon}}^{\mu\nu}$.}
\label{fig:BFMphotontad}
\end{figure}

\begin{figure}
\begin{center}
\includegraphics[width=6cm]{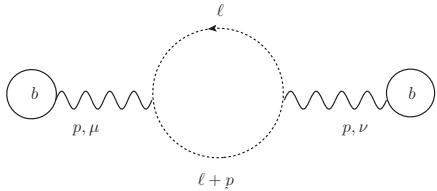}
\end{center}
\caption{Photon-ghost bubble contribution to the photon two-point function $B_{\rm BFG_{\rm ghost}}^{\mu\nu}$ .}
\label{fig:BFMphotonghostbubble}
\end{figure}

\begin{figure}
\begin{center}
\includegraphics[width=6cm]{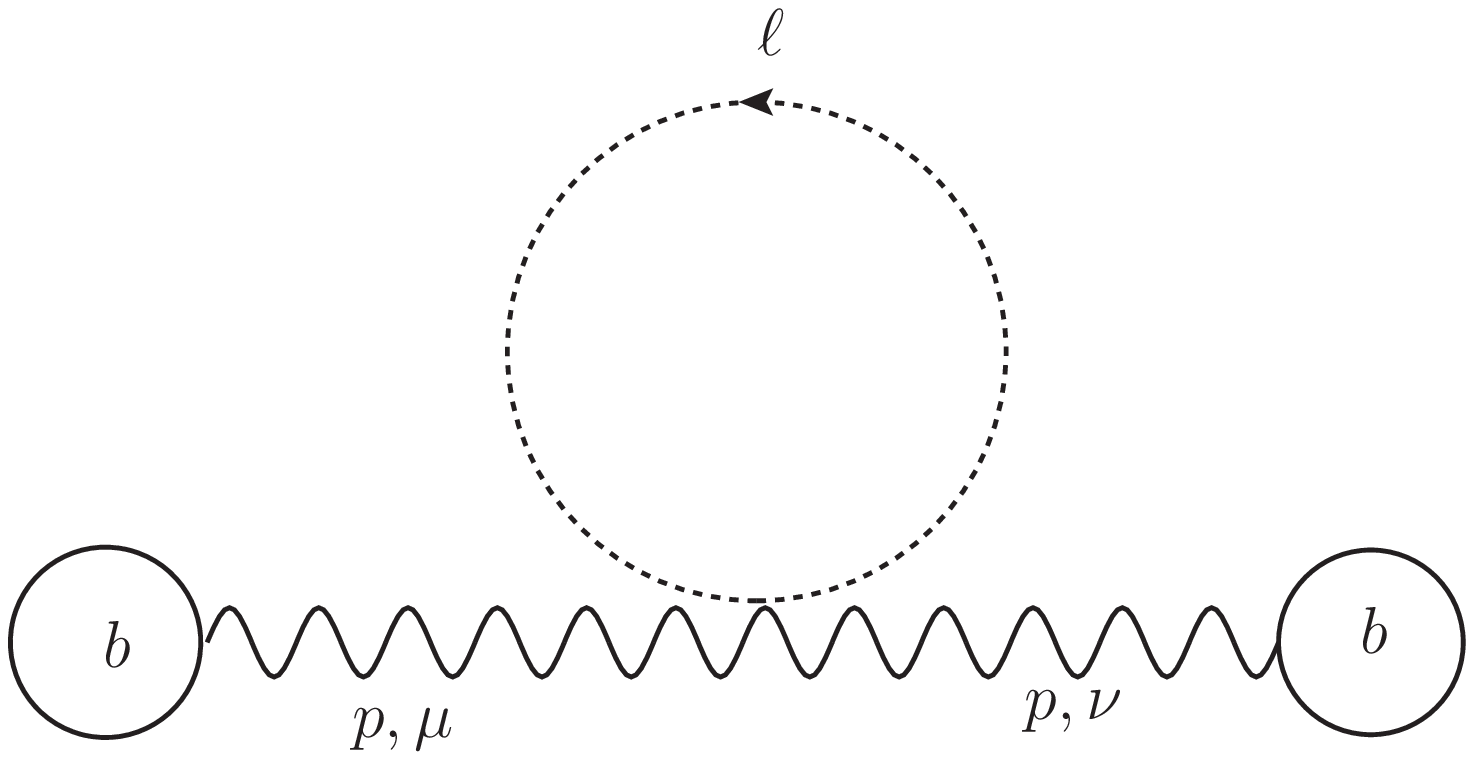}
\end{center}
\caption{Photon-ghost tadpole contribution to the photon two-point function $T_{\rm BFG_{\rm ghost}}^{\mu\nu}$.}
\label{fig:BFMphotonghosttad}
\end{figure}

The one-loop photon 1-PI two-point function computation in the background-field gauge consists four diagrams (Figures~\ref{fig:BFMphotonbubble}-\ref{fig:BFMphotonghosttad}): the photon self-interacting bubble $B_{\rm BFG_{\rm photon}}^{\mu\nu}$ and tadpole $T_{\rm BFG_{\rm photon}}^{\mu\nu}$, as well as the ghost bubble $B_{\rm BFG_{\rm ghost}}^{\mu\nu}$ and tadpole $T_{\rm BFG_{\rm ghost}}^{\mu\nu}$:
\begin{figure}
\begin{center}
\includegraphics[width=6cm]{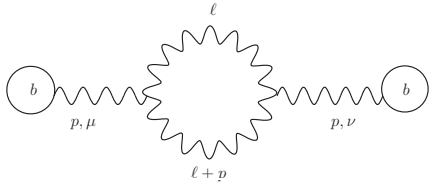}
\end{center}
\caption{Three-photon bubble contribution to the photon two-point function $B_{\rm BFG_{\rm photon}}^{\mu\nu}$.}
\label{fig:BFMphotonbubble}
\end{figure}
\begin{equation}
\Gamma_{\rm BFG}^{\mu\nu}=B_{\rm BFG_{\rm photon}}^{\mu\nu}+T_{\rm BFG_{\rm photon}}^{\mu\nu}+B_{\rm BFG_{\rm ghost}}^{\mu\nu}+T_{\rm BFG_{\rm ghost}}^{\mu\nu},
\end{equation}
with
\begin{equation}
\begin{split}
B_{\rm BFG_{\rm photon}}^{\mu\nu}=&\frac{1}{2}\int\frac{d^D\ell}{(2\pi)^D}\frac{-ig_{\rho_1\rho_2}}{\ell^2}
\frac{-ig_{\sigma_1\sigma_2}}{(\ell+p)^2}\Gamma_{BQQ_{BFG}}^{\mu\rho_1\sigma_2}\left(p;\ell,-p-\ell\right)\Gamma_{BQQ_{BFG}}^{\nu\rho_2\sigma_2}\left(-p;-\ell,p+\ell\right)
\\=&\mathcal{B}_{1_{\rm BFG}}^{\mu\nu}+\mathcal{B}_{2_{\rm BFG}}^{\mu\nu},
\end{split}
\end{equation}
\begin{equation}
\begin{split}
\mathcal{B}_{1_{\rm BFG}}^{\mu\nu}=\frac{1}{2}\int\frac{d^D\ell}{(2\pi)^D}\frac{-ig_{\rho_1\rho_2}}{\ell^2}&
\frac{-ig_{\sigma_1\sigma_2}}{(\ell+p)^2}\Big(\Gamma_{BQQ_{BFG_1}}^{\mu\rho_1\sigma_2}\left(p;\ell,-p-\ell\right)\Gamma_{BQQ_{BFG_1}}^{\nu\rho_2\sigma_2}\left(-p;-\ell,p+\ell\right)
\\&+\Gamma_{BQQ_{BFG_1}}^{\mu\rho_1\sigma_2}\left(p;\ell,-p-\ell\right)\Gamma_{BQQ_{BFG_2}}^{\nu\rho_2\sigma_2}\left(-p;-\ell,p+\ell\right)
\\&+\Gamma_{BQQ_{BFG_2}}^{\mu\rho_1\sigma_2}\left(p;\ell,-p-\ell\right)\Gamma_{BQQ_{BFG_1}}^{\nu\rho_2\sigma_2}\left(-p;-\ell,p+\ell\right)\Big),
\end{split}
\end{equation}
\begin{equation}
\mathcal{B}_{2_{\rm BFG}}^{\mu\nu}=\frac{1}{2}\int\frac{d^D\ell}{(2\pi)^D}\frac{-ig_{\rho_1\rho_2}}{\ell^2}
\frac{-ig_{\sigma_1\sigma_2}}{(\ell+p)^2}\Gamma_{BQQ_{BFG_2}}^{\mu\rho_1\sigma_2}\left(p;\ell,-p-\ell\right)\Gamma_{BQQ_{BFG_2}}^{\nu\rho_2\sigma_2}\left(-p;-\ell,p+\ell\right),
\end{equation}
\begin{equation}
\begin{split}
T_{\rm BFG_{\rm photon}}^{\mu\nu}=&\frac{1}{2}\int\frac{d^D\ell}{(2\pi)^D}\frac{-ig_{\rho_1\rho_2}}{\ell^2}\Gamma_{BBQQ_{BFG}}^{\mu\nu\rho_1\rho_2}\left(p,-p;\ell,-\ell\right)
\\=&\mathcal{T}_{1_{\rm BFG}}^{\mu\nu}+\mathcal{T}_{2_{\rm BFG}}^{\mu\nu},
\end{split}
\end{equation}
\begin{equation}
\mathcal{T}_{1_{\rm BFG}}^{\mu\nu}=\frac{1}{2}\int\frac{d^D\ell}{(2\pi)^D}\frac{-ig_{\rho_1\rho_2}}{\ell^2}\Gamma_{BBQQ_{BFG_1}}^{\mu\nu\rho_1\rho_2}\left(p,-p;\ell,-\ell\right),
\end{equation}
\begin{equation}
\mathcal{T}_{2_{\rm BFG}}^{\mu\nu}=\frac{1}{2}\int\frac{d^D\ell}{(2\pi)^D}\frac{-ig_{\rho_1\rho_2}}{\ell^2}\Gamma_{BBQQ_{BFG_2}}^{\mu\nu\rho_1\rho_2}\left(p,-p;\ell,-\ell\right),
\end{equation}
\begin{equation}
B_{\rm BFG_{\rm ghost}}^{\mu\nu}=-\int\frac{d^D\ell}{(2\pi)^D}\frac{i}{\ell^2}\frac{i}{(\ell+p)^2}\Gamma_{Bc\bar c_{BFG}}^\mu\left(p;\ell\right)\Gamma_{Bc\bar c_{BFG}}^\nu\left(-p;p+\ell\right),
\end{equation}
\begin{equation}
T_{\rm BFG_{\rm ghost}}^{\mu\nu}=-\int\frac{d^D\ell}{(2\pi)^D}\frac{i}{\ell^2}\Gamma_{BBc\bar c_{BFG}}^{\mu\nu}\left(p,-p;\ell,\ell\right).
\end{equation}
One can prove that\footnote{This cancellation actually indicates that the stand-alone gauge-fixing contribution to the 1PI photon two point function vanishes. Gauge fixing contributions still exist via the products of $\Gamma_1$ and $\Gamma_2$ in $\mathcal{B}_1$. However if one replaces $S_{\rm loop}$ by $S'_{\rm loop}$, this effect also disappears because $\Gamma_1$ is orthogonal to $\Gamma_2$ in that case, the final result for $S'_{\rm loop}$ then goes back to~\cite{Martin:2016zon} because the background-field splitting becomes trivial in that case.}
\begin{equation}
\mathcal{B}_{2_{\rm BFG}}^{\mu\nu}+\mathcal{T}_{2_{\rm BFG}}^{\mu\nu}+B_{\rm BFG_{\rm ghost}}^{\mu\nu}+T_{\rm BFG_{\rm ghost}}^{\mu\nu}=0.
\end{equation}
So
\begin{equation}
\Gamma_{\rm BFG}^{\mu\nu}=\mathcal{B}_{1_{\rm BFG}}^{\mu\nu}+\mathcal{T}_{1_{\rm BFG}}^{\mu\nu}.
\end{equation}
Explicit computation then yields
\begin{equation}
\begin{split}
\mathcal{B}_{1_{\rm BFG}}^{\mu\nu}=&\frac{1}{(4\pi)^2}\Bigg(\Big(g^{\mu\nu}p^2-p^\mu p^\nu\Big)
\\&
\cdot\bigg((4\pi\mu^2)^{2-\frac{D}{2}}(p^2)^{\frac{D}{2}-2}2(6-7D){\rm\Gamma}\left(1-\frac{D}{2}\right){\rm B}\left(\frac{D}{2}, \frac{D}{2}\right)\bigg|_{D\to 4-\epsilon}
\\&-12I_{K_0}-16I_{K_1}\bigg)-g^{\mu\nu}p^2(\theta p)^2 T_{-2}
-\frac{(\theta p)^\mu(\theta p)^\nu}{(\theta p)^2}\bigg(\frac{16}{3}T_0+8I_K^0-48p^2I_K^1\bigg)\Bigg),
\end{split}
\end{equation}
\begin{equation}
\mathcal{T}_{1_{\rm BFG}}^{\mu\nu}=\frac{1}{(4\pi)^2}\left(g^{\mu\nu}p^2(\theta p)^2(\theta p)^2 T_{-2}-\frac{(\theta p)^\mu(\theta p)^\nu}{(\theta p)^2}\frac{32}{3}T_0\right).
\end{equation}
Thus
\begin{equation}
\begin{split}
\Gamma_{\rm BFG}^{\mu\nu}=&\frac{1}{(4\pi)^2}\Bigg(\Big(g^{\mu\nu}p^2-p^\mu p^\nu\Big)
\\&
\cdot\bigg((4\pi\mu^2)^{2-\frac{D}{2}}(p^2)^{\frac{D}{2}-2}2(6-7D){\rm\Gamma}\left(1-\frac{D}{2}\right){\rm B}\left(\frac{D}{2}, \frac{D}{2}\right)\bigg|_{D\to 4-\epsilon}
\\&-12I_{K_0}-16I_{K_1}\bigg)
-\frac{(\theta p)^\mu(\theta p)^\nu}{(\theta p)^2}\bigg(16T_0+8I_K^0-48p^2I_K^1\bigg)\Bigg).
\end{split}
\end{equation}
This result exactly matches $\hat \Gamma_{\rm BFG}^{\mu\nu}$, eqs. (\ref{NCSYMPhotontotal}-\ref{NCSYMPhotontadpoles}). Using the fact that $T_0=-2/(\theta p)^2$ \cite{Martin:2016zon} one can immediately recover the same quadratic IR divergence equals to $32(\theta p)^\mu(\theta p)^\nu/(\theta p)^4$, which is the same as noncommutative $\rm U(1)$ theory~\cite{Hayakawa:1999yt,Hayakawa:1999zf}. Now the UV divergent part of $\Gamma_{\rm BFG}^{\mu\nu}$ at the $D\to 4-\epsilon$ limit reads
\begin{equation}
\Gamma_{\rm BFG}^{\mu\nu}\big|_{\rm UV}=\frac{1}{(4\pi)^2}\Big(g^{\mu\nu}p^2-p^\mu p^\nu\Big)\frac{22}{3}\left(\frac{2}{\epsilon}+\ln(\mu^2(\theta p)^2)\right).
\end{equation}
This coefficient $22/3$ matches exactly the coefficient for $\beta(g)$ of the NC $\rm U(1)$ theory~\cite{Martin:1999aq,Hayakawa:1999zf}.

\subsection{One-loop corrections in the noncommutative Feynman gauge}

We perform a second test on the gauge-fixing (in-)dependence by shifting from the background field gauge fixing $\hat D_\mu[\hat B_\mu] \hat Q^\mu$, to the NC Feynman gauge fixing (NCFG) $\partial_\mu\hat Q^\mu$. The standard background field method procedure then leads us to a modification to the following action
\begin{equation}
\begin{split}
S^{(1)}_{\rm U(1)_{\rm NCFG}}=&-\frac{1}{4}\int \left(\hat D_\mu\big[\hat B_\mu\big] \hat{\hat Q}_\nu-\hat D_\nu\big[\hat B_\mu\big] \hat{\hat Q}_\mu\right)^2 -\frac{i}{2}\int \hat F^{\mu\nu}\big[\hat B_\mu\big]\Big[\hat{\hat Q}_\mu\stackrel{\star}{,}\hat{\hat Q}_\nu\Big]
\\&-\int\left(\frac{1}{2}\left(\partial_\mu\big[\hat B_\mu\big]\hat{\hat Q}^{\mu}\right)^2+\bar C\partial_\mu\hat D^\mu\big[\hat B_\mu\big]\hat{\hat C}\right).
\end{split}
\label{S1U1}
\end{equation}
The resulted Feynman rules are listed in the appendix E.2. In analogy to the background-field gauge, we have the following one-loop contributions
\begin{equation}
\Gamma_{\rm NCFG-BFM}^{\mu\nu}=B_{\rm NCFG-BFM_{\rm photon}}^{\mu\nu}+T_{\rm NCFG-BFM_{\rm photon}}^{\mu\nu}+B_{\rm NCFG-BFM_{\rm ghost}}^{\mu\nu}+T_{\rm NCFG-BFM_{\rm ghost}}^{\mu\nu},
\end{equation}
with
\begin{equation}
\begin{split}
&B_{\rm NCFG-BFM_{\rm photon}}^{\mu\nu}
\\=&\frac{1}{2}\int\frac{d^D\ell}{(2\pi)^D}\frac{-ig_{\rho_1\rho_2}}{\ell^2}
\frac{-ig_{\sigma_1\sigma_2}}{(\ell+p)^2}\Gamma_{BQQ_{NCFG-BFM}}^{\mu\rho_1\sigma_2}\left(p;\ell,-p-\ell\right)\Gamma_{BQQ_{NCFG-BFM}}^{\nu\rho_2\sigma_2}\left(-p;-\ell,p+\ell\right)
\\=&\mathcal{B}_{1_{\rm NCFG-BFM}}^{\mu\nu}+\mathcal{B}_{2_{\rm NCFG-BFM}}^{\mu\nu},
\end{split}
\end{equation}
\begin{equation}
\begin{split}
\mathcal{B}_{1_{\rm NCFG-BFM}}^{\mu\nu}=\frac{1}{2}&\int\frac{d^D\ell}{(2\pi)^D}\frac{-ig_{\rho_1\rho_2}}{\ell^2}
\frac{-ig_{\sigma_1\sigma_2}}{(\ell+p)^2}
\\&\cdot\Big(\Gamma_{BQQ_{NCFG-BFM_1}}^{\mu\rho_1\sigma_2}\left(p;\ell,-p-\ell\right)\Gamma_{BQQ_{NCFG-BFM_1}}^{\nu\rho_2\sigma_2}\left(-p;-\ell,p+\ell\right)
\\&+\Gamma_{BQQ_{NCFG-BFM_1}}^{\mu\rho_1\sigma_2}\left(p;\ell,-p-\ell\right)\Gamma_{BQQ_{NCFG-BFM_2}}^{\nu\rho_2\sigma_2}\left(-p;-\ell,p+\ell\right)
\\&+\Gamma_{BQQ_{NCFG-BFM_2}}^{\mu\rho_1\sigma_2}\left(p;\ell,-p-\ell\right)\Gamma_{BQQ_{NCFG-BFM_1}}^{\nu\rho_2\sigma_2}\left(-p;-\ell,p+\ell\right)\Big),
\end{split}
\end{equation}
\begin{equation}
\begin{split}
&\mathcal{B}_{2_{\rm NCFG-BFM}}^{\mu\nu}
\\
=&\frac{1}{2}\int\frac{d^D\ell}{(2\pi)^D}\frac{-ig_{\rho_1\rho_2}}{\ell^2}
\frac{-ig_{\sigma_1\sigma_2}}{(\ell+p)^2}\Gamma_{BQQ_{NCFG-BFM_2}}^{\mu\rho_1\sigma_2}\left(p;\ell,-p-\ell\right)\Gamma_{BQQ_{NCFG-BFM_2}}^{\nu\rho_2\sigma_2}\left(-p;-\ell,p+\ell\right),
\end{split}
\end{equation}
\begin{equation}
\begin{split}
T_{\rm NCFG-BFM_{\rm photon}}^{\mu\nu}=&\frac{1}{2}\int\frac{d^D\ell}{(2\pi)^D}\frac{-ig_{\rho_1\rho_2}}{\ell^2}\Gamma_{BBQQ_{NCFG-BFM}}^{\mu\nu\rho_1\rho_2}\left(p,-p;\ell,-\ell\right)
\\=&\mathcal{T}_{1_{\rm NCFG-BFM}}^{\mu\nu}+\mathcal{T}_{2_{\rm NCFG-BFM}}^{\mu\nu},
\end{split}
\end{equation}
\begin{equation}
\mathcal{T}_{1_{\rm NCFG-BFM}}^{\mu\nu}=\frac{1}{2}\int\frac{d^D\ell}{(2\pi)^D}\frac{-ig_{\rho_1\rho_2}}{\ell^2}\Gamma_{BBQQ_{NCFG-BFM_1}}^{\mu\nu\rho_1\rho_2}\left(p,-p;\ell,-\ell\right),
\end{equation}
\begin{equation}
\mathcal{T}_{2_{\rm NCFG-BFM}}^{\mu\nu}=\frac{1}{2}\int\frac{d^D\ell}{(2\pi)^D}\frac{-ig_{\rho_1\rho_2}}{\ell^2}\Gamma_{BBQQ_{NCFG-BFM_2}}^{\mu\nu\rho_1\rho_2}\left(p,-p;\ell,-\ell\right),
\end{equation}
\begin{equation}
B_{\rm NCFG-BFM_{\rm ghost}}^{\mu\nu}=-\int\frac{d^D\ell}{(2\pi)^D}\frac{i}{\ell^2}\frac{i}{(\ell+p)^2}\Gamma_{Bc\bar c_{NCFG-BFM}}^\mu\left(p;\ell\right)\Gamma_{Bc\bar c_{NCFG-BFM}}^\nu\left(-p;p+\ell\right),
\end{equation}
\begin{equation}
T_{\rm NCFG-BFM_{\rm ghost}}^{\mu\nu}=-\int\frac{d^D\ell}{(2\pi)^D}\frac{i}{\ell^2}\Gamma_{BBc\bar c_{NCFG-BFM}}^{\mu\nu}\left(p,-p;\ell,\ell\right).
\end{equation}
Again
\begin{equation}
\mathcal{B}_{2_{\rm NCFG-BFM}}^{\mu\nu}+\mathcal{T}_{2_{\rm NCFG-BFM}}^{\mu\nu}+B_{\rm NCFG-BFM_{\rm ghost}}^{\mu\nu}+T_{\rm NCFG-BFM_{\rm ghost}}^{\mu\nu}=0,
\end{equation}
so
\begin{equation}
\Gamma_{\rm NCFG-BFM}^{\mu\nu}=\mathcal{B}_{1_{\rm NCFG-BFM}}^{\mu\nu}+\mathcal{T}_{1_{\rm NCFG-BFM}}^{\mu\nu}.
\end{equation}
Explicit computation then yields
\begin{equation}
\begin{split}
\mathcal{B}_{1_{\rm NCFG-BFM}}^{\mu\nu}=&\frac{1}{(4\pi)^2}\Bigg(\Big(g^{\mu\nu}p^2-p^\mu p^\nu\Big)
\\&\cdot\bigg((4\pi\mu^2)^{2-\frac{D}{2}}(p^2)^{\frac{D}{2}-2}2(2-3D){\rm\Gamma}\left(1-\frac{D}{2}\right){\rm B}\left(\frac{D}{2}, \frac{D}{2}\right)\bigg|_{D\to 4-\epsilon}
\\&-8I_{K_0}-16I_{K_1}\bigg)+p^\mu p^\nu(\theta p)^2 T_{-2}
-\frac{(\theta p)^\mu(\theta p)^\nu}{(\theta p)^2}\bigg(\frac{16}{3}T_0+8I_K^0-48p^2I_K^1\bigg)\Bigg),
\end{split}
\end{equation}
\begin{equation}
\mathcal{T}_{1_{\rm NCFG-BFM}}^{\mu\nu}=-\frac{1}{(4\pi)^2}\left(p^\mu p^\nu(\theta p)^2 T_{-2}+\frac{(\theta p)^\mu(\theta p)^\nu}{(\theta p)^2}\frac{32}{3}T_0\right).
\end{equation}
Consequently
\begin{equation}
\begin{split}
\Gamma_{\rm NCFG-BFM}^{\mu\nu}=&\frac{1}{(4\pi)^2}\Bigg(\Big(g^{\mu\nu}p^2-p^\mu p^\nu\Big)
\\&\cdot
\bigg((4\pi\mu^2)^{2-\frac{D}{2}}(p^2)^{\frac{D}{2}-2}2(2-3D){\rm\Gamma}\left(1-\frac{D}{2}\right){\rm B}\left(\frac{D}{2}, \frac{D}{2}\right)\bigg|_{D\to 4-\epsilon}
\\&-8I_{K_0}-16I_{K_1}\bigg)
-\frac{(\theta p)^\mu(\theta p)^\nu}{(\theta p)^2}\bigg(16T_0+8I_K^0-48p^2I_K^1\bigg)\Bigg).
\end{split}
\end{equation}
This result matches the computations in the Feynman gauge without Seiberg-Witten map in the literature~\cite{Martin:1999aq,Hayakawa:1999yt,Hayakawa:1999zf}. Since the result without Seiberg-Witten map is equivalent to the background field gauge result on shell~\cite{Kallosh:1974yh,Ichinose:1992np}, we conclude that the Seiberg-Witten mapped result here fulfills this equivalence too.

\subsection{One-loop corrections in the noncommutative U(1) Super Yang-Mils}

We also investigate whether our method can be used to remove non-polynomial UV divergences in the 1-PI two point functions of the superpartners, i.e. the photinos and adjoint scalars. Our starting actions are as follows
\begin{equation}
S_{\rm photino}=\int i\hat{\bar\lambda}\bar\sigma^\mu\hat D_\mu\hat\lambda,
\end{equation}
\begin{equation}
S_{\rm scalar}=\int\frac{1}{2}\hat D_\mu\hat\phi \hat D^\mu\hat\phi.
\end{equation}
In this case after the background field splitting $\hat\lambda=\hat\lambda_B+\hat\lambda_Q$ and $\hat\phi=\hat\phi_B+\hat\phi_Q$ we must subtract both the equations of motion of superpartner fields, and their contributions as source of the photon equations of motion, the resulted action for loop computation is listed below
\begin{equation}
\begin{split}
S_{\rm photino}^{(1)}&=\int i\Big(\hat{\bar\lambda}_Q\bar\sigma^\mu\hat D_\mu[\hat B_\mu]\hat\lambda_Q+i\hat{\bar\lambda}_Q\bar\sigma^\mu\left[\hat{\hat Q}_\mu\stackrel{\star}{,}\hat\lambda_B\right]+i\hat{\bar\lambda}_B\bar\sigma^\mu\left[\hat{\hat Q}_\mu\stackrel{\star}{,}\hat\lambda_Q\right]\Big),
\\
S_{\rm scalar}^{(1)}&=\int\frac{1}{2} \bigg(\hat D_\mu[\hat B_\mu]\hat\phi_Q \hat D^\mu[\hat B_\mu]\hat\phi_Q
+2i\left(\hat D_\mu[\hat B_\mu]\hat\phi_B\left[\hat{\hat Q}_\mu\stackrel{\star}{,}\hat\phi_Q\right]+\hat D_\mu[\hat B_\mu]\hat\phi_Q\left[\hat{\hat Q}_\mu\stackrel{\star}{,}\hat\phi_B\right]\right)
\\&-\left[\hat{\hat Q}_\mu\stackrel{\star}{,}\hat\phi_B\right]\left[\hat{\hat Q}^\mu\stackrel{\star}{,}\hat\phi_B\right]\bigg).
\end{split}
\label{SPartnerActions}
\end{equation}
The relevant SW map can be derived using the background-field splitting method in the subsection 2.2 and results~\cite{Martin:2016zon}. Once we start reading out Feynman rules our first observation is that the superpartner's contribution to the photon effective action is identical to the results in~\cite{Martin:2016zon}. Therefore we have the same quadratic IR divergence cancellation. The total UV divergence in the background-field gauge is now
\begin{equation}
\Gamma_{\rm BFG-total}^{\mu\nu}\big|_{\rm UV}=\frac{1}{(4\pi)^2}\Big(g^{\mu\nu}p^2-p^\mu p^\nu\Big)\left(\frac{22}{3}-\frac{4}{3}{\rm n_f}-\frac{1}{3}{\rm n_s}\right)\left(\frac{2}{\epsilon}+\ln(\mu^2(\theta p)^2)\right).
\end{equation}
Therefore it vanishes for $\mathcal{N}=4$ SUSY, i.e. when $n_f=4,n_s=6$, as expected. The results we have obtained is in full harmony with the results obtained in \cite{Ferrari:2003vs, AlvarezGaume:2003mb,Ferrari:2004ex,Jack:2001cr, Santambrogio:2000rs,Pernici:2000va, Buchbinder:2001at,Hanada:2014ima} by formulating the theory in terms of noncommutative fields.

\begin{figure}
\begin{center}
\includegraphics[width=6cm]{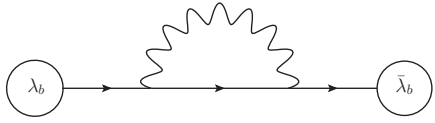}
\end{center}
\caption{Photino-photon BFM-bubble.}
\label{fig:BFM5}
\end{figure}

\begin{figure}
\begin{center}
\includegraphics[width=6cm]{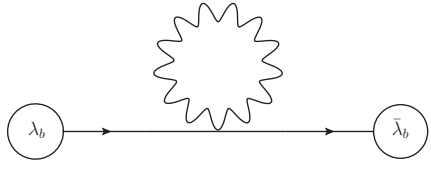}
\end{center}
\caption{Photino-photon BFM-tadpole.}
\label{fig:BFM6}
\end{figure}

\begin{figure}
\begin{center}
\includegraphics[width=6cm]{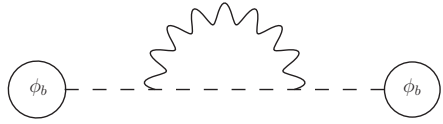}
\end{center}
\caption{Scalar-photon BFM-bubble.}
\label{fig:BFM7}
\end{figure}

\begin{figure}
\begin{center}
\includegraphics[width=6cm]{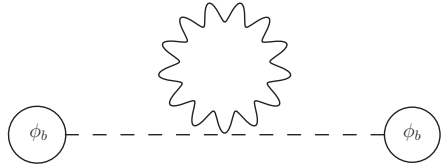}
\end{center}
\caption{Scalar-photon BFM-tadpole.}
\label{fig:BFM8}
\end{figure}

We then use the same action to derive the Feynman rules for computing the one-loop 1-PI two point functions of the superpartners. The FR results are listed in the appendix E.3. These Feynman rules produce the two diagrams Fig.~\ref{fig:BFM5} and Fig.~\ref{fig:BFM6} for 1-loop photino, as well as two diagrams Fig.~\ref{fig:BFM7} and Fig.~\ref{fig:BFM8} for adjoint scalar two-point functions.

Out of Figures~\ref{fig:BFM5}-\ref{fig:BFM8} we read out the following loop integrals for
$\mathcal{N}=1$ photino
\begin{equation}
\Sigma^{\dot{\alpha}\alpha}_{\rm BFM}=\Sigma^{\dot{\alpha}\alpha}_{\rm BFM_{\rm bubble}}+\Sigma^{\dot{\alpha}\alpha}_{\rm BFM_{\rm tadpole}},
\label{lambdaSE}
\end{equation}
\begin{equation}
\Sigma^{\dot{\alpha}\alpha}_{\rm BFM_{\rm bubble}}=\int\frac{d^D\ell}{(2\pi)^D}\Gamma_{\lambda_B\bar\lambda_QQ}^\mu\left(p,\ell+p;\ell\right)\frac{i\sigma^\rho (\ell+p)_\rho}{(\ell+p)^2}\Gamma_{\lambda_Q\bar\lambda_BQ}^\nu\left(p,\ell+p;-\ell\right)\frac{-ig_{\mu\nu}}{\ell^2},
\end{equation}
\begin{equation}
\Sigma^{\dot{\alpha}\alpha}_{\rm BFM_{\rm tadpole}}=\frac{1}{2}\int\frac{d^D\ell}{(2\pi)^D}\frac{-ig_{\mu\nu}}{\ell^2}\Gamma_{\lambda_B\bar\lambda_BQQ}^{\mu\nu}\left(p,p;\ell,-\ell\right),
\end{equation}
and the following for the minimally coupled adjoint scalar
\begin{equation}
\Sigma_{{(\phi)}_{\rm BFM}}=\Sigma_{{(\phi)}_{\rm BFM_{\rm bubble}}}+\Sigma_{{(\phi)}_{\rm BFM_{\rm tadpole}}},
\label{phiSE}
\end{equation}
\begin{equation}
\Sigma_{{(\phi)}_{\rm BFM_{\rm bubble}}}=\int\frac{d^D\ell}{(2\pi)^D}\frac{i}{\ell^2}\frac{-ig_{\mu\nu}}{(\ell+p)^2}\Gamma_{\phi_B\phi_QQ}^\mu\left(p,\ell;-p-\ell\right)\Gamma_{\phi_B\phi_QQ}^\nu\left(-p,-\ell;p+\ell\right),
\end{equation}
\begin{equation}
\Sigma_{{(\phi)}_{\rm BFM_{\rm tadpole}}}=\frac{1}{2}\int\frac{d^D\ell}{(2\pi)^D}\frac{-ig_{\mu\nu}}{\ell^2}\Gamma_{\phi_B\phi_BQQ}^{\mu\nu}\left(p,-p;\ell,-\ell\right).
\end{equation}
Explicit computation then yields
\begin{equation}
\begin{split}
\Sigma^{\dot{\alpha}\alpha}_{\rm BFM_{\rm bubble}}=&\bar\sigma^\mu p_\mu\frac{1}{(4\pi)^2}\bigg((4\pi\mu^2)^{2-\frac{D}{2}}(p^2)^{\frac{D}{2}-2}(2-D){\rm\Gamma}\left(2-\frac{D}{2}\right){\rm B}\left(\frac{D}{2}-1, \frac{D}{2}-1\right)\bigg|_{D\to 4-\epsilon}
\\&-(\theta p)^2 T_{-2}-4I_K^0\bigg),
\end{split}
\end{equation}
\begin{equation}
\Sigma^{\dot{\alpha}\alpha}_{\rm BFM_{\rm tadple}}=\bar\sigma^\mu p_\mu\frac{1}{(4\pi)^2}(\theta p)^2 T_{-2},
\end{equation}
thus
\begin{equation}
\begin{split}
\Sigma^{\dot{\alpha}\alpha}_{\rm BFM}&=\bar\sigma^\mu p_\mu\frac{1}{(4\pi)^2}
\\\cdot&
\left((4\pi\mu^2)^{2-\frac{D}{2}}(p^2)^{\frac{D}{2}-2}(2-D){\rm\Gamma}\left(2-\frac{D}{2}\right){\rm B}\left(\frac{D}{2}-1, \frac{D}{2}-1\right)\bigg|_{D\to 4-\epsilon}+4I_K^0\right),
\end{split}
\label{lambdatotal}
\end{equation}
also
\begin{equation}
\begin{split}
&\Sigma_{{(\phi)}_{\rm BFM_{\rm bubble}}}=p^2\frac{1}{(4\pi)^2}
\\&\cdot \left(-4(4\pi\mu^2)^{2-\frac{D}{2}}(p^2)^{\frac{D}{2}-2}{\rm\Gamma}\left(2-\frac{D}{2}\right){\rm B}\left(\frac{D}{2}-1, \frac{D}{2}-1\right)\bigg|_{D\to 4-\epsilon}+T_{-2}+4T_0+2 I_K^0\right),
\end{split}
\label{phibubble}
\end{equation}
\begin{equation}
\Sigma_{{(\phi)}_{\rm BFM_{\rm tadpole}}}=p^2\frac{1}{(4\pi)^2}\left(-T_{-2}+8T_0\right),
\label{phitadpole}
\end{equation}
so
\begin{equation}
\begin{split}
&\Sigma_{{(\phi)}_{\rm BFM}}=p^2\frac{4}{(4\pi)^2}
\\&\cdot\left(-(4\pi\mu^2)^{2-\frac{D}{2}}(p^2)^{\frac{D}{2}-2}{\rm\Gamma}\left(2-\frac{D}{2}\right){\rm B}\left(\frac{D}{2}-1, \frac{D}{2}-1\right)\bigg|_{D\to 4-\epsilon}+3T_0+2 I_K^0\right).
\end{split}
\label{phitotal}
\end{equation}
Comparing \eqref{lambdatotal} and \eqref{phitotal} with their unexpanded counterparts \eqref{NCSYMPhotino} and (\ref{NCSYMStotal}-\ref{NCSYMStadpole}), one can immediately observe an exact match. On the other hand, this match only occurs when all contributing diagrams are summed together. Individual diagrams, for example \eqref{phibubble} and \eqref{NCSYMSbubble}, or \eqref{phitadpole} and \eqref{NCSYMStadpole}, do not match each other.

Since all other diagrams in the superpartner two point function computation in the SW mapped U(1) NCSYM are identical to the diagrams in the unexpanded theory~\cite{Martin:2016zon} (see also the short summary in the appendix F.2), we conclude that the full 1-PI two point functions/quadratic part of the background field effective actions are identical up to one-loop in $\rm U(1)$ NCSYM with and without SW map.


\section{Discussion and conclusions}

We have shown that at the quantum level the $\theta$-exact Seiberg-Witten map provides --at least in perturbative theory with respect to the coupling constant-- a dual description, in terms of ordinary fields, of the noncommutative U(N) Yang-Mills theory with or without Supersymmetry. We have shown that by performing appropriate changes of variables  in the path integral defining the on-shell DeWitt effective action in dimensional regularization. We have explicitly computed, by using the Feynmann rules derived from the classsical action, the one-loop two-point contribution to the on-shell DeWitt action for U(1) SuperYang-Mills with ${\cal N}$=0,\,1,\,2 and 4 Supersymmetry and found complete agreement with general result obtained by carrying out changes of variables in the path integral. We have also shown that all the nasty non-local noncommutative UV divergences which occur in the one-loop 1PI functional in the Feynman gauge, computed in  \cite{Horvat:2011bs,Horvat:2013rga,Horvat:2015aca,Martin:2016zon} are merely off-shell gauge artifacts since they do not occur in the one-loop two-point contribution to the on-shell DeWitt action --which is a gauge-fixing independent object-- and therefore they do not contribute to any physical quantity. We have also shown that the same quadratic noncommutative IR divergences that occur in nonsupersymmetric noncommutative U(N) gauge theories formulated in terms of noncommutative fields occur in the ordinary theory obtained from the former by using the $\theta$-exact Seiberg-Witten map and that this UV/IR mixing effect --signaling a vacuum instability-- is a gauge-fixing independent characteristic of the ordinary gauge theory, in keeping with the duality statement. We have also  seen that those quadratic noncommutative IR diverges can be removed by considering  supersymmetric versions of the  theory, a nontrivial  effect since  supersymmetry is not linearly realized in terms of the ordinary fields \cite{Martin:2008xa}. Finally, there remain to be seen how the results presented here carry over to the nonpertubative regime in the coupling constant. In this regard the analysis of the nonperturbative features of ${\cal N}=2$ and $4$  supersymmetric gauge theories looks particularly interesting.

\section{Acknowledgments}
The work by C.P. Martin has been financially supported in part by the Spanish MINECO through grant FPA2014-54154-P. J.Y. has been fully supported by Croatian Science Foundation under Project No. IP-2014-09-9582. The work  J.T. is conducted under the European Commission and the Croatian Ministry of Science, Education and Sports Co-Financing Agreement No. 291823. In particular, J.T. acknowledges project financing by the Marie Curie FP7-PEOPLE-2011-COFUND program NEWFELPRO: Grant Agreement No. 69, and  Max-Planck-Institute for Physics, and W. Hollik for hospitality.  We would like to acknowledge L. Alvarez-Gaume, J. Ellis and P. Minkowski for fruitful discussions and CERN Theory Division, where part of this work was conducted, for hospitality.  We would like to acknowledge the COST Action MP1405  (QSPACE). 
We would also like to thank J. Erdmenger and W. Hollik, for fruitful discussions. J.Y. would like to acknowledge the Center of Theoretical Physics, College of Physical Science and Technology, Sichuan University, China, for hospitality during his visit, as well as Yan He, Xiao Liu, Hiroaki Nakajima, Bo Ning, Rakibur Rahman, Zheng Sun, Peng Wang, Houwen Wu, Haitang Yang and Shuxuan Ying for fruitful discussions. J.Y. would also like to acknowledge H2020 CSA Twinning project No. 692194, ``RBI-T-WINNING'' for financially supporting his trip to Max-Planck-Institute for Physics, Munich, Germany and CERN. A great deal of computation was done by using MATHEMATICA 8.0 \cite{mathematica} plus the tensor algebra package xACT~\cite{xAct}. Special thanks to A. Ilakovac and D. Kekez for the computer software and hardware support.

\appendix

\section{Classical equations of motion for the noncommutative and ordinary fields}

In this subsection we prove that the equations of motion are equivalent for the noncommutative and ordinary fields in the NC U(N) gauge theories. We start with the noncommutative fields. The action reads
\begin{equation}
S_{\rm NCYM}=-\frac{1}{4g^2}\int \tr\left(\hat F_{\mu\nu}\big[\hat B_\mu\big]\hat F^{\mu\nu}\big[\hat B_\mu\big]\right),
\end{equation}
where
\begin{equation}
\hat F_{\mu\nu}\big[\hat B_\mu\big]=\partial_\mu \hat B_\nu-\partial_\nu \hat B_\mu+i\left[\hat B_\mu\stackrel{\star}{,}\hat B_\nu\right].
\end{equation}
If, in terms of the component fields $\hat B_\mu=\hat B_\mu^a T^a$, than $T^a$ is in the fundamental representation of U(N). The equations of motion for $\hat B_\mu^a$ read
\begin{equation}
\tr\left(T^a\hat D^\mu\big[\hat B_\mu\big]\hat F_{\mu\nu}\big[\hat B_\mu\big]\right)=0,
\end{equation}
which is equivalent to
\begin{equation}
\hat D^\mu\big[\hat B_\mu\big]\hat F_{\mu\nu}\big[\hat B_\mu\big]=0.
\end{equation}
Now, if $B_\nu^b$ and $\hat B_\mu^a$ are related by the SW map
\begin{equation}
\begin{split}
\hat B_\mu^a\Big[B_\nu^b\Big]=B_\mu^a+\sum\limits_{n=2}^\infty\int\prod\limits_{i=1}^n\frac{d^4 p_i}{(2\pi)^4} e^{i\left(\sum\limits_{i=1}^n p_i\right)x}
&\tr\left(T^a\mathfrak{A}^{(n)}_\mu\left[(a_1,\mu_1,p_1),......,(a_n,\mu_n,p_n);\theta\right]\right)
\\&\cdot\tilde B_{\mu_1}^{a_1}(p_1)......\tilde B_{\mu_n}^{a_n}(p_n),
\end{split}
\end{equation}
and
\begin{equation}
\det\frac{\delta\hat B_\mu^a\Big[B_\nu^b\Big](x)}{B_\nu^b(y)}\neq 0,
\end{equation}
i.e.
\begin{equation}
\begin{split}
0=\delta B_\mu^a+\sum\limits_{n=2}^\infty\int\prod\limits_{i=1}^n\frac{d^4 p_i}{(2\pi)^4} e^{i\left(\sum\limits_{i=1}^n p_i\right)x}
&\tr\left(T^a\mathfrak{A}^{(n)}_\mu\left[(a_1,\mu_1,p_1),......,(a_n,\mu_n,p_n);\theta\right]\right)
\\&\cdot n\cdot\tilde B_{\mu_1}^{a_1}(p_1)......\delta\tilde B_{\mu_n}^{a_n}(p_n),
\end{split}
\end{equation}
has no zero modes (nonzero solutions), than
$\hat B_\mu^a=\hat B_\mu^a\Big[B_\nu^b\Big]$
can be inverted into
$B_\mu^a=B_\mu^a\Big[\hat B_\nu^b\Big]$.

We have that the equation of motion for $B_\mu^a$ with action
\begin{equation}
S_{\rm NCYM}=-\frac{1}{4g^2}\int \tr\left(\hat F_{\mu\nu}\left[\hat B_\mu\left[B_\mu\right]\right]\hat F^{\mu\nu}\left[\hat B_\mu\left[B_\mu\right]\right]\right),
\end{equation}
reads
\begin{equation}
\begin{split}
&0=\frac{\delta S_{\rm NCYM}}{\delta B_\mu^a(x)}=\int d^4y \frac{\delta S_{\rm NCYM}}{\delta\hat B_\nu^b(y)}\frac{\delta\hat B_\nu^b(y)}{\delta B_\mu^a(x)}\Bigg|_{\hat B_\mu^a=\hat B_\mu^a\left[B_\nu^b\right]}\Longleftrightarrow \frac{\delta S_{\rm NCYM}}{\delta\hat B_\mu^a}\Bigg|_{\hat B_\mu^a\left[B_\nu^b\right]}=0
\\
&\Longleftrightarrow \hat D^\mu\left[\hat B_\mu\left[B_\mu\right]\right]\hat F_{\mu\nu}\left[\hat B_\mu\left[B_\mu\right]\right]=0.
\end{split}
\label{EOM}
\end{equation}
Notice however:

\begin{enumerate}
\item For SU(N), SO(N) etc. groups \eqref{EOM} is not the equation of motion of $B_\mu^a$ since the dependence of $S_{\rm NCYM}$ on $B_\mu^a$ is not exhausted by the dependence of $\hat B_\mu^a$ on $B_\mu^a$.

\item While the equations of motion of noncommutative and ordinary fields are equivalent, they are not exactly identical. This would affect the subtraction of EOM proportional terms when evaluating the background field effective action and lead to nonidentical off-shell results. As we described in the main text, one can obtain exactly identical results in direct computations using noncommutative or ordinary fields only by subtracting the identical EOM proportional terms.
\end{enumerate}

\section{Some detailed computations}

From \eqref{Qhat} --see also (\ref{SW1}), (\ref{ordinarysplit}), and (\ref{Qdefinition})-- one gets
\begin{equation*}
\begin{array}{l}
{\frac{\delta\hat Q^a_\mu(x)}{\delta Q^b_\nu(y)}\,=\,\frac{1}{\hbar^{\frac{1}{2}}}\,\frac{\delta\hat A^a_\mu(x)}{\delta Q^b_\nu(y)}}\\[8pt]
=\delta^a_b\delta^\nu_\mu\,\delta(x-y)+\sum\limits_{n=2}^{\infty}\,
\int\prod\limits_{i=1}^n\frac{d^4 p_i}{(2\pi)^4}\Big[e^{i\left(\sum\limits_{i=1}^{n} p_i\right)x}
\cdot{n\,\hbar^{-\frac{1}{2}}\tr\Big(T^a{\mathfrak{A}^{(n)}}_\mu\big[(a_1,\mu_1,p_1),}\\[8pt]
{.....,(a_{n-1},\mu_{n-1},p_{n-1}),(a_n,\mu_n,p_n);\theta\big]\Big)
\cdot\tilde A_{\mu_1}^{a_1}(p_1)......\tilde A_{\mu_{n-1}}^{a_{n-1}}(p_{n-1})\,\frac{\delta\tilde A^a_{\mu_n}(p_n)}{\delta Q^b_\nu(y)}\Big].}
\end{array}
\end{equation*}
Taking into account \eqref{functionaldelta} and using $\tilde A_{\mu_n}^{a_n}(p_n)=\tilde B_{\mu_n}^{a_n}(p_n)+\hbar^{\frac{1}{2}}\tilde B_{\mu_n}^{a_n}(p_n)$ 
one obtains (\ref{dethQQ}) and (\ref{Mcaldef}).

Let us introduce the following definition
\begin{equation}
\begin{array}{l}
{\cal M}^{a\,\nu}_{b\,\mu}(x;y)=\sum\limits_{n=2}^{\infty}\,
\int\prod\limits_{i=1}^n\frac{d^4 p_i}{(2\pi)^4} e^{i\left(\sum\limits_{i=1}^{n-1} p_i\right)x}\,e^{ip_n (x-y)}
{\cal M}^{(n)\,a\,\nu}_{\phantom{(n)\,}b\,\mu}(p_1,p_2,....p_{n-1};p_n;\theta),
\end{array}
\label{Mdef}
\end{equation}
where ${\cal M}^{(n)\,a\,\nu}_{\phantom{(n)\,}b\,\mu}(p_1,p_2,....p_{n-1};p_n;\theta)$ has been given in (\ref{Mcaldef}). Then,
\begin{equation}
\begin{array}{l}
{\ln\,J_1[B,Q]={\rm Tr}\ln\,\left(\frac{\delta\hat Q^a_\mu(x)}{\delta Q^b_\nu(y)}\right)={\rm Tr}\ln\,\Big[\delta^{a}_{b}\delta^{\nu}_{\mu}\delta(x-y)+{\cal M}^{a\,\nu}_{b\,\mu}(x;y)\Big]}\\[8pt]
={\int d^4x {\cal M}^{a\,\mu}_{a\,\mu}(x;x)}\\[8pt]
+{\sum\limits_{m=1}^{\infty}\frac{(-1)^m}{m+1}\int d^4x \int\prod\limits_{i=1}^m  d^4x_i\; {\cal M}^{a\,\mu_1}_{a_1\,\mu}(x;x_1){\cal M}^{a_1\,\mu_2}_{a_2\,\mu_1}(x_1;x_2)\cdots
{\cal M}^{a_m\,\mu}_{a\,\mu_m}(x_m;x).}\\[8pt]
\end{array}
\label{B.2}
\end{equation}

The substitution of (\ref{Mdef}) in the previous equation (\ref{B.2}) yields
\begin{equation}
\begin{array}{l}
{\ln\,J_1[B,Q]=\sum\limits_{n=2}^{\infty}\,\int d^4x\int\prod\limits_{i=1}^n\frac{d^4 p_i}{(2\pi)^4}e^{i\left(\sum\limits_{i=1}^{n-1} p_i\right)x}\,e^{ip_n (x-x)}
{\cal M}^{(n)\,a\,\mu}_{\phantom{(n)\,}a\,\mu}(p_1,p_2,....p_{n-1};p_n;\theta)}\\[8pt]
+{\sum\limits_{m=1}^{\infty}\frac{(-1)^m}{m+1}\sum\limits_{n_1=2}^{\infty}\sum\limits_{n_2=2}^{\infty}\cdots\sum\limits_{n_{m}=2}^{\infty}\sum\limits_{n_{m+1}=2}^{\infty}\int d^4x \int\prod\limits_{i=1}^m  d^4x_i}\\[8pt]
\Big\{{\Big[\int\prod\limits_{i_1=1}^{n_1}\frac{d^4 p_{1,i_1}}{(2\pi)^4}e^{i\left(\sum\limits_{i_1=1}^{n_1-1} p_{1,i_1}\right)x}\,e^{ip_{1,n_1} (x-x_1)}
{\cal M}^{(n_1)\,a\,\mu_1}_{\phantom{(n_1)\,}a_1\,\mu}\left(p_{1,1},p_{1,2},....,p_{1,n_1-1};p_{1,n_1};\theta\right)\Big]}
\\[8pt]
{\cdot\Big[\int\prod\limits_{i_2=1}^{n_2}\frac{d^4 p_{2,i_2}}{(2\pi)^4}e^{i\left(\sum\limits_{i_2=1}^{n_2-1} p_{2,i_2}\right)x_1}\,e^{ip_{2,n_2} (x_1-x_2)}
{\cal M}^{(n_2)\,a_1\,\mu_2}_{\phantom{(n_2)\,}a_2\,\mu_1}\left(p_{2,1},p_{2,2},....,p_{2,n_2-1};p_{2,n_2};\theta\right)\Big]}\\[8pt]
{\cdot.......\cdot}\\[8pt]
{\cdot\Big[\int\prod\limits_{i_{m}=1}^{n_{m}}\frac{d^4 p_{m,i_{m}}}{(2\pi)^4}e^{i\left(\sum\limits_{i_{m}=1}^{n_{m}-1} p_{m,i_{m}}\right)x_{m-1}}\,e^{ip_{m,n_{m}} (x_{m-1}-x_m)}}\\[8pt]
{\quad\quad\quad\quad\quad\quad\quad\quad\quad\quad\quad\cdot{\cal M}^{(n_{m})\,a_m-1\,\mu_m}_{\phantom{(n_m)\,}a_m\,\mu_{m-1}}\left(p_{m,1},p_{m,2},....,p_{m,n_{m}-1};p_{m,n_{m}}
;\theta\right)\Big]}\\[8pt]
{\cdot\Big[\int\prod\limits_{i_{m+1}=1}^{n_{m+1}}\frac{d^4 p_{m+1,i_{m+1}}}{(2\pi)^4}e^{i\left(\sum\limits_{i_{m+1}=1}^{n_{m+1}-1} p_{m+1,i_{m+1}}\right)x_{m}}\,e^{ip_{m+1,n_{m+1}} (x_{m}-x)}}\\[8pt]
{\quad\quad\quad\quad\quad\quad\quad\quad\quad\cdot{\cal M}^{(n_{m+1})\,a_m\,\mu}_{\phantom{(n_{m+1})\,}a\,\mu_m}\left(p_{m+1,1},p_{m+1,2},....,p_{m+1,n_{m+1}-1};p_{m+1,n_{m+1}}
;\theta\right)\Big]\Big\}.}
\end{array}
\label{B.3}
\end{equation}
Introducing the following definitions 
\begin{equation*}
l_1=\sum\limits_{i_1=1}^{n_1-1}\, p_{1,i_1}, l_2=\sum\limits_{i_2=1}^{n_2-1}\,p_{2,i_2},..., l_{m+1}=\sum\limits_{i_{m+1}=1}^{n_{m+1}}\,p_{m+1,i_{m+1}},
\end{equation*}
and carrying out the integration over $x$ and $x_i$, $i=1,...,m$, one obtains the following product of Dirac deltas
\begin{equation*}
\begin{array}{l}
{\delta(l_1+p_{1,n_1}-p_{m+1,n_{m+1}})\delta(l_2-p_{1,n_1}+p_{2,n_2})\delta(l_3-p_{2,n_2}+p_{3,n_3})\cdots\cdots}\\[8pt]
{\cdot\delta(l_m-p_{m-1,n_{m-1}}+p_{m,n_m})\delta(l_{m+1}-p_{m,n_m}+p_{m+1,n_{m+1}}).}
\end{array}
\end{equation*}
Renaming $p_{1,n_1}$ as $q$ and integrating out $p_{2,n_2}$, $p_{3,n_3}$,.... and $p_{m+1,n_{m+1}}$, one removes all Dirac deltas but one, which turns out to be $\delta(\sum\limits_{i=1}^{m+1}\,l_i)$, and obtains (\ref{genexpln}).

\section{Vanishing integrals in dimensional regularization}

In this appendix we shall discuss why integrals over the internal momentum $q$, that arise in the computation of the Jacobian determinants in sections 3.1 and 3.2, vanish in dimensional regularization. These integrals are of the following type
\begin{equation}
\mathfrak{I}\,=\,\int\frac{d^D q}{(2\pi)^D}\,\mathbb{Q}(q)\,\mathbb{I}(q\theta k_i,k_i\theta k_j),
\label{thevanishingintegral}
\end{equation}
where $\mathbb{Q}(q)=q^{\rho_1}q^{\rho_2}\cdots q^{\rho_n}$, $q\theta k_i=q_\mu\theta^{\mu\nu}k_{i\nu}$, $i=1,.....s$, and $k_i\theta k_j=k_{i\mu}\theta^{\mu\nu}k_{j\nu}$, $i,j=1,.....s$. Indices $n$ and $s$ run over all relevant momenta other than $q$, in general. The function $\mathbb{I}$ in the integrand of the previous integral is a function of variables $q\theta k_i$ and $k_i\theta k_j$ ${\it only}$.

We shall define the integral (\ref{thevanishingintegral}) by Wick rotating the corresponding integral defined for Euclidean signature, a signature which we shall assume for the time being.

The first problem one has to face when defining, in dimensional regularization, the object in (\ref{thevanishingintegral}) is the definition of $\theta^{\mu\nu}$ in the infinite dimensional space, $E_{\infty}$ -see section 4.1 of ref. \cite{Collins:1984xc}-- of which the momenta $q^\mu$, $k^\mu_i$ are elements in dimensional regularization. Let us recall  that, to avoid problems with unitarity, our $\theta^{\mu\nu}$ in four dimensions is such that $\theta^{0i}=0$, $i=1,2,3$. Hence, by a rotation, this $\theta^{\mu\nu}$ in four dimensions can be transformed into an object whose only non-vanishing components are $\theta^{23}$ and $\theta^{32}$. Then, without loss of generality, we shall assume this latter
$\theta^{\mu\nu}$ to be our object in four dimensions.

Now, since $\theta^{\mu\nu}$ is an antisymmetric object, its properties depend on the dimension of spacetime. So, as happens with the Levi-Civita tensor and the $\gamma_5$ matrix \cite{Collins:1984xc}, the only consistent way to define it in dimensional regularization is to keep it essentially four-dimensional, since our physical theory is in four dimensions. This amounts to defining $\theta^{\mu\nu}$ in the infinite dimensional space --see section 4.1 of ref. \cite{Collins:1984xc}-- of dimensional regularization:
\begin{equation*}
\begin{array}{l}
{\theta^{\mu\nu}=\theta,\quad{\rm if}\quad \mu=2, \nu=3},\\[2pt]
{\theta^{\mu\nu}=-\theta,\quad{\rm if}\quad \mu=3, \nu=2},\\[2pt]
{\theta^{\mu\nu}=0,\quad{\rm otherwise}.}
\end{array}
\end{equation*}

With this definition of our $\theta^{\mu\nu}$-object in dimensional regularization, one comes to the conclusion that all the vectors $\frac{1}{\theta}\theta^{\mu\nu}k_{i\nu}$, $i=1,...,n$, belong to the same two-dimensional subspace, $E_2$,  of the infinite dimensional space $E_{\infty}$. Let us follow ref. \cite{Collins:1984xc} and split the vector $q^\mu\in E_{\infty} $ into two components:
\begin{equation}
q^\mu=q_{\bot}^\mu\,+\,q_{\|}^\mu ,
\end{equation}
where $q_{\|}^\mu\in E_2$ and  $q_{\bot}^\mu\in E_{\bot}$, $E_{\bot}$ being the subspace orthogonal to $E_2$. Then, using \cite{Collins:1984xc}, we define the following object in (\ref{thevanishingintegral})
\begin{equation}
\int\frac{d^D q}{(2\pi)^D}\,\mathbb{Q}(q)\,\mathbb{I}(q\theta k_i,k_i\theta k_j)=\frac{1}{(2\pi)^D}\int dl^1 dl^2\Big\{\int\,d^{D-2}q_{\bot}\,\mathbb{Q}(q)\,\mathbb{I}(q_{\|}\theta k_i,k_i\theta k_j)\Big\},
\label{azeroint}
\end{equation}
where $l^1$ and $l^2$ are the coordinates of $q_{\|}^\mu$ in an orthonormal basis of $E_2$ and we have taken into account that $q\theta k_i=q_{\|}\theta k_i$.

 Now, $\mathbb{I}(q_{\|}\theta k_i,k_i\theta k_j)$ does not depend on $q_{\bot}^\mu$, so that
 \begin{equation}
 \int\,d^{D-2}q_{\bot}\,\mathbb{Q}(q)\,\mathbb{I}(q_{\|}\theta k_i,k_i\theta k_j)=\mathbb{I}(q_{\|}\theta k_i,k_i\theta k_j)\int\,d^{D-2}q_{\bot}\,\mathbb{Q}(q).
\end{equation}
But in dimensional regularization tadpole-type integrals --see \cite{Collins:1984xc}-- vanish:
\begin{equation}
\int\,d^{D-2}q_{\bot}\,\mathbb{Q}(q)\,=\,0,
\end{equation}
recall that $\mathbb{Q}(q)$ is a monomial. We thus conclude that
\begin{equation}
\int\,d^{D-2}q_{\bot}\,\mathbb{Q}(q)\,\mathbb{I}(q_{\|}\theta k_i,k_i\theta k_j)=0,
\end{equation}
so that the right hand side of equation (\ref{azeroint}) vanishes, which in turn implies that
\begin{equation}
\int\frac{d^D q}{(2\pi)^D}\,\mathbb{Q}(q)\,\mathbb{I}(q\theta k_i,k_i\theta k_j)\,=\,0.
\end{equation}

\section{Expansion of the action $\hbar^{-1}S_{\rm NCYM}\big[\hat B_\mu+\hbar^\frac{1}{2}\hat Q_\mu\big]$ in terms of $\hbar$}

We shall assume that $\hat D^\mu\Big[\hat B_\mu\big[B_\mu\big]\Big]\hat F_{\mu\nu}\Big[\hat B_\mu\big[B_\mu\big]\Big]=0$, then the action is
\begin{equation}
\begin{split}
\frac{1}{\hbar}S_{\rm NCYM}&\big[\hat B_\mu+\hbar^\frac{1}{2}\hat Q_\mu\big]=-\frac{1}{4g^2\hbar}\int \tr\left(\hat F_{\mu\nu}\big[\hat B_\mu+\hbar^\frac{1}{2}\hat Q_\mu\big]\hat F^{\mu\nu}\big[\hat B_\mu+\hbar^\frac{1}{2}\hat Q_\mu\big]\right)
\\=&-\frac{1}{4g^2\hbar}\int \left(\hat F_{\mu\nu}\big[\hat B_\mu\big]+\hbar^\frac{1}{2}\left(\hat D_\mu\big[\hat B_\mu\big] \hat Q_\nu-\hat D_\nu\big[\hat B_\mu\big] \hat Q_\mu\right)
-\hbar\left[\hat Q_\mu\stackrel{\star}{,}\hat Q_\nu\right]\right)^2
\\=&-\frac{1}{4g^2\hbar}\int \tr\left(\hat F_{\mu\nu}\big[\hat B_\mu\big]\hat F^{\mu\nu}\big[\hat B_\mu\big]\right)
-\frac{1}{2g^2\hbar^\frac{1}{2}}\int \tr\left(\hat D^\mu\hat F_{\mu\nu}\big[\hat B_\mu\big]\hat Q^\nu\right)
\\&-\frac{1}{4g^2}\int \tr\left(\hat D_\mu\big[\hat B_\mu\big] \hat Q_\nu-\hat D_\nu\big[\hat B_\mu\big] \hat Q_\mu\right)^2
+\frac{1}{2g^2}\int \tr\hat F^{\mu\nu}\big[\hat B_\mu\big]\left[\hat Q_\mu\stackrel{\star}{,}\hat Q_\nu\right]
+\mathcal{O}\big(\hbar^\frac{1}{2}\big).
\end{split}
\end{equation}
The second line after the third equality can be neglected because the background field satisfies the equations of motion \eqref{EOM} (Kallosh formalism). Therefore
\begin{equation}
\begin{split}
S_{\rm NCYM}\big[\hat B_\mu+\hbar^\frac{1}{2}\hat Q_\mu\big]=&S_{\rm NCYM}\big[\hat B_\mu\big]-\frac{1}{4g^2}\int \tr\left(\hat D_\mu\big[\hat B_\mu\big] \hat Q_\nu-\hat D_\nu\big[\hat B_\mu\big] \hat Q_\mu\right)^2
\\&+\frac{1}{2g^2}\int \tr\hat F^{\mu\nu}\big[\hat B_\mu\big]\left[\hat Q_\mu\stackrel{\star}{,}\hat Q_\nu\right]+\mathcal{O}\big(\hbar^\frac{1}{2}\big).
\end{split}
\end{equation}
Now, one extracts the $\mathcal{O}(\hbar^0)$ order terms of $\hat Q_\mu$ from \eqref{Qhat} and deduces that
\begin{equation}
\begin{split}
S_{\rm NCYM}\big[\hat B_\mu+\hbar^\frac{1}{2}\hat Q_\mu\big]=&S_{\rm NCYM}\big[\hat B_\mu\big]-\frac{1}{4g^2}\int \tr\left(\hat D_\mu\big[\hat B_\mu\big] \hat{\hat Q}_\nu-\hat D_\nu\big[\hat B_\mu\big] \hat{\hat Q}_\mu\right)^2
\\&-\frac{i}{2g^2}\int \tr\hat F^{\mu\nu}\big[\hat B_\mu\big]\left[\hat{\hat Q}_\mu\stackrel{\star}{,}\hat{\hat Q}_\nu\right]+\mathcal{O}\big(\hbar^\frac{1}{2}\big),
\end{split}
\label{NCYMexpanded}
\end{equation}
where
\begin{equation}
\begin{split}
\hat{\hat Q}_\mu=Q_\mu+\sum\limits_{n=2}^\infty\int\prod\limits_{i=1}^n\frac{d^4 p_i}{(2\pi)^4} e^{i\left(\sum\limits_{i=1}^n p_i\right)x}&
\mathfrak{A}^{(n)}_\mu\left[(a_1,\mu_1,p_1),......,(a_n,\mu_n,p_n);\theta\right]
\\&\cdot n\cdot \tilde B_{\mu_1}^{a_1}(p_1)......\tilde B_{\mu_{n-1}}^{a_{n-1}}(p_{n-1})\tilde Q_{\mu_n}^{a_n}(p_n).
\end{split}
\label{Qhathatdefinition}
\end{equation}
i.e. $\hat Q_\mu=\hat{\hat Q}_\mu+\mathcal O\big(\hbar^\frac{1}{2}\big)$.
Similarly, we can expand the gauge fixing action \eqref{gaugefix} up to the $\hbar^{0}$ order
\begin{equation}
S_{\rm gf}=\frac{1}{g^2}\int\tr\left(\alpha \hat F^2+\hat F\hat D_\mu\big[\hat B_\mu\big]\hat{\hat Q}^{\mu}-\hat{\bar C}\hat D_\mu\big[\hat B_\mu\big]\hat D^\mu\big[\hat B_\mu\big]\hat{\hat C}\right),
\label{gaugefixexpanded}
\end{equation}
where
\begin{equation}
\hat{\hat C}=\hat C\left[B_\mu, C; \theta \right].
\label{Chathatdefinition}
\end{equation}

\section{Feynman rules in the background field formalism}

We list here all Feynman rules the relevant to the computation in section 4. We use the Fourier transformation rule
\begin{equation}
f(x)=\int\frac{d^4 p}{(2\pi)^4} \tilde f(p)e^{ipx},
\end{equation}
and a convention in the vertex diagrams that sets all photon momenta as incoming. The SW map expansion of $\hat{\hat Q}_\mu$ for $\rm U(1)$ gauge theory is derived from SW map for unsplitted field
\begin{equation}
\begin{split}
\hat A_\mu=&A_\mu+\frac{1}{2}\theta^{ij}A_i\star_2(\partial_j
A_\mu+A_{j\mu})
\\&-\frac{1}{8}\theta^{ij}\theta^{kl}\big[(\partial_i A_\mu+A_{i\mu})A_k(\partial_l A_j+A_{lj})-A_i\partial_j(A_k(\partial_l A_\mu+A_{l\mu}))
\\&+2A_i(A_{jk}A_{\mu l}-A_k\partial_l A_{j\mu})\big]_{\star_{3'}}+\mathcal{O}(A^3),\;\;A_{\mu\nu}\equiv\partial_\mu A_\nu-\partial_\nu A_\mu,
\end{split}
\label{E2}
\end{equation}
the background-field splitting \eqref{Qdefinition} and the expansion \eqref{Qhathatdefinition}. The leading order (in $\hbar$) ghost Seiberg-Witten map $\hat{\hat C}$ in $\rm U(1)$ theory is defined as follows
\begin{equation}
\begin{split}
\hat{\hat C}=\hat C\Big[B_\mu,C;\theta\Big]
&=C+\frac{1}{2}\theta^{ij} a_i\star_2\partial_j C
\\&+\frac{1}{8}\theta^{ij}\theta^{kl}\Big[A_i\partial_j(A_k\partial_l C-\partial_i C A_k(\partial_l A_j)+A_{lj}\Big]_{\star_{3'}}+\mathcal{O}\left(A^3\right)C.
\end{split}
\label{E3}
\end{equation}
The generalized star products $\star_2$ and $\star_{3'}$ here and the corresponding nonlocal factors $f_{\star_2}$ and $f_{\star_{3'}}$ below are the same as defined in~\cite{Martin:2016zon}. Employing all these ingredients we obtain the Feynman rules below for (one) loop computation in the background field formalism in section 4.\footnote{Note that there is a sign change in front of the first order SW map expansion terms in \eqref{E2} and \eqref{E3} with respect to~\cite{Martin:2016zon}, which is due to the change of signature in the covariant derivative definition.}

\begin{figure}
\begin{center}
\includegraphics[width=6cm]{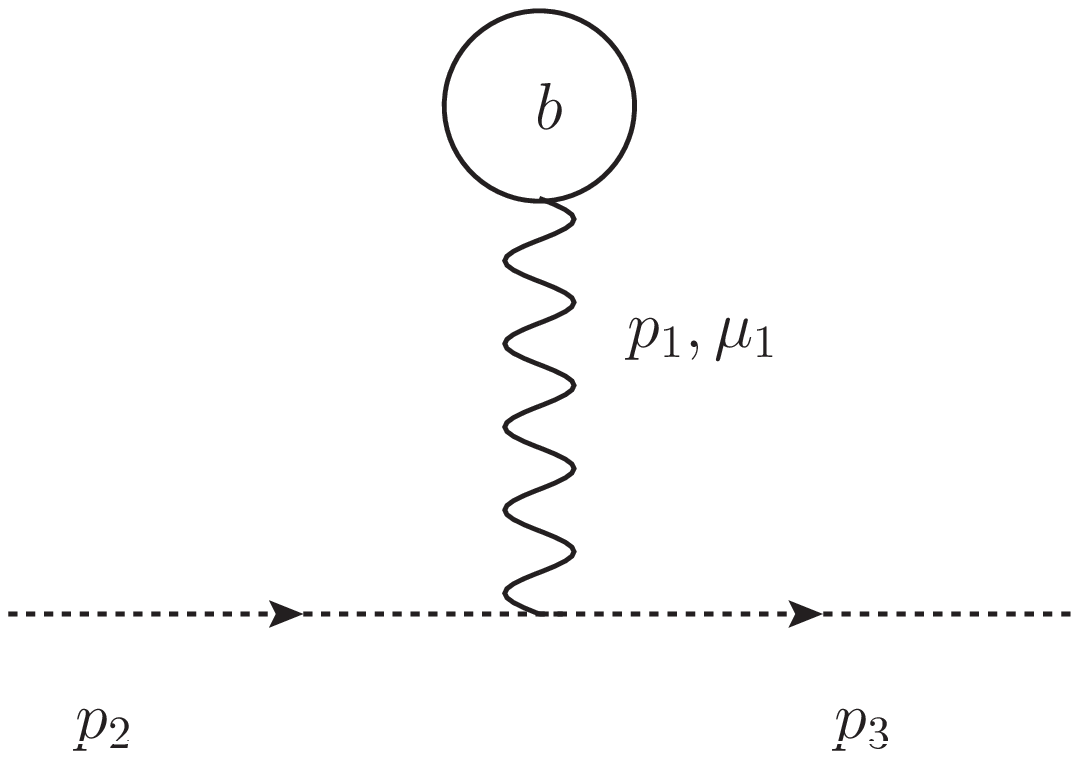}
\end{center}
\caption{Ghost-photon BFM FR.}
\label{fig:BFMFR7}
\end{figure}

\begin{figure}
\begin{center}
\includegraphics[width=6cm]{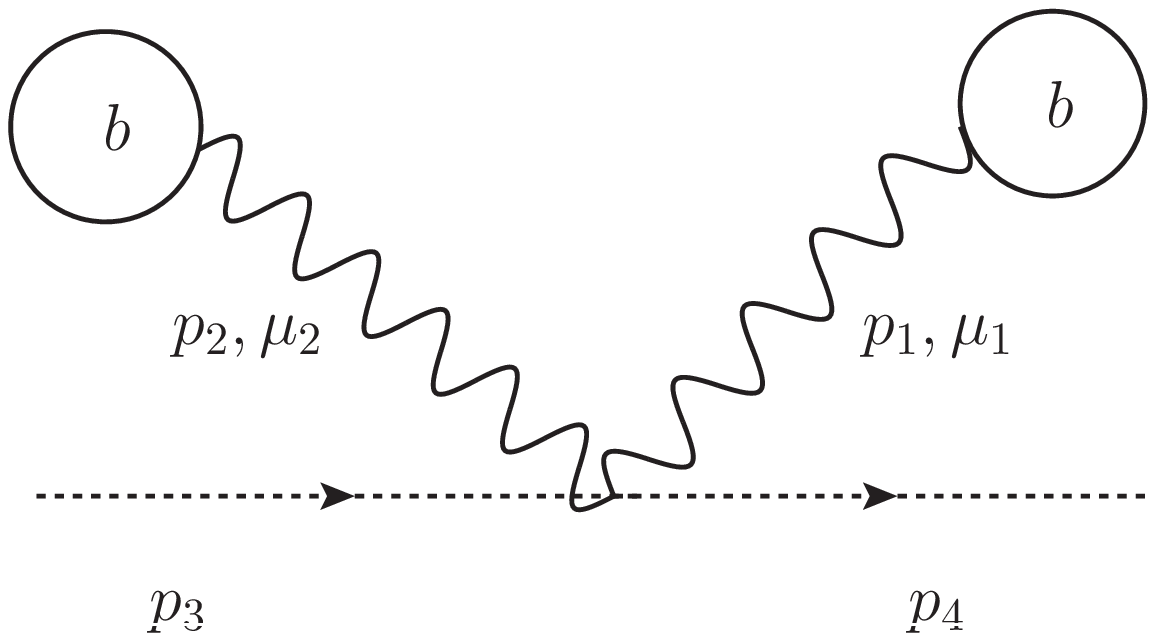}
\end{center}
\caption{Ghost-2photons BFM FR.}
\label{fig:BFMFR8}
\end{figure}

In the next two subsections we are giving Feynman rules which generically correspond to the following figures:
Fig. \ref{fig:BFMFR5}, Fig. \ref{fig:BFMFR6}, Fig. \ref{fig:BFMFR7}, and Fig. \ref{fig:BFMFR8}.

\subsection{The background field gauge}

\begin{figure}
\begin{center}
\includegraphics[width=6cm]{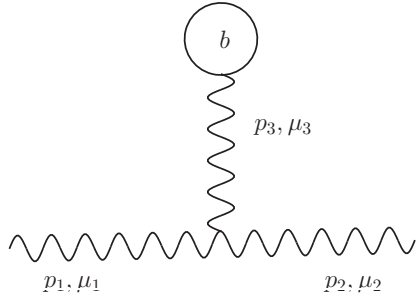}
\end{center}
\caption{Three-photon BFM FR.}
\label{fig:BFMFR5}
\end{figure}

\begin{figure}
\begin{center}
\includegraphics[width=6cm]{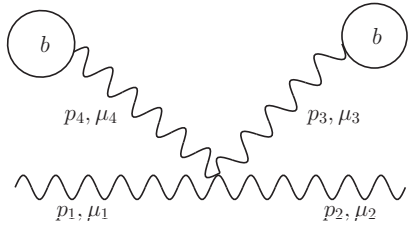}
\end{center}
\caption{Four-photon BFM FR.}
\label{fig:BFMFR6}
\end{figure}

\begin{equation}
\Gamma_{BQQ_{BFG}}^{\mu\nu_1\nu_2}\left(p;q_1,q_2\right)=\Gamma_{BQQ_{BFG_1}}^{\mu\nu_1\nu_2}\left(p;q_1,q_2\right)+\Gamma_{BQQ_{BFG_2}}^{\mu\nu_1\nu_2}\left(p;q_1,q_2\right),
\end{equation}
\begin{equation}
\begin{split}
\Gamma_{BQQ_{BFG_1}}^{\mu\nu_1\nu_2}=&\frac{1}{2}f_{\star_2}(p,q_1)\bigg(\Big((p\theta q_1)(P^{\nu_1 } g^{\mu\nu_2}-P^{\nu_2} g^{\mu\nu_1})+(p\theta q_2)(P^{\nu_2 } g^{\mu\nu_1}-P^{\nu_1} g^{\mu\nu_2})\Big)
\\&-2\Big((\theta p)^{\nu_1}((q_1\cdot p)g^{\mu\nu_2}-q_2^\mu p^{\nu_2})+(\theta p)^{\nu_2}((q_1\cdot p)g^{\mu\nu_1}-q_1^\mu p^{\nu_1})\Big)
\\&+2\Big((\theta p)^{\mu}((q_1\cdot q_2)g^{\nu_1 \nu_2}-q_2^{\nu_1}q_1^{\nu_2})\Big)
\\&+2\Big(q_1^\mu(p\theta q_2)g^{\nu_1\nu_2}+q_1^\mu(\theta p)^{\nu_2}q_2^{\nu_1}-(q_1\cdot p)(\theta q_2)^\mu g^{\nu_1\nu_2}+(q_1\cdot p)q_2^{\nu_1}\theta^{\mu\nu_2}
\\&-q_1^{\nu_2}(p\theta q_2)g^{\mu\nu_1}-(q_1\cdot q_2)(\theta p)^{\nu_2} g^{\mu\nu_1}+q_1^{\nu_2}p^{\nu_1}(\theta q_2)^{\mu}-(q_1\cdot q_2)p^{\nu_1}\theta^{\mu\nu_2}
\\&+q_2^\mu(p\theta q_1)g^{\nu_1\nu_2}+q_2^\mu(\theta p)^{\nu_1}q_1^{\nu_2}-(q_2\cdot p)(\theta q_1)^\mu g^{\nu_1\nu_2}+(q_2\cdot p)q_1^{\nu_2}\theta^{\mu\nu_1}
\\&-q_2^{\nu_1}(p\theta q_1)g^{\mu\nu_2}-(q_1\cdot q_2)(\theta p)^{\nu_1} g^{\mu\nu_2}+q_2^{\nu_1}p^{\nu_2}(\theta q_1)^{\mu}-(q_1\cdot q_2)p^{\nu_2}\theta^{\mu\nu_1}\Big)\bigg),
\end{split}
\end{equation}
\begin{equation}
\begin{split}
\Gamma_{BQQ_{BFG_2}}^{\mu\nu_1\nu_2}=&f_{\star_2}(p,q_1)\bigg(\left((p\theta q_1)g^{\mu\nu_1}q_2^{\nu_2}+(p\theta q_2)g^{\mu\nu_2}q_1^{\nu_1}\right)
\\&+\frac{1}{2}\Big(q_1^{\nu_1}\left(2q_1^{\nu_2}(\theta q_2)^\mu-(q_1\cdot q_2)\theta^{\mu\nu_2}+2q_1^\mu(\theta p)^{\nu_2}+(q_1\cdot p)\theta^{\mu\nu_2}\right)
\\&+q_2^{\nu_2}\left(2q_2^{\nu_1}(\theta q_1)^\mu-(q_1\cdot q_2)\theta^{\mu\nu_1}+2q_2^\mu(\theta p)^{\nu_1}+(q_2\cdot p)\theta^{\mu\nu_1}\right)\Big)\bigg),
\end{split}
\end{equation}
\begin{equation}
\Gamma_{BBQQ_{BFG}}^{\mu_1\mu_2\nu_1\nu_2}\left(p_1,p_2;q_1,q_2\right)=\Gamma_{BBQQ_{BFG_1}}^{\mu_1\mu_2\nu_1\nu_2}\left(p_1,p_2;q_1,q_2\right)+\Gamma_{BBQQ_{BFG_2}}^{\mu_1\mu_2\nu_1\nu_2}\left(p_1,p_2;q_1,q_2\right),
\end{equation}

\begin{equation}
\Gamma_{BBQQ_{BFG_1}}^{\mu_1\mu_2\nu_1\nu_2}=\Gamma_A^{\mu_1\mu_2\nu_1\nu_2}+\Gamma_B^{\mu_1\mu_2\nu_1\nu_2}+\Gamma_C^{\mu_1\mu_2\nu_1\nu_2},
\end{equation}
\begin{equation}
\Gamma_{BBQQ_{BFG_2}}^{\mu_1\mu_2\nu_1\nu_2}=\Gamma_{BBQQ_{NCFG-BFM_2}}^{\mu_1\mu_2\nu_1\nu_2}+\Gamma_F^{\mu_1\mu_2\nu_1\nu_2},
\end{equation}

\begin{equation}
\begin{split}
&\Gamma_A^{\mu_1\mu_2\nu_1\nu_2}=V_A^{\mu_1\nu_1\nu_2\mu_2}(p_1,q_1,q_2,p_2)+V_A^{\mu_1\nu_2\nu_1\mu_2}(p_1,q_2,q_1,p_2)
\\&+V_A^{\mu_2\nu_1\nu_2\mu_1}(p_2,q_1,q_2,p_1)+V_A^{\mu_2\nu_2\nu_1\mu_1}(p_2,q_2,q_1,p_1)+V_A^{\nu_1\mu_1\nu_2\mu_2}(q_1,p_1,q_2,p_2)
\\&+V_A^{\nu_1\mu_2\nu_2\mu_1}(q_1,p_2,q_2,p_1)+V_A^{\nu_2\mu_1\nu_1\mu_2}(q_2,p_1,q_1,p_2)+V_A^{\nu_2\mu_2\nu_1\mu_1}(q_2,p_2,q_1,p_1)
\\&+V_A^{\mu_2\nu_1\mu_1\nu_2}(p_2,q_1,p_1,q_2)+V_A^{\mu_2\nu_2\mu_1\nu_1}(p_2,q_2,p_1,q_1)+V_A^{\nu_1\mu_1\mu_2\nu_2}(q_1,p_1,p_2,q_2)
\\&+V_A^{\nu_1\mu_2\mu_1\nu_2}(q_1,p_2,p_1,q_2)+V_A^{\nu_2\mu_1\mu_2\nu_1}(q_2,p_1,p_2,q_1)+V_A^{\nu_2\mu_2\mu_1\nu_1}(q_2,p_2,p_1,q_1)
\\&+V_A^{\mu_1\nu_1\mu_2\nu_2}(p_1,q_1,p_2,q_2)+V_A^{\mu_1\nu_2\mu_2\nu_1}(p_1,q_2,p_2,q_1),
\end{split}
\end{equation}

\begin{equation}
\begin{split}
&\Gamma_B^{\mu_1\mu_2\nu_1\nu_2}=V_B^{\mu_1\nu_1\nu_2\mu_2}(p_1,q_1,q_2,p_2)+V_B^{\mu_1\nu_2\nu_1\mu_2}(p_1,q_2,q_1,p_2)
\\&+V_B^{\nu_1\mu_1\nu_2\mu_2}(q_1,p_1,q_2,p_2)+V_B^{\nu_2\mu_1\nu_1\mu_2}(q_2,p_1,q_1,p_2)+V_B^{\nu_1\mu_1\mu_2\nu_2}(q_1,p_1,p_2,q_2)
\\&+V_B^{\nu_2\mu_1\mu_2\nu_1}(q_2,p_1,p_2,q_1)+V_B^{\mu_1\nu_1\mu_2\nu_2}(p_1,q_1,p_2,q_2)+V_B^{\mu_1\nu_2\mu_2\nu_1}(p_1,q_2,p_2,q_1),
\end{split}
\end{equation}

\begin{equation}
\begin{split}
&\Gamma_C^{\mu_1\mu_2\nu_1\nu_2}=V_C^{\nu_1\mu_1\mu_2\nu_2}(q_1,p_1,p_2,q_2)+V_C^{\nu_1\mu_2\mu_1\nu_2}(q_1,p_2,p_1,q_2)
\\&+V_C^{\nu_1\mu_1\nu_2\mu_2}(q_1,p_1,q_2,p_2)+V_C^{\nu_1\mu_2\nu_2\mu_1}(k_1,k_2,k_3,k_4)+V_C^{\nu_2\mu_1\mu_2\nu_1}(q_2,p_1,p_2,q_1)
\\&+V_C^{\nu_2\mu_2\mu_1\nu_1}(q_2,p_2,p_1,q_1)+V_C^{\nu_2\mu_1\nu_1\mu_2}(q_2,p_1,q_1,p_2)+V_C^{\nu_2\mu_2\nu_1\mu_1}(q_2,p_2,q_1,p_1)
\\&+{\rm irrelevant},
\end{split}
\end{equation}

\begin{equation}
\Gamma_F^{\mu_1\mu_2\nu_1\nu_2}=V_F(p_1,q_1,p_2,q_2)+V_F(p_2,q_1,p_1,q_2)+V_F(p_1,q_2,p_2,q_1)+V_F(p_2,q_2,p_1,q_1),
\end{equation}

\begin{equation}
\begin{split}
&V_A^{\mu_1\mu_2\mu_3\mu_4}(k_1,k_2,k_3,k_4)=\frac{i}{2}f_{\star_2}(k_1,k_2)f_{\star_2}(k_3,k_4)(k_3\theta k_4)
\\&\cdot\Big(2(\theta k_2)^{\mu_1}k_4^{\mu_2}g^{\mu_3\mu_4}-2(\theta k_2)^{\mu_1}k_4^{\mu_3}g^{\mu_2\mu_4}-(k_2\cdot k_4)\theta^{\mu_1\mu_2}g^{\mu_3\mu_4}+k_2^{\mu_4}k_4^{\mu_3}g^{\mu_1\mu_2}\Big),
\end{split}
\end{equation}

\begin{equation}
\begin{split}
&V_B^{\mu_1\mu_2\mu_3\mu_4}(k_1,k_2,k_3,k_4)=-\frac{i}{4}f_{\star_2}(k_1,k_2)f_{\star_2}(k_3,k_4)\Big((k_1\theta k_2)\big((k_3\theta k_4)g^{\mu_1\mu_3}g^{\mu_2\mu_4}
\\&+(\theta k_3)^{\mu_4}k_4^{\mu_2}g^{\mu_1\mu_3}-k_3^{\mu_1}(\theta k_4)^{\mu_3}g^{\mu_2\mu_4}+k_3^{\mu_1}k_4^{\mu_2}\big)
\\&+(\theta k_1)^{\mu_2}\big(k_2^{\mu_4}(k_3\theta k_4)g^{\mu_1\mu_3}+(k_2\cdot k_4)(\theta k_3)^{\mu_4}g^{\mu_1\mu_3}-k_2^{\mu_4}k_3^{\mu_1}(\theta k_4)^{\mu_3}+(k_2\cdot k_4)k_3^{\mu_1}\theta^{\mu_3\mu_4}\big)
\\&-k_1^{\mu_3}\big((\theta k_2)^{\mu_1}(k_3\theta k_4)g^{\mu_2\mu_4}+(\theta k_2)^{\mu_1}(\theta k_3)^{\mu_4}k_4^{\mu_2}-k_2^{\mu_4}(k_3\theta k_4)\theta^{\mu_1\mu_2}-(k_2\cdot k_4)(\theta k_3)^{\mu_4}\theta^{\mu_1\mu_2}\big)
\\&+(k_1\cdot k_3)\big((\theta k_2)^{\mu_1}(\theta k_4)^{\mu_3}g^{\mu_2\mu_4}-(\theta k_2)^{\mu_1}k_4^{\mu_2}\theta^{\mu_3\mu_4}-(\theta k_4)^{\mu_3}k_2^{\mu_4}\theta^{\mu_1\mu_2}+(k_2\cdot k_4)\theta^{\mu_1\mu_2}\theta^{\mu_3\mu_4}\big)
\\&-(\theta k_2)^{\mu_1}\big(k_2^{\mu_3}(k_3\theta k_4)g^{\mu_2\mu_4}+k_2^{\mu_3}(\theta k_3)^{\mu_4}k_4^{\mu_2}
\\&-(k_2\cdot k_3)(\theta k_4)^{\mu_3} g^{\mu_2\mu_4}+(k_2\cdot k_3)k_4^{\mu_2}\theta^{\mu_3\mu_4}-k_2^{\mu_4}k_3^{\mu_2}(\theta k_4)^{\mu_3}-(k_2\cdot k_4)k_3^{\mu_2}\theta^{\mu_3\mu_4}\big)
\\&-(\theta k_4)^{\mu_3}\big(k_4^{\mu_1}(k_1\theta k_2)g^{\mu_2\mu_4}+k_4^{\mu_1}(\theta k_1)^{\mu_2}k_2^{\mu_4}
\\&-(k_1\cdot k_4)(\theta k_2)^{\mu_1} g^{\mu_2\mu_4}+(k_1\cdot k_4)k_2^{\mu_4}\theta^{\mu_1\mu_1}-k_4^{\mu_2}k_1^{\mu_4}(\theta k_2)^{\mu_1}-(k_1\cdot k_3)k_1^{\mu_4}\theta^{\mu_1\mu_2}\big)
\\&+2(\theta k_2)^{\mu_1}(\theta k_4)^{\mu_3}\big((k_2\cdot k_4)g^{\mu_2\mu_4}-k_2^{\mu_4}k_4^{\mu_2}\big)\Big),
\end{split}
\end{equation}

\begin{equation}
\begin{split}
&V_C^{\mu_1\mu_2\mu_3\mu_4}(k_1,k_2,k_3,k_4)=\frac{i}{8}f_{\star_{3'}}(k_2,k_3,k_4)\Big(k_1^2\big(-3(\theta k_3)^{\mu_2}k_4^{\mu_1}\theta^{\mu_3\mu_4}
\\&+4(\theta k_4)^{\mu_2}(\theta k_4)^{\mu_3}g^{\mu_1\mu_4}-k_4^{\mu_1}(\theta k_4)^{\mu_2}\theta^{\mu_3\mu_4}+2\theta^{\mu_2\mu_3}(k_3\theta k_4)g^{\mu_1\mu_4}+2(\theta k_3)^{\mu_4}k_4^{\mu_1}\theta^{\mu_2\mu_3}
\\&-2(\theta k_4)^{\mu_3}k_4^{\mu_1}\theta^{\mu_2\mu_4}+4(\theta k_2)^{\mu_4}(\theta k_4)^{\mu_3}g^{\mu_1\mu_2}-2(k_2\theta k_4)g^{\mu_1\mu_2}\theta^{\mu_3\mu_4}
\\&-2 k_2^{\mu_1}(\theta k_4)^{\mu_3}\theta^{\mu_2\mu_4}-k_2^{\mu_1}(\theta k_4)^{\mu_2}\theta^{\mu_3\mu_4}\big)-k_1^{\mu_1}\big(-3(k_1\cdot k_4)(\theta k_3)^{\mu_2}\theta^{\mu_3\mu_4}
\\&+4k_1^{\mu_4}(\theta k_4)^{\mu_2}(\theta k_4)^{\mu_3}-(k_1\cdot k_4)(\theta k_4)^{\mu_2}\theta^{\mu_3\mu_4}
-2k_1^{\mu_4}(k_3\theta k_4)\theta^{\mu_2\mu_3}\\&-2(k_1\cdot k_4)(\theta k_3)^{\mu_4}\theta^{\mu_2\mu_3}-2(k_1\cdot k_4)(\theta k_4)^{\mu_3}\theta^{\mu_2\mu_4}+4k_1^{\mu_2}(\theta k_2)^{\mu_4}(\theta k_4)^{\mu_3}\\&+2k_1^{\mu_2}(k_2\theta k_4)\theta^{\mu_3\mu_4}+2(k_1\cdot k_2)(\theta k_4)^{\mu_3}\theta^{\mu_2\mu_4}-(k_1\cdot k_2)(\theta k_4)^{\mu_2}\theta^{\mu_3\mu_4}\big)\Big),
\end{split}
\end{equation}

\begin{equation}
\begin{split}
&V_F^{\mu_1\mu_2\mu_3\mu_4}(k_1,k_2,k_3,k_4)=-\frac{i}{2}f_{\star_2}(k_1,k_2)f_{\star_2}(k_3,k_4)
\cdot\Big((k_3\theta k_4)k_4^{\mu_4}\big(2(\theta k_2)^{\mu_1}g^{\mu_2\mu_3}
\\&-k_2^{\mu_3}\theta^{\mu_1\mu_2}+2(\theta k_1)^{\mu_2}g^{\mu_1\mu_3}+\theta^{\mu_1\mu_2}k_1^{\mu_3}\big)-(k_1\theta k_2)g^{\mu_1\mu_2}\big(2(k_3+k_4)^{\mu_4}(\theta k_4)^{\mu_3}
\\&-(k_3+k_4)\cdot k_4\theta^{\mu_3\mu_4}+2(k_3+k_4)^{\mu_3}(\theta k_3)^{\mu_4}+k_3\cdot(k_3+k_4)\theta^{\mu_3\mu_4}\big)
\\&+(k_1\theta k_2)(k_3\theta k_4)g^{\mu_1\mu_2}g^{\mu_3\mu_4}\Big),
\end{split}
\end{equation}

\begin{equation}
\Gamma_{Bc\bar c_{BFG}}^\mu\left(p;q\right)=f_{\star_2}(p,q)\left(-\frac{1}{2}(p+q)^2(\theta q)^\mu+(p\theta q)(p+2q)^\mu\right),
\end{equation}

\begin{equation}
\begin{split}
\Gamma_{BBc\bar c_{BFG}}^{\mu_1\mu_2}\left(p_1,p_2;q,q'\right)&=\Gamma_{BBc\bar c_{NCFG-BFM}}^{\mu\nu_1\nu_2}\left(p_1,p_2;q,q'\right)
\\&-if_{\star_2}(p_1,q')f_{\star_2}(p_2,q)(p_1\theta q')\left(-\frac{1}{2}(\theta q)^{\mu_2}(p_2+q)^{\mu_1}+g^{\mu_1\mu_2}(p_2\theta q)\right)
\\&-if_{\star_2}(p_2,q')f_{\star_2}(p_1,q)(p_2\theta q')\left(-\frac{1}{2}(\theta q)^{\mu_2}(p_1+q)^{\mu_1}+g^{\mu_1\mu_2}(p_1\theta q)\right)
\\&+{\rm irrelevant}.
\end{split}
\end{equation}

\subsection{The noncommutative Feynman gauge}

\begin{equation}
\Gamma_{BQQ_{NCFG-BFM}}^{\mu\nu_1\nu_2}\left(p;q_1,q_2\right)=\Gamma_{BQQ_{NCFG-BFM_1}}^{\mu\nu_1\nu_2}\left(p;q_1,q_2\right)+\Gamma_{BQQ_{NCFG-BFM_2}}^{\mu\nu_1\nu_2}\left(p;q_1,q_2\right),
\end{equation}

\begin{equation}
\Gamma_{BQQ_{NCFG-BFM_1}}^{\mu\nu_1\nu_2}=\Gamma_{BQQ_{BFG_1}}^{\mu\nu_1\nu_2},
\end{equation}

\begin{equation}
\begin{split}
\Gamma_{BQQ_{NCFG-BFM_2}}^{\mu\nu_1\nu_2}=&\frac{1}{2}f_{\star_2}(p,q_1)\Big(q_1^{\nu_1}\left(2q_1^{\nu_2}(\theta q_2)^\mu-(q_1\cdot q_2)\theta^{\mu\nu_2}+2q_1^\mu(\theta p)^{\nu_2}+(q_1\cdot p)\theta^{\mu\nu_2}\right)
\\&+q_2^{\nu_2}\left(2q_2^{\nu_1}(\theta q_1)^\mu-(q_1\cdot q_2)\theta^{\mu\nu_1}+2q_2^\mu(\theta p)^{\nu_1}+(q_2\cdot p)\theta^{\mu\nu_1}\right)\Big),
\end{split}
\end{equation}

\begin{equation}
\Gamma_{BBQQ_{NCFG-BFM}}^{\mu_1\mu_2\nu_1\nu_2}=\Gamma_{BBQQ_{NCFG-BFM_1}}^{\mu_1\mu_2\nu_1\nu_2}+\Gamma_{BBQQ_{NCFG-BFM_2}}^{\mu_1\mu_2\nu_1\nu_2},
\end{equation}

\begin{equation}
\Gamma_{BBQQ_{NCFG-BFM_1}}^{\mu_1\mu_2\nu_1\nu_2}=\Gamma_{BBQQ_{BFG_1}}^{\mu_1\mu_2\nu_1\nu_2},
\end{equation}

\begin{equation}
\Gamma_{BBQQ_{NCFG-BFM_2}}^{\mu_1\mu_2\nu_1\nu_2}=\Gamma_D^{\mu_1\mu_2\nu_1\nu_2}+\Gamma_E^{\mu_1\mu_2\nu_1\nu_2},
\end{equation}

\begin{equation}
\begin{split}
&\Gamma_D^{\mu_1\mu_2\nu_1\nu_2}=V_D^{\nu_1\mu_1\mu_2\nu_2}(q_1,p_1,p_2,q_2)+V_D^{\nu_1\mu_2\mu_1\nu_2}(q_1,p_2,p_1,q_2)
\\&+V_D^{\nu_1\mu_1\nu_2\mu_2}(q_1,p_1,q_2,p_2)+V_D^{\nu_1\mu_2\nu_2\mu_1}(k_1,k_2,k_3,k_4)+V_D^{\nu_2\mu_1\mu_2\nu_1}(q_2,p_1,p_2,q_1)
\\&+V_D^{\nu_2\mu_2\mu_1\nu_1}(q_2,p_2,p_1,q_1)+V_D^{\nu_2\mu_1\nu_1\mu_2}(q_2,p_1,q_1,p_2)+V_D^{\nu_2\mu_2\nu_1\mu_1}(q_2,p_2,q_1,p_1)
\\&+{\rm irrelevant},
\end{split}
\end{equation}

\begin{equation}
\Gamma_E^{\mu_1\mu_2\nu_1\nu_2}=V_E(p_1,q_1,p_2,q_2)+V_E(p_2,q_1,p_1,q_2)+V_E(p_1,q_2,p_2,q_1)+V_E(p_2,q_2,p_1,q_1),
\end{equation}

\begin{equation}
\begin{split}
&V_D^{\mu_1\mu_2\mu_3\mu_4}(k_1,k_2,k_3,k_4)=\frac{i}{8} f_{\star_{3'}}(k_2,k_3,k_4)k_1^{\mu_1}\big(-3(k_1\cdot k_4)(\theta k_3)^{\mu_2}\theta^{\mu_3\mu_4}
\\&+4k_1^{\mu_4}(\theta k_4)^{\mu_2}(\theta k_4)^{\mu_3}-(k_1\cdot k_4)(\theta k_4)^{\mu_2}\theta^{\mu_3\mu_4}
-2k_1^{\mu_4}(k_3\theta k_4)\theta^{\mu_2\mu_3}\\&-2(k_1\cdot k_4)(\theta k_3)^{\mu_4}\theta^{\mu_2\mu_3}-2(k_1\cdot k_4)(\theta k_4)^{\mu_3}\theta^{\mu_2\mu_4}+4k_1^{\mu_2}(\theta k_2)^{\mu_4}(\theta k_4)^{\mu_3}\\&+2k_1^{\mu_2}(k_2\theta k_4)\theta^{\mu_3\mu_4}+2(k_1\cdot k_2)(\theta k_4)^{\mu_3}\theta^{\mu_2\mu_4}-(k_1\cdot k_2)(\theta k_4)^{\mu_2}\theta^{\mu_3\mu_4}\big),
\end{split}
\end{equation}

\begin{equation}
\begin{split}
&V_E^{\mu_1\mu_2\mu_3\mu_4}(k_1,k_2,k_3,k_4)=-\frac{i}{8}f_{\star_2}(k_1,k_2)f_{\star_2}(k_3,k_4)\\&\cdot((k_1+k_2)^{\mu_2}(\theta k_2)^{\mu_1}-(k_1+k_2)\cdot k_2\theta^{\mu_1\mu_2})
((k_3+k_4)^{\mu_4}(\theta k_4)^{\mu_3}-(k_3+k_4)\cdot k_4\theta^{\mu_3\mu_4}),
\end{split}
\end{equation}

\begin{equation}
\Gamma_{Bc\bar c_{NCFG-BFM}}^\mu\left(p;q\right)=f_{\star_2}(p,q)\left(-\frac{1}{2}(p+q)^2(\theta q)^\mu+(p\theta q)(p+q)^\mu\right)
\end{equation}

\begin{equation}
\begin{split}
\Gamma_{BBc\bar c_{NCFG-BFM}}^{\mu_1\mu_2}\left(p_1,p_2;q,q'\right)=&\frac{i}{2}\Big(
f_{\star_2}(p_1,q')f_{\star_2}(p_2,q)q'^{\mu_1}(p_1\theta q')(\theta q)^{\mu_2}
\\&+f_{\star_2}(p_2,q')f_{\star_2}(p_1,q)q'^{\mu_2}(p_2\theta q')(\theta q)^{\mu_1}\Big)+{\rm irrelevant}.
\end{split}
\end{equation}

\subsection{Feynman rules for the noncommutative U(1) Super Yang-Mils}

\begin{figure}
\begin{center}
\includegraphics[width=6cm]{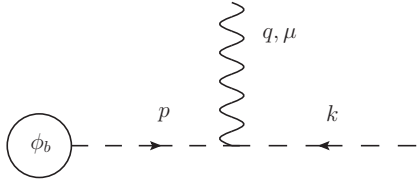}
\end{center}
\caption{Scalar-photon BFM FR.}
\label{fig:BFMFR9}
\end{figure}

\begin{figure}
\begin{center}
\includegraphics[width=6cm]{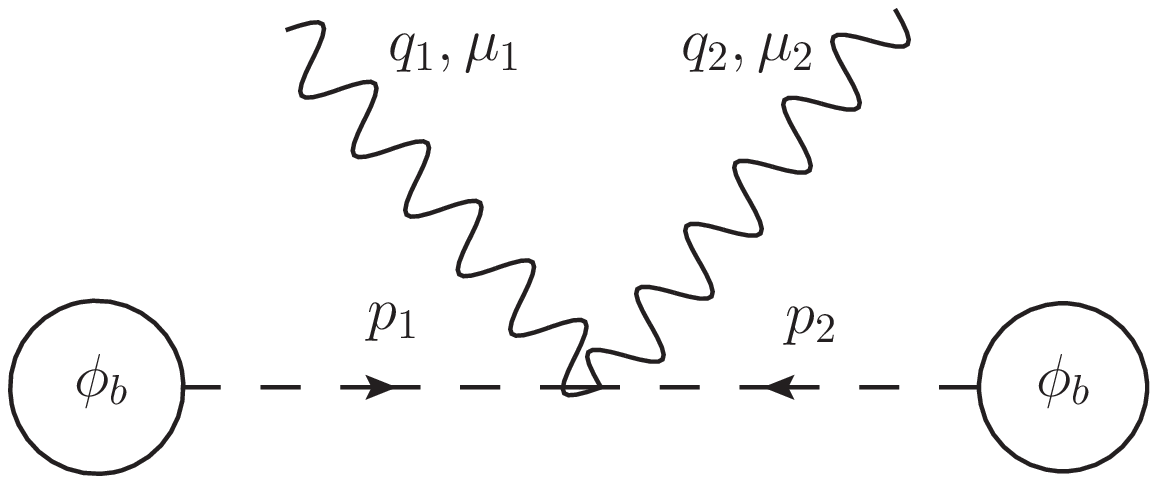}
\end{center}
\caption{Scalar-2photons BFM FR.}
\label{fig:BFMFR10}
\end{figure}

\begin{figure}
\begin{center}
\includegraphics[width=6cm]{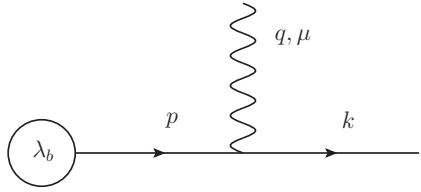}
\end{center}
\caption{Fermion-photon BFM FR.}
\label{fig:BFMFR11}
\end{figure}

\begin{figure}
\begin{center}
\includegraphics[width=6cm]{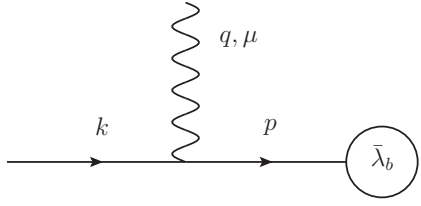}
\end{center}
\caption{Antifermion-photon BFM FR.}
\label{fig:BFMFR12}
\end{figure}

\begin{figure}
\begin{center}
\includegraphics[width=6cm]{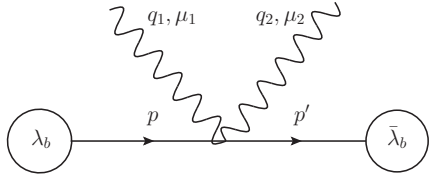}
\end{center}
\caption{Fermions-2photons BFM FR.}
\label{fig:BFMFR13}
\end{figure}

We list here only the coupling involving background photino and antiphotino fields $\lambda_B$ and $\bar\lambda_B$ as well as background adjoint scalar field(s) $\phi_B$, since the coupling between background photon field $B_\mu$ and quantum fluctuations of photino, antiphotino and adjoint scalar(s) are identical to those in~\cite{Martin:2016zon}. Figures corresponding to the Feynman rules in this subsection are: Fig. \ref{fig:BFMFR9}, Fig. \ref{fig:BFMFR10}, Fig. \ref{fig:BFMFR11}, Fig. \ref{fig:BFMFR12}, and Fig. \ref{fig:BFMFR13}.

\begin{equation}
\Gamma_{\lambda_B\bar\lambda_QQ}^\mu\left(p,k;q\right)=if_{\star_2}(p,q)\left(\gamma^\mu(p\theta q)-(\theta q)^\mu(/\!\!\! p+/\!\!\! q)\right),
\end{equation}

\begin{equation}
\Gamma_{\lambda_Q\bar\lambda_BQ}^\mu\left(p,k;q\right)=if_{\star_2}(p,q)\left(\gamma^\mu(p\theta k)+(\theta p+\theta k)^\mu /\!\!\! q\right),
\end{equation}

\begin{equation}
\begin{split}
\Gamma_{\lambda_B\bar\lambda_BQQ}^{\mu_1\mu_2}\left(p,p';q_1,q_2\right)
&=if_{\star_2}(p',q_1)f_{\star_2}(p,q_2)\Big(-(/\!\!\! q_2+/\!\!\! p)(\theta p')^{\mu_1}(\theta p)^{\mu_2}
\\&
+(q_1\theta(q_2+p))(\theta p)^{\mu_2}\bar\sigma^{\mu_1}-(\theta p')^{\mu_1}(q_2\theta p_2)\bar\sigma^{\mu_2}\Big)
\\&+if_{\star_2}(p',q_2)f_{\star_2}(p,q_1)\Big(-(/\!\!\! q_1+/\!\!\! p)(\theta p')^{\mu_2}(\theta p)^{\mu_1}
\\&
+(q_2\theta(q_1+p))(\theta p)^{\mu_1}\bar\sigma^{\mu_2}-(\theta p')^{\mu_2}(q_1\theta p)\bar\sigma^{\mu_1}\Big),
\end{split}
\end{equation}

\begin{equation}
\Gamma_{\phi_B\phi_QQ}^\mu\left(p,k;q\right)=-f_{\star_2}(p,q)(k^2(\theta p)^\mu+p^\mu(q\theta k)+k^\mu(p\theta k)),
\end{equation}

\begin{equation}
\begin{split}
&\Gamma_{\phi_B\phi_BQQ}^{\mu_1\mu_2}\left(p_1,p_2;q_1,q_2\right)
\\&=\frac{i}{2}f_{\star_2}(p_1,q_1)f_{\star_2}(p_2,q_2)\Big((q_1\theta p_1)(q_2\theta p_2)g^{\mu_1\mu_2}+(p_1+q_1)\cdot(p_2+q_2)(\theta p_1)^{\mu_1}(\theta p_2)^{\mu_2}
\\&+2(\theta p_1)^{\mu_1}(q_2\theta p_2)p_2^{\mu_2}-2(p_1+q_1)^{\mu_2}(\theta p_1)^{\mu_1}(q_2\theta p_2)\Big)
\\&+\frac{i}{2}f_{\star_2}(p_2,q_1)f_{\star_2}(p_1,q_2)\Big((q_1\theta p_2)(q_2\theta p_1)g^{\mu_1\mu_2}+(p_2+q_1)\cdot(p_1+q_2)(\theta p_2)^{\mu_1}(\theta p_1)^{\mu_2}
\\&+2(\theta p_2)^{\mu_1}(q_2\theta p_1)p_1^{\mu_2}-2(p_2+q_1)^{\mu_2}(\theta p_2)^{\mu_1}(q_2\theta p_1)\Big)
\\&+\frac{i}{2}f_{\star_2}(p_1,q_2)f_{\star_2}(p_2,q_1)\Big((q_2\theta p_1)(q_1\theta p_2)g^{\mu_1\mu_2}+(p_1+q_2)\cdot(p_2+q_1)(\theta p_1)^{\mu_2}(\theta p_2)^{\mu_1}
\\&+2(\theta p_1)^{\mu_2}(q_1\theta p_2)p_2^{\mu_1}-2(p_1+q_2)^{\mu_1}(\theta p_1)^{\mu_2}(q_1\theta p_2)\Big)
\\&+\frac{i}{2}f_{\star_2}(p_1,q_1)f_{\star_2}(p_2,q_2)\Big((q_1\theta p_1)(q_2\theta p_2)g^{\mu_1\mu_2}+(p_1+q_1)\cdot(p_2+q_2)(\theta p_1)^{\mu_1}(\theta p_2)^{\mu_2}
\\&+2(\theta p_1)^{\mu_1}(q_1\theta p_1)p_2^{\mu_2}-2(p_2+q_2)^{\mu_1}(\theta p_2)^{\mu_2}(q_1\theta p_1)\Big).
\end{split}
\end{equation}

\section{Evaluation of DeWitt effective action in terms of noncommutative fields}

We accumulate the reference results for section 4, i.e. the one-loop quantum corrections to the quadratic part of the  DeWitt effective action of the $\rm U(1)$ Super Yang Mills. We first give the model setting, then the results of relevant one-loop diagrams.

\subsection{The noncommutative Yang-Mils theory}
Let's first handle the $\rm U(N)$ NCYM only, then extend the results to its supersymmetrization. The NCYM action is the usual one 
\begin{gather}
S_{\rm NCYM}=-\frac{1}{4g^2}\int \tr\Big(\hat F_{\mu\nu}\hat F^{\mu\nu}\Big),
\\
\hat F_{\mu\nu}=\partial_\mu \hat A_\nu-\partial_\nu \hat A_\mu+i\Big[\hat A_\mu\stackrel{\star}{,}\hat A_\nu\Big], \:\:\hat A_\mu=\hat A_\mu^a T^a,
\\
\hat\delta_{\rm BRS}\hat A_\mu=\hat D_\mu \hat C=\partial_\mu \hat C+i\Big[\hat A_\mu\stackrel{\star}{,}\hat C\Big], \:\:\hat C=\hat C^a T^a.
\end{gather}

Background field quantization follows the BRST procedure below:
\begin{gather}
\hat A_\mu\Longrightarrow\hat B_\mu+\hbar^\frac{1}{2}\hat Q_\mu,\:\:\hat Q_\mu=\hat Q_\mu^a T^a,
\\
\hat\delta_{\rm BRS}\hat B_\mu=0,\:\:
\hbar\hat\delta_{\rm BRS}\hat Q_\mu=\hat D_\mu\big[\hat B_\mu+\hbar^\frac{1}{2}\hat Q_\mu\big]\hat C
=\partial_\mu \hat C+i\Big[\hat B_\mu+\hbar^\frac{1}{2}\hat Q_\mu\stackrel{\star}{,}\hat C\Big].
\end{gather}

Next we introduce the DeWitt effective action $\hat\Gamma_{\rm DeW}\big[\hat B_\mu\big]$ in the background field gauge 
\begin{equation}
e^{\frac{i}{\hbar}\hat\Gamma_{\rm BFG}\big[\hat B_\mu\big]}=\int d\hat Q_\mu^a d\hat C^a d{\bar C}^a d F^a e^{\frac{i}{\hbar}S_{\rm NCYM}\big[\hat B_\mu+\hbar^{\frac{1}{2}}\hat Q_\mu\big]+i S_{\rm BFG}\big[\hat B_\mu,\hat Q_\mu\big]},
\label{NCDeWitt}
\end{equation}
with
\begin{gather}
S_{\rm NCYM}\big[\hat B_\mu+\hbar^\frac{1}{2}\hat Q_\mu\big]=-\frac{1}{4g^2}\int \tr\left(\hat F_{\mu\nu}\big[\hat B_\mu+\hbar^\frac{1}{2}\hat Q_\mu\big]\hat F^{\mu\nu}\big[\hat B_\mu+\hbar^\frac{1}{2}\hat Q_\mu\big]\right),
\\
S_{\rm BFG}\big[\hat B_\mu,\hat Q_\mu\big]=\frac{\hbar}{g^2}\int \tr \hat\delta_{\rm BRS} \hat{\bar C}
\left(\alpha F+ \hat D_\mu\big[\hat B_\mu\big]\hat Q^\mu\right),
\label{NCBFG}
\end{gather}
and
\begin{equation}
\hat\delta_{\rm BRS}\bar C=\hbar^{-\frac{1}{2}}\hat F,\:\:\hat\delta_{\rm BRS} F=0.
\end{equation}

The one-loop contribution $\hat\Gamma_{\rm BFG}^{(1)}\big[\hat B_\mu\big]$ to $\hat\Gamma_{\rm DeW}\big[\hat B_\mu\big]$ corresponds to the $\hbar$ order expansion of the latter
\begin{equation}
\hat\Gamma_{\rm BFG}\big[\hat B_\mu\big]=\hat\Gamma_{\rm BFG}^{(0)}\big[\hat B_\mu\big]+\hbar\hat\Gamma_{\rm BFG}^{(1)}\big[\hat B_\mu\big]+......
\label{NCDeWitthbar}
\end{equation}
To evaluate it we first expand the corresponding classical actions to the appropriate order
\begin{equation}
\begin{split}
\hbar^{-1}S_{\rm NCYM}\big[\hat B_\mu+\hbar^\frac{1}{2}\hat Q_\mu\big]=&-\frac{1}{4g^2\hbar}\int \tr\left(\hat F_{\mu\nu}\big[\hat B_\mu\big]\hat F^{\mu\nu}\big[\hat B_\mu\big]\right)
\\&-\frac{1}{2g^2\hbar^\frac{1}{2}}\int \tr\left(\hat D^\mu\hat F_{\mu\nu}\big[\hat B_\mu\big]\hat Q^\nu\right)
\\&-\frac{1}{4g^2}\int \tr\left(\hat D_\mu\big[\hat B_\mu\big] \hat Q_\nu-\hat D_\nu\big[\hat B_\mu\big] \hat Q_\mu\right)^2
\\&+\frac{1}{2g^2}\int \tr\hat F^{\mu\nu}\big[\hat B_\mu\big]\left[\hat Q_\mu\stackrel{\star}{,}\hat Q_\nu\right]
+\mathcal{O}\big(\hbar^\frac{1}{2}\big),
\end{split}
\label{NCYMhbar0}
\end{equation}
\begin{equation}
S_{\rm BFG}=\int\tr\left(\alpha \hat F^2+\hat F\hat D_\mu\big[\hat B_\mu\big]\hat Q^{\mu}-\bar C\hat D_\mu\big[\hat B_\mu\big]\hat D^\mu\big[\hat B_\mu\big]\hat C+\mathcal{O}\big(\hbar^{\frac{1}{2}}\big)\right).
\label{NCBFGhbar0}
\end{equation}

Now, let's choose $\hat B_\mu$ on-shell, i.e. $\hat D^\mu\big[\hat B_\mu\big]\hat F_{\mu\nu}\big[\hat B_\mu\big]=0$. Then, substituting \eqref{NCDeWitthbar}, \eqref{NCYMhbar0} and \eqref{NCBFGhbar0} in \eqref{NCDeWitt}, one gets
\begin{equation}
\Gamma_{\rm BFG}^{(0)}\big[\hat B_\mu\big]=S_{\rm NCYM}\big[\hat B_\mu\big],
\end{equation}
\begin{equation}
\hat\Gamma_{\rm BFG}^{(1)}\big[\hat B_\mu\big]=-i\ln\int d\hat Q_\mu^a d\hat C^a d{\bar C}^a d F^a e^{\frac{i}{\hbar}S^{(1)}_{\rm NC}},
\label{NCeffectivehbar1}
\end{equation}
with
\begin{equation}
\begin{split}
S^{(1)}_{\rm NC}=&-\frac{1}{4g^2}\int \tr\left(\hat D_\mu\big[\hat B_\mu\big] \hat Q_\nu-\hat D_\nu\big[\hat B_\mu\big] \hat Q_\mu\right)^2
+\frac{1}{2g^2}\int \tr\hat F^{\mu\nu}\big[\hat B_\mu\big]\left[\hat Q_\mu\stackrel{\star}{,}\hat Q_\nu\right]
\\&+\int\tr\left(\alpha \hat F^2+\hat F\hat D_\mu\big[\hat B_\mu\big]\hat Q^{\mu}-\bar C\hat D_\mu\big[\hat B_\mu\big]\hat D^\mu\big[\hat B_\mu\big]\hat C\right).
\end{split}
\label{NCtotalactionhbar1}
\end{equation}
Restrict \eqref{NCtotalactionhbar1} to $\rm U(1)$ and $\alpha=1$, the 1-loop 1PI photon two point function is then evaluated as the sum over 1-loop 1PI diagrams with all $\hat B_\mu$ external lines. There are four diagrams in total, which can be separated into two parts: the bubble part which sums over the photon and ghost bubble diagrams and tadpole part which sums over the photon and ghost tadpole diagrams. Consequently the final result is as follows
\begin{gather}
\hat\Gamma^{\mu\nu}_{\rm BFG}=\hat B^{\mu\nu}_{\rm BFG}+\hat T^{\mu\nu}_{\rm BFG},
\label{NCSYMPhotontotal}
\\
\begin{split}
\hat B^{\mu\nu}_{\rm BFG}=&\frac{1}{(4\pi)^2}\Bigg(\Big(g^{\mu\nu}p^2-p^\mu p^\nu\Big)
\\&
\cdot\bigg((4\pi\mu^2)^{2-\frac{D}{2}}(p^2)^{\frac{D}{2}-2}2(6-7D){\rm\Gamma}\left(1-\frac{D}{2}\right){\rm B}\left(\frac{D}{2}, \frac{D}{2}\right)\bigg|_{D\to 4-\epsilon}
\\&-12I_{K_0}-16I_{K_1}\bigg)
-\frac{(\theta p)^\mu(\theta p)^\nu}{(\theta p)^2}\bigg(16T_0+8I_K^0-48p^2I_K^1\bigg)+g_{\mu\nu}8T_0\Bigg),
\label{NCSYMPhotonbubbles}
\end{split}
\\
\hat T^{\mu\nu}_{\rm BFG}=-\frac{1}{(4\pi)^2}g_{\mu\nu}8T_0.
\label{NCSYMPhotontadpoles}
\end{gather}

\subsection{The $\rm U(1)$ noncommutative Super Yang-Mils theory}

\begin{figure}
\begin{center}
\includegraphics[width=7cm]{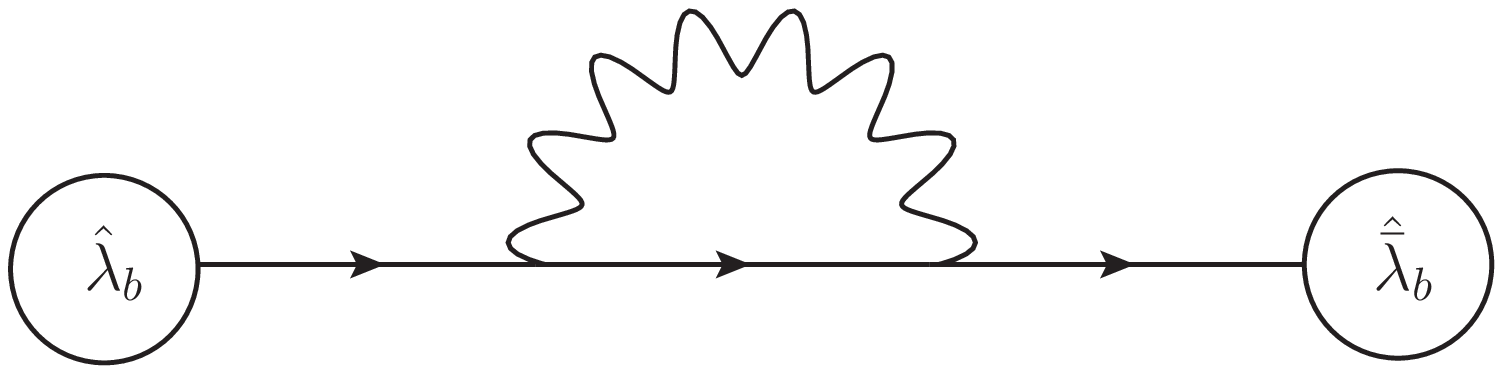}
\end{center}
\caption{$\cal N$=1 photino-photon bubble: $\Sigma^{\dot{\alpha}\alpha}_{({\hat\lambda}_b)}(p)_{\rm bub}$.}
\label{fig:N=1 fermion-photon bubble}
\end{figure}

\begin{figure}
\begin{center}
\includegraphics[width=7cm]{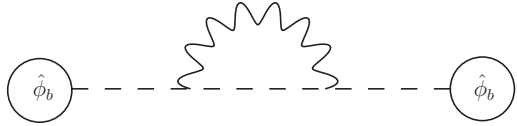}
\end{center}
\caption{$\cal N$=2 scalar-photon bubble: $\Sigma_{({\hat\phi}_b)}(p)_{\rm bub}$.}
\label{fig:N=2 scalar-photon bubble}
\end{figure}

\begin{figure}
\begin{center}
\includegraphics[width=7cm]{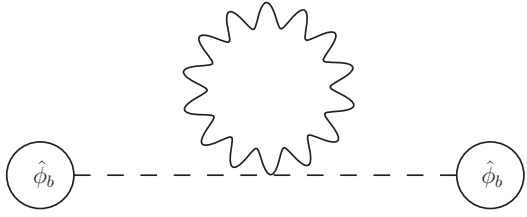}
\end{center}
\caption{$\cal N$=2 scalar-photon tadpole: $\Sigma_{({\hat\phi}_b)}(p)_{\rm tad}$.}
\label{fig:N=2 scalar-photon tadpole}
\end{figure}

Now we shift to the supersymmetrization of the $\rm U(1)$ theory. As discussed in~\cite{Martin:2016zon}, this sector contains the photino(s) for $\mathcal N=1,2,4$ and adjoint scalars for $\mathcal N=2,4$. The interaction between photinos and adjoint scalars remain the same before and after SW map~\cite{Martin:2016zon}, therefore we are not going to repeat them here. Using \eqref{SPartnerActions} without SW map we obtain one self-energy/bubble diagram Fig.~\ref{fig:N=1 fermion-photon bubble} for photino, as well as a bubble diagram Fig.~\ref{fig:N=2 scalar-photon bubble} and a tadpole diagram Fig.~\ref{fig:N=2 scalar-photon tadpole} for adjoint scalar.

Explicit computation based on these diagrams then gives the following two point functions ($\Sigma^{\dot{\alpha}\alpha}_{\rm NCSYM}$ and $\Sigma_{{(\hat\phi)}_{\rm NCSYM}}$) for noncommutative photino $\hat\lambda$ and adjoint scalar $\hat\phi$:
\begin{equation}
\begin{split}
&\Sigma^{\dot{\alpha}\alpha}_{\rm NCSYM}=\Sigma^{\dot{\alpha}\alpha}_{\rm NCSYM_{\rm bubble}}
\\&=\bar\sigma^\mu p_\mu\frac{1}{(4\pi)^2}\left((4\pi\mu^2)^{2-\frac{D}{2}}(p^2)^{\frac{D}{2}-2}(2-D){\rm\Gamma}\left(2-\frac{D}{2}\right){\rm B}\left(\frac{D}{2}-1, \frac{D}{2}-1\right)\bigg|_{D\to 4-\epsilon}+4I_K^0\right),
\end{split}
\label{NCSYMPhotino}
\end{equation}
\begin{equation}
\Sigma_{{(\hat\phi)}_{\rm NCSYM}}=\Sigma_{{(\hat\phi)}_{\rm NCSYM_{\rm bubble}}}+\Sigma_{{(\hat\phi)}_{\rm NCSYM_{\rm tadpole}}},
\label{NCSYMStotal}
\end{equation}
\begin{equation}
\begin{split}
&\Sigma_{{(\hat\phi)}_{\rm NCSYM_{\rm bubble}}}=p^2 \frac{4}{(4\pi)^2}
\\&\cdot\left(-(4\pi\mu^2)^{2-\frac{D}{2}}(p^2)^{\frac{D}{2}-2}{\rm\Gamma}\left(2-\frac{D}{2}\right){\rm B}\left(\frac{D}{2}-1, \frac{D}{2}-1\right)\bigg|_{D\to 4-\epsilon}-T_0+2 I_K^0\right),
\end{split}
\label{NCSYMSbubble}
\end{equation}
\begin{equation}
\Sigma_{{(\hat\phi)}_{\rm NCSYM_{\rm tadpole}}}=p^2\frac{16}{(4\pi)^2}T_0.
\label{NCSYMStadpole}
\end{equation}


\newpage

\begin{thebibliography}{999}

\bibitem{Seiberg:1999vs}
  N.~Seiberg and E.~Witten,
 {\it String theory and noncommutative geometry},
  JHEP {\bf 9909} (1999) 032,
  doi:10.1088/1126-6708/1999/09/032
  [hep-th/9908142].

\bibitem{Mehen:2000vs}
T.~Mehen and M.~B. Wise,
{\it Generalized *-products, Wilson lines and the solution of the Seiberg-Witten equations},
 JHEP {\bf 12} (2000) 008,  [\href{http://xxx.lanl.gov/abs/hep-th/0010204}{{\tt hep-th/0010204}}].

\bibitem{Jurco:2000fb}
B.~Jurco and P.~Schupp,
{\it Noncommutative Yang-Mills from equivalence of star products},
{\em Eur.\ Phys.\ J.}  {\bf C14} (2000) 367.

\bibitem{Martin:1999aq}
 C.~P.~Martin and D.~Sanchez-Ruiz,
  {\it The one loop UV divergent structure of U(1) Yang-Mills theory on noncommutative R**4},
  Phys.\ Rev.\ Lett.\  {\bf 83} (1999) 476
  [hep-th/9903077].

\bibitem{Minwalla:1999px}
  S.~Minwalla, M.~Van Raamsdonk and N.~Seiberg,
  {\it Noncommutative perturbative dynamics},
  JHEP {\bf 0002} (2000) 020
  doi:10.1088/1126-6708/2000/02/020
  [hep-th/9912072].

\bibitem{Hayakawa:1999yt}
  M.~Hayakawa,
  {\it Perturbative analysis on infrared aspects of noncommutative QED on R**4},
  Phys.\ Lett.\ B {\bf 478} (2000) 394
  [hep-th/9912094].

\bibitem{Hayakawa:1999zf}
  M.~Hayakawa,
  {\it Perturbative analysis on infrared and ultraviolet aspects of noncommutative QED on R**4},
  hep-th/9912167.

\bibitem{Schupp:2008fs}
  P.~Schupp and J.~You,
  {\it UV/IR mixing in noncommutative QED defined by Seiberg-Witten map},
  JHEP {\bf 0808} (2008) 107
  doi:10.1088/1126-6708/2008/08/107
  [arXiv:0807.4886].

\bibitem{Trampetic:2015zma}
  J.~Trampetic and J.~You,
 {\it The theta-exact Seiberg-Witten maps at the e3 order,}
  Phys.\ Rev.\ D {\bf 91} 125027 (2015),  arXiv:1501.00276 [hep-th].

\bibitem{Horvat:2011bs}
  R.~Horvat, A.~Ilakovac, J.~Trampetic and J.~You,
  {\it On UV/IR mixing in noncommutative gauge field theories},
  JHEP {\bf 1112} (2011) 081
  doi:10.1007/JHEP12(2011)081
  [arXiv:1109.2485].

\bibitem{Horvat:2013rga}
  R.~Horvat, A.~Ilakovac, J.~Trampetic and J.~You,
  {\it Self-energies on deformed spacetimes},
  JHEP {\bf 1311} (2013) 071,
  doi:10.1007/JHEP11(2013)071
  [arXiv:1306.1239 [hep-th]].

  
\bibitem{Horvat:2015aca}
  R.~Horvat, J.~Trampetic and J.~You,
  {\it Photon self-interaction on deformed spacetime},
  Phys.\ Rev.\ D {\bf 92} (2015) no.12,  125006
  doi:10.1103/PhysRevD.92.125006,  [arXiv:1510.08691].

\bibitem{Armoni:2001uw}
  A.~Armoni and E.~Lopez,
  {\it UV / IR mixing via closed strings and tachyonic instabilities},
  Nucl.\ Phys.\ B {\bf 632} (2002) 240
  doi:10.1016/S0550-3213(02)00290-0
  [hep-th/0110113].

\bibitem{Zanon:2000nq}
  D.~Zanon,  {\it Noncommutative N=1, N=2 super U(N) Yang-Mills:
 UV / IR mixing and effective action results at one loop,}
  Phys.\ Lett.\ B {\bf 502} (2001) 265,  [hep-th/0012009].

\bibitem{Ruiz:2000hu}
  F.~R.~Ruiz,
{\it Gauge fixing independence of IR divergences in noncommutative U(1), perturbative tachyonic instabilities and supersymmetry,}  Phys.\ Lett.\ B {\bf 502} (2001) 274,  [hep-th/0012171].

\bibitem{Martin:2008xa}
  C.~P.~Martin and C.~Tamarit,
 {\it The Seiberg-Witten map and supersymmetry,}
  JHEP {\bf 0811} (2008) 087,  [arXiv:0809.2684 [hep-th]].

  
\bibitem{Martin:2016zon}
  C.~P.~Martin, J.~Trampetic and J.~You,
  ``Super Yang-Mills and $\theta$-exact Seiberg-Witten map: absence of quadratic noncommutative IR divergences,''
  JHEP {\bf 1605} (2016) 169
  doi:10.1007/JHEP05(2016)169
  [arXiv:1602.01333 [hep-th]].

\bibitem{DeWitt}
  B. S. DeWitt, {\it Quantum Theory of Gravity. II. The Manifestly Covariant Theory},
  Phys. Rev. {\bf 162} (1967) 1195.

\bibitem{Kallosh:1974yh}
  R.~E.~Kallosh,
  {\it The Renormalization in Nonabelian Gauge Theories},
  Nucl.\ Phys.\ B {\bf 78} (1974) 293.
  doi:10.1016/0550-3213(74)90284-3.

\bibitem{DeWitt:1980jv}
  B.~S.~DeWitt,
  {\it A Gauge Invariant Effective Action}, 
  NSF-ITP-80-31.
  
\bibitem{DeWitt:1988fm}
  B.~S.~DeWitt,
 {\it Dynamical Theory Of Groups And Fields},
Modern Kaluza-Klein Theories, edited by T. Appelquist {\it et al.} 
(Addison-Wesley, Reading, MA, 1987), p.114; Relativity, groups and 
 topology, edited by C. DeWitt (Gordon and Breach, New York, 1965), p.725.


  
\bibitem{Gomis:2000zz}
  J.~Gomis and T.~Mehen,
  {\it Space-time noncommutative field theories and unitarity},
  Nucl.\ Phys.\ B {\bf 591} (2000) 265
  doi:10.1016/S0550-3213(00)00525-3
  [hep-th/0005129].
  
\bibitem{Martin:2016hji}
  C.~P.~Martin, J.~Trampetic and J.~You,
 {\it Equivalence of quantum field theories related by the theta-exact Seiberg-Witten map}, 
  Phys.\ Rev.\ D {\bf 94} (2016) 041703
  doi:10.1103/PhysRevD.94.041703
  [arXiv:1606.03312 [hep-th]].

\bibitem{Ichinose:1992np}
  S.~Ichinose,
  {\it BRS symmetry on background field, Kallosh theorem and renormalization},
  Nucl.\ Phys.\ B {\bf 395} (1993) 433 
  doi:10.1016/0550-3213(93)90224-D.

\bibitem{Martin:2012aw}
  C.~P.~Martin,
  {\it Computing the $\theta$-exact Seiberg-Witten map for arbitrary gauge groups},
  Phys.\ Rev.\ D {\bf 86} (2012) 065010
  doi:10.1103/PhysRevD.86.065010
  [arXiv:1206.2814 [hep-th]].

\bibitem{Martin:2015nna}
  C.~P.~Martin and D.~G.~Navarro,
  {\it The hybrid Seiberg-Witten map, its $\theta$-exact expansion and the antifield formalism},
  Phys.\ Rev.\ D {\bf 92} (2015) no.6,  065026
  doi:10.1103/PhysRevD.92.065026
  [arXiv:1504.06168 [hep-th]].

\bibitem{Ferrari:2003vs}
  A.~F.~Ferrari, H.~O.~Girotti, M.~Gomes, A.~Y.~Petrov, A.~A.~Ribeiro, V.~O.~Rivelles and A.~J.~da Silva,
  {\it Superfield covariant analysis of the divergence structure of noncommutative supersymmetric QED(4),}
  Phys.\ Rev.\ D {\bf 69} 025008 (2004),  [hep-th/0309154].

\bibitem{AlvarezGaume:2003mb}
  L.~Alvarez-Gaume and M.~A.~Vazquez-Mozo,
{\it General properties of noncommutative field theories,}
  Nucl.\ Phys.\  B {\bf 668}, 293 (2003), [arXiv:hep-th/0305093].

\bibitem{Ferrari:2004ex}
  A.~F.~Ferrari, H.~O.~Girotti, M.~Gomes, A.~Y.~Petrov, A.~A.~Ribeiro, V.~O.~Rivelles and A.~J.~da Silva,
 {\it Towards a consistent noncommutative supersymmetric Yang-Mills theory: Superfield covariant analysis,}
  Phys.\ Rev.\ D {\bf 70} 085012 (2004),  [hep-th/0407040].

\bibitem{Jack:2001cr}
  I.~Jack and D.~R.~T.~Jones,
  {\it Ultraviolet finiteness in noncommutative supersymmetric theories},
  New J.\ Phys.\  {\bf 3} (2001) 19
  doi:10.1088/1367-2630/3/1/319
  [hep-th/0109195].

\bibitem{Santambrogio:2000rs}
  A.~Santambrogio and D.~Zanon,
  {\it One loop four point function in noncommutative N=4 Yang-Mills theory},
  JHEP {\bf 0101} (2001) 024
  doi:10.1088/1126-6708/2001/01/024
  [hep-th/0010275].

\bibitem{Pernici:2000va}
  M.~Pernici, A.~Santambrogio and D.~Zanon,
  {\it The One loop effective action of noncommutative N=4 superYang-Mills is gauge invariant},
  Phys.\ Lett.\ B {\bf 504} (2001) 131
  doi:10.1016/S0370-2693(01)00279-9
  [hep-th/0011140].

\bibitem{Buchbinder:2001at}
  I.~L.~Buchbinder and I.~B.~Samsonov,
  {\it Noncommutative N=2 supersymmetric theories in harmonic superspace},
  Grav.\ Cosmol.\  {\bf 8} (2002) 17
  [hep-th/0109130].

\bibitem{Hanada:2014ima}
  M.~Hanada and H.~Shimada,
  {\it On the continuity of the commutative limit of the 4d N=4 non-commutative super Yang –Mills theory},
  Nucl.\ Phys.\ B {\bf 892} (2015) 449
  doi:10.1016/j.nuclphysb.2015.01.016
  [arXiv:1410.4503 [hep-th]].
  
\bibitem{Collins:1984xc}
  J.~C.~Collins,
  {\it Renormalization: an introduction to renormalization, the renormalization group, and the operator product expansion},
  380 pgs., Cambridge University Press 1985, Cambridge monographs on mathematical physics 26, ISBN: 0521311772.

\bibitem{mathematica}
Wolfram Research, Inc.,
{\it Mathematica, Version 8.0,} Champaign, IL (2010).

\bibitem{xAct}
J. Martin-Garcia, {\it xAct,} \href{http://www.xact.es/}
{http://www.xact.es/}.

\end{thebibliography}
\end{document}